\documentclass{layout_paper}

\graphicspath{{Fig/}}

\begin{document}

\title{Synthetic Control Methods and Big Data\footnote{I would like to thank Fabio Canova, Christian Brinch, Ragnar Juelsrud, Hilde Bjørnland, Leif Anders Thorsrud and Gudmund Hermansen for valuable suggestions and comments.}}
\author{Daniel Kinn\footnote{Department of Economics, BI Norwegian Business School, Nydalsveien 37, N-0484 Oslo. E-mail: daniel.o.kinn@bi.no.}}
\date{February 2018}
\maketitle

\begin{abstract}
Many macroeconomic policy questions may be assessed in a case study framework, where the time series of a treated unit is compared to a counterfactual constructed from a large pool of control units. I provide a general framework for this setting, tailored to predict the counterfactual by minimizing a tradeoff between underfitting (bias) and overfitting (variance). The framework nests recently proposed structural and reduced form machine learning approaches as special cases. Furthermore, difference-in-differences with matching and the original synthetic control are restrictive cases of the framework, in general not minimizing the bias-variance objective. Using simulation studies I find that machine learning methods outperform traditional methods when the number of potential controls is large or the treated unit is substantially different from the controls. Equipped with a toolbox of approaches, I revisit a study on the effect of economic liberalisation on economic growth. I find effects for several countries where no effect was found in the original study. Furthermore, I inspect how a systematically important bank respond to increasing capital requirements by using a large pool of banks to estimate the counterfactual. Finally, I assess the effect of a changing product price on product sales using a novel scanner dataset.
\end{abstract}

\section{Introduction}\label{sec_retail_intro}
The increasing availability of high dimensional data offers opportunities for assessing the effects of macroeconomic policies in a case study format. Central to such studies is estimation of the counterfactual, the outcome of the time series of interest in absence of the policy change. For example, the effect of regulation on economic growth in a specific country may be estimated by predicting growth had the regulation not taken place. If a specific bank is exposed to new regulation (e.g. capital requirements or quantitative easing), its lending rate in absence of regulation can be predicted using a set of banks not exposed to the policy. The propensity to consume could be assessed by comparing a houshold receiving tax rebate to a counterfactual based on housholds not receiving such rebates. In all of these studies, the estimated causal effect of the policy is simply the difference between the actual time series and the predicted time series at each point in time where the policy takes place. The counterfactual is estimated using the pretreatment period, where none of the units recive treatment. This is known as the synthetic control approach, according to \cite{athey2017state} one of the most important innovations in the evaluation literature in the last fifteen years.\footnote{I use synthetic control as a general term referring to a counterfactual constructed from a weighted combination of control units. This terminology is due to \cite{abadie2010synthetic}. I will refer to their method as the ADH synthetic control.}

\par In any of the above applications, choosing the controls and their relative contribution is an important and difficult estimation problem. In many cases there is no economic theory to guide the selection process. Ad hoc selections may ``throw away'' important information leading to underfitted (biased) counterfactual estimates, while naively including the full set of controls may result in highly imprecise estimates due to overfitting (variance). In the event that the number of control units exceed the number of pretreatment periods, linear regression is not feasible. In that case selection must be done prior to estimation, or as part of the estimation procedure. With several hundred control units and possibly a relatively short estimation period prior to the intervention (e.g. yearly or quarterly data), this issue is predominant in many macroeconomic applications. 

\par In this paper I provide a general machine learning framework tailored to assess policy questions in high dimensional problems. The framework rests on the conditional independence assumption, see e.g.  \cite{angrist2008mostly}, aiming to balance the treated unit outcomes with a weighted combination of control unit outcomes. I argue that choosing a subset of the large set of controls should be done to balance the tradeoff between bias and variance of the synthetic control estimator. This is a key success factor in order to capture the true counterfactual in high dimensional problems, and by using the framework, I explain why machine learning methods are tailored for this purpose. In contrast, traditional methods such as difference-in-differences and the \cite{abadie2010synthetic} (ADH) synthetic control are shown to be restricted versions of the framework, in general not optimally balancing the bias-variance objective. 

\par The framework distinguish between two different approaches; either model the counterfactual directly as a linear combination of the control units (``reduced-form approach'') or explicitly model components of the treated unit such as e.g. the trend and use the control units to pick up any residual variation (``structural approach''). Several reduced form methods are discussed by \cite{doudchenko2016balancing} and \cite{brodersen2015inferring} introduced the first method that fit the structural approach.\footnote{\cite{doudchenko2016balancing} show that elastic net, difference-in-differences and ADH fit the reduced form framework.} 

\par Second, I assess the performance of machine learning and traditional approaches using simulation studies. I find that the structural approach may improve on the reduced approach when the noise is large. When trends are heterogeneous, both machine learning and ADH are able to pick the correct control units in high dimensional problems. However, machine learning can significantly improve estimates of the treatment effect when the treated unit lies outside the range of relevant control units, a setting where it is well known that ADH fails. Finally, I find that short estimation periods combined with substantial noise is difficult regardless of method chosen.

\par Third, I illustrate the use of machine learning on macroeconomic case studies. The first case is taken from \cite{billmeier2013assessing}, analysing the effect of economic liberalisation on economic growth. The country under study is compared to a counterfactual constructed from the worldwide pool of closed economies. Contrary to the orignal paper using the ADH synthetic control, I find effects of liberalisation for several countries where no effect was found in the previous study. In the second application I consider a bank reform studied by \cite{getz2016consequences}. I exploit the fact that one systematically important bank was given a higher capital requirement than other banks, and assess the effect of increasing capital requirements on capital ratio. Since the important bank is relatively low-capitalised compared to the rest of the banks in the sample, negative weights are needed to ensure a good fit. Machine learning methods significantly improve predictions compared to ADH in this setting, as ADH restricts weights to the $[0,1]$ interval. In the third application, I use a novel daily scanner dataset to assess the effect of price changes on sales. I observe sold quantities of a product in a specific store, and estimate a counterfactual based on several hundred control stores not exposed to price discounts in the relevant campaign period. I find large, robust changes in sales during the price campaigns, and find that the observed data is returning to the counterfactual estimates once the campaign ends.
 
\par In many econometric studies it is common to either apply difference-in-difference analysis or the ADH synthetic control. Based on the proposed framework in this paper, it can be argued that neither of these approaches are particularly well suited for estimating the counterfactual in the high dimensional setting.

\par In the case of difference-in-differences, the outcomes of the treated unit is compared to simple average outcome of a set of chosen control units. Selection of these controls is often carried out using an ad hoc procedure. \cite{ashenfelter1985using} base selection on the age distribution for treated and control units. Similarly, \cite{card1990impact} use aspects of the population and economic growth in the 1970s and 1980s to choose Atlanta, Los Angeles, Houston and Petersburg to represent the control for Miami. Central to these studies is that both ad hoc selection and equal contribution of each control are unnecessary restrictions. When the objective is to estimate the counterfactual, including too few units will lead to bias and including too many will lead to low precision. There is no guarantee that ad hoc selection followed by an equal contribution of the selected control units will balance these criteria. On the other hand, machine learning is tailored to choose controls and their coefficients to balance bias and variance, thus better capture the true data generating process for the outcomes.

\par A potential improvement of the above approach is to substitute ad hoc selection with matching.\footnote{In matching, a subset of controls are chosen to minimize some distance measure between the treated unit and the controls in terms of covariates and propensity scores, see e.g. \cite{abadie2006large,abadie2011bias} and \cite{diamond2013genetic}.} Matching is data-driven, and is promising in the high dimensional setting because a large pool of controls increases the chance of finding units similar to the treated unit. Still, the number of matches must be determined prior to the analysis, raising the same problem that the chosen number of matches must balance bias with precision of the counterfactual estimates. Even if the number of matches is chosen based on the data, it is not straightforward to allow the matched controls to contribute with different weights in subsequent difference-in-differences analysis. I show that matching followed by difference-in-differences is a restricted version of best subset selection, which is not optimally balancing bias and variance.

\par In the ADH synthetic control, non-negative weights between zero and one are assigned to each control store such that the weighed combination of control units is similar to the treated unit. \cite{peri2015labor} revisit the study by \cite{card1990impact}, allowing each of the cities to contribute with unequal weights to form a synthetic Miami. In applications, the synthetic control has proven to work well in relatively high dimensional problems, see e.g. \cite{billmeier2013assessing} and \cite{abadie2010synthetic}, reducing the number of relevant control units substantially compared to the total pool of units. However, as argued by \cite{athey2017state}, it is not obvious that restricting the weights to be positive and sum to one is the best option for constructing the counterfactual. I show that the ADH approach is equivalent to a restricted form of a machine learning approach known as Lasso, where the penalty parameter is exogenously given. Although this observation suggest that the ADH method is in fact a regularization approach, see e.g. \cite{hastie2011elements}, it uses a constant value for the penalty parameter. This observation motivates the use of the more flexible machine learning approaches, where the penalty is chosen based on the data to balance the bias-variance objective.

\par The main focus of the paper is the case of a single treated time series and a high dimensional pool of control time series, potentially larger than the number of time periods. In other words, any synthetic control approach outlined in this paper estimate treatment effects, but not average effects across multiple units.\footnote{\cite{xu2017generalized} generalizes the ADH method to allow for several treated units.} In addition, my focus is restricted to the outcomes of the observed units, without considering underlying covariates. There are two reasons for this restriction. First, recent studies point out that covariates are of less relevance in applications. Both \cite{athey2017state} and \cite{doudchenko2016balancing} argue that the outcomes tend to be substantially more important than covariates in terms of predictive power. Thus, minimizing the difference between treated outcomes and control outcomes prior to treatment is often sufficient to construct the synthetic control. Second, \cite{kaul2015synthetic} has shown that covariates become redundant when all lagged outcomes are included in the ADH approach.

\par The paper is structured as follows. The general framework for synthetic control methods in the high dimensional setting, identification strategies and estimation is outlined in Section \ref{sec_retail_framework}. Details on several machine learning methods and comparisons with traditional econometric methods are provided in Section \ref{sec_retail_methods}, and the performance of each method is assessed in simulation studies in Section \ref{sec_retail_sim}. Several examples of macroeconomic policy questions suited for the case study format are provided in Section \ref{sec_retail_applications}. 

\section{Framework}\label{sec_retail_framework}

\subsection{Potential outcomes and notation}\label{sec_retail_framework_notation}
Consider $J+1$ units observed in time periods $t=1,\hdots,T$. Let $j=0$ index the treated unit and $j=1,\hdots,J$ index the control units. In the notation of e.g. \cite{imbens2015causal}, the potential outcomes for unit $j$ at time $t$ are given by $(y_{jt}(1),y_{jt}(0))$, representing the outcome with and without the treatment, respectively. Furthermore, the actual observed outcome for unit $j$ at time $t$ is denoted by $y_{jt}$ and is given in terms of the potential outcomes as follows
\begin{equation}\label{eq_retail_framework_notation_y}
y_{0t} =
\begin{cases} 
y_{0t}(0) & \text{ for } t=1,\hdots,T_0 \\
y_{0t}(1) & \text{ for } t=T_0+1,\hdots,T
\end{cases} \hspace{1cm} y_{jt} = y_{jt}(0) \text{ } \text{ }\text{ for } t=1,\hdots,T
\end{equation}
The observed outcome for the treated unit is the potential outcome without treatment in the first $T_0$ time periods and the potential outcome with treatment in the remaining periods. For the control units, observed outcome is equal to the potential outcome without treatment for all time periods and all control units, $j=1,\hdots,J$. The main objective of this paper is the \emph{treatment effect on the treated}, which is given by
\begin{equation}\label{eq_retail_framework_notation_att}
\tau_h = y_{0h}(1) - y_{0h}(0) \text{ for } h=T_0+1,\hdots,T
\end{equation}
The counterfactual $y_{0h}(0)$ is never observed for the relevant time period, so the treatment effect (\ref{eq_retail_framework_notation_att}) is not observed in the data. Estimation require that the researcher imputes a value for the missing counterfactual.

\subsection{Identification}\label{sec_retail_framework_id}
I use the conditional independence assumption for identification of the treatment effect in (\ref{eq_retail_framework_notation_att}).\footnote{The conditional independence assumption is discussed in e.g. \cite{angrist2008mostly} and \cite{o2016estimating} relate it to synthetic control studies.} Let $d_{jt}$ be an indicator variable equal to 1 if unit $j$ is treated at time $t$ and zero otherwise. Furthermore, let the potential outcomes be given by some function of an unobserved common effect $\gamma_t$, unobserved time-varying unit specific effects $\phi_{jt}$ and a set of covariates $\z_{jt}$, i.e. $y_{jt}(0) = f(\gamma_t,\phi_{jt},\z_{jt})$. By definition, potential outcomes without treatment are independent of treatment assignment conditional on $\gamma_t$, $\phi_{jt}$ and $\z_{jt}$, commonly written as $y_{jt}(0) \independent d_{jt} | \gamma_t,\phi_{jt},\z_{jt}$. The conditional independence assumption that is used in this paper is
\begin{equation}\label{eq_retail_framework_id_conditional}
y_{jh}(0) \independent d_{jt} |\y_{j},\z_{jt}
\end{equation}
where $h>T_0$. In words, the conditional independence assumption uses the full set of pretreatment outcomes $\y_j=(y_{j1},\hdots,y_{jT_0})$ to proxy for the unobserved confounding factors $\gamma_t$ and $\phi_{jt}$. This assumption is likely to remove bias if the unobserved confounders are strongly correlated with the outcomes and the noise is small relative to the length of the pretreatment period. \cite{abadie2010synthetic} argue that since $y_{jh}(0)$ is influenced by both unobserved and observed confounders, units with similar values of the outcomes in the pretreatment period, are also likely to have similar values of the time-varying unobserved confounding factors. This suggest that if the synthetic control succeed in balancing the treated unit outcome with a weighted combination of control units in the pretreatment period, the time-varying confounding factor component will also be balanced. An alternative assumption is to impose that the unobserved confounding factor is constant in time, $\phi_{jt}=\phi_j$. This restriction is known as the parallel trend assumption, and stated formally as $y_{jt}(0) \independent d_{jt} | \gamma_t,\phi_{j},\z_{jt}$. When unobserved confounders are constant in time, difference-in-differences and fixed effects estimators may be applied.

\subsection{Estimation and inference} \label{sec_retail_framework_est}
Since $y_{0h}(0)$ is never observed in (\ref{eq_retail_framework_notation_att}), all estimates of $\tau_{h}$ require imputation of the missing value. In general the researcher has three different sources of information available for the imputation task; i) the treated unit before undergoing treatment, ii) the control units in the pretreatment period and iii) the control units after treatment has been assigned. Obtaining estimates of $y_{0h}(0)$ may be formulated as a two-stage prediction problem. The first stage is to estimate a relationship between the treated unit and the control units by using the data from the pretreatment period and second, predict the counterfactual of interest by using the estimated model on the treatment period. Formally, specify a statistical model relating the treated unit to the control units
\begin{equation}\label{eq_retail_framework_est_model}
y_{0t} = f_{\boldsymbol{\theta}}(\y_t) + \varepsilon_t
\end{equation}
where $\varepsilon_t$ is Gaussian with mean zero and variance $\sigma^2$, $f_{\boldsymbol{\theta}}(.)$ is some function parametrized by $\boldsymbol{\theta}$ and $\y_t$ denotes the vector of outcomes for the control units.\footnote{By a slight abuse of notation, I refer to $\y_j=(y_{j1},\hdots,y_{jT_0})$ as the $T_0\times 1$ vector of pretreatment outcomes for unit $j$, while $\y_t=(y_{1t},\hdots,y_{Jt})$ refers to the $J\times 1$ vector of outcomes at time $t$.} Let $f_{\hat{\boldsymbol{\theta}}}(.)$ be the function obtained by estimating (\ref{eq_retail_framework_est_model}) on the pretreatment period $t=1,\hdots,T_0$. In the terminology of \cite{abadie2010synthetic}, $f_{\hat{\boldsymbol{\theta}}}(.)$ represents a synthetic control, i.e. a synthetic version of the treated unit constructed from the control units.

\par The second stage of the prediction problem is to predict $y_{0h}(0)$ for some $h>T_0$ by using the synthetic control and the observed vector of outcomes for the control units. Imputed values are obtained from $\hat{y}_{0h}(0) = f_{\hat{\boldsymbol{\theta}}}(\y_h)$ and the treatment effect estimate is $\hat{\tau}_h=y_{0h}(1)-\hat{y}_{0h}(0)$. To evaluate how well the synthetic control imputes the missing value I suggest to assess the tradeoff between bias and variance
\begin{equation}\label{eq_retail_framework_est_mse}
L=\E[(y_{0h}(0)-\hat{y}_{0h}(0))^2] = \underbrace{\left(y_{0h}(0)-\E[\hat{y}_{0h}(0)]\right)^2}_\text{squared bias}+ \underbrace{\V[\hat{y}_{0h}(0)]}_\text{variance} + \sigma^2
\end{equation}
where $(y_{0h}(0),\y_h)$ is considered fixed and the expectation is with respect to the training set that produced the synthetic control, conditional on the treatment assignment. The bias-variance tradeoff (\ref{eq_retail_framework_est_mse}) is useful for assessing the quality of the counterfactual predictions in the high dimensional setting for three reasons.

\par
\begin{figure}
\centering
\begin{minipage}{1\textwidth}
\begin{subfigure}{0.45\textwidth}
\includegraphics[width=\linewidth]{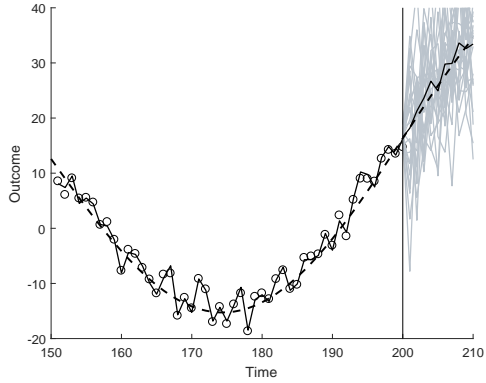}
\caption{OLS}
\label{fig_retail_intro_ols}
\end{subfigure}\hspace*{\fill}
\begin{subfigure}{0.45\textwidth}
\includegraphics[width=\linewidth]{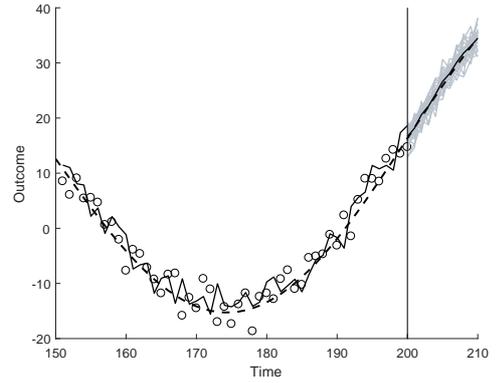}
\caption{ADH}
\label{fig_retail_intro_sc}
\end{subfigure}
\medskip
\begin{subfigure}{0.45\textwidth}
\includegraphics[width=\linewidth]{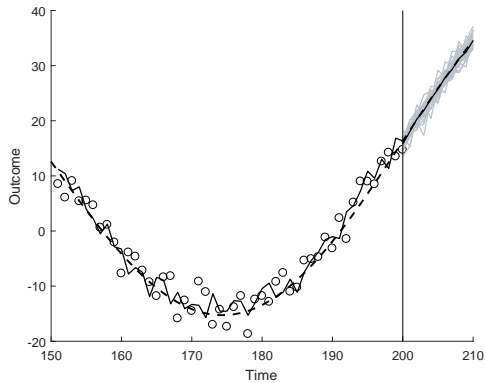}
\caption{LASSO}
\label{fig_retail_intro_reg}
\end{subfigure}\hspace*{\fill}
\begin{subfigure}{0.45\textwidth}
\includegraphics[width=\linewidth]{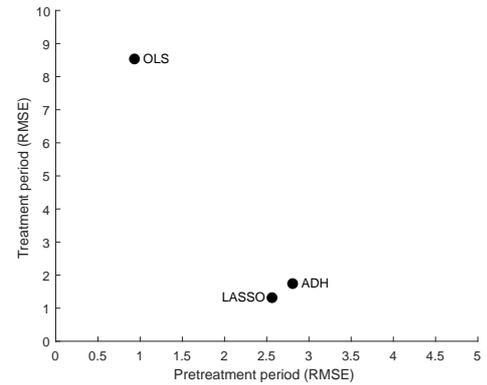}
\caption{RMSE}
\label{fig_retail_intro_errors}
\end{subfigure}\\
\caption{\footnotesize Synthetic controls in high dimensions. Figures (a)-(c) show synthetic control estimators $\hat{y}_{0t}(0)$ (thick line), the true potential outcome $y_{0t}(0)$ (stipulated), one data draw (dots) and 25 predictions based on different datasets (gray lines, thick line is mean). Figure (d) report in and out of sample root mean squared errors (RMSE). The data generating process is $y_{0t} = \sum_{j=1}^J \omega_j y_{jt} + \varepsilon_t$ with $T_0=200$ and $J=180$ where $\omega_1=0.3$, $\omega_2=0.7$, $\omega_3 = -0.01$ and $\omega_j=0$ for $j=4,\hdots,180$. The three first control units follow one specific trend and seasonality pattern plus noise, while the remaining $177$ control units follow a different trend and seasonality pattern.} 
\label{fig_retail_intro}
\end{minipage}
\end{figure}

 First, when the number of control units is large, controlling the variance of the estimator is important. In short, synthetic controls based on too many control units may yield low bias, but are likely to exhibit high variance (``overfitting''). Figure \ref{fig_retail_intro} illustrates this phenomenon in a simulated example based on $T_0=200$ pretreatment periods and $J=180$ control units, where only the three first control units are relevant for predicting the treated unit according to the data generating process. In general I focus on the setting where $J$ is larger than $T_0$, but the case presented here is useful for making a comparison to standard linear regression.\footnote{It is well known that OLS does not work in a setting where the number of variables $J$ is larger than the number of observations $T_0$, see \cite{bajari2015machine} for a related discussion.} Figure \ref{fig_retail_intro_ols} reports the results for OLS where all 180 control units are included without any restrictions. The large set of parameters in this specification make the synthetic control estimator highly flexible, providing a close match to the treated unit (i.e. the dotted observations) in the pretreatment period. However, since OLS almost perfectly captures the data for the treated unit, it is also fitting noise, leading to counterfactual predictions that are unstable and unreliable. This is illustrated from 25 draws of the data and the average prediction, shown by the light red and dark red lines, respectively. Figure \ref{fig_retail_intro_sc} show the same setting for the ADH synthetic control where weights are restricted to be non-negative and sum to one, see Section \ref{sec_retail_methods} for details. The figure suggest that ADH is less subject to overfitting. Indeed, an average of only 3.9 control units receive positive weights. Compared to OLS, the estimated function is less dependent on individual observations, closer to the data generating process and more precise. This improvement is due to the weight restriction - if none of the weights are allowed to exceed one, weights are constrained not to largely follow individual observations and variance will be bounded. However, the weight restriction is somewhat ad hoc, it is not obvious that weights in the range $[0,1]$ optimally controls the variance of the predictions. Although this study suggest that ADH is capable of performing control unit selection in high dimensional problems, it is in many cases not the choice that minimizes the bias-variance tradeoff. To see this, consider Lasso in Figure \ref{fig_retail_intro_reg}. Lasso is a specific machine learning algorithm where the sum of the absolute value of the weights is bounded by some threshold. Based on the data, the threshold is chosen to empirically minimize the bias-variance tradeoff. Although not easily seen from the figures, Lasso reduces the variance of the counterfactual predictions by 5\% in this particular simulation study. The conclusion is that the data driven selection of the threshold value is more efficient than bounding the weights to the $[0,1]$ interval in this case. I show that ADH is a special case of a constrained Lasso in Section \ref{sec_retail_methods}, where the threshold parameter is fixed.

\par Second, although weight restrictions such as in ADH and Lasso may lead to relatively precise estimators, they also lead to bias (``underfitting''). Without regularization (i.e. OLS), estimates vary severely, but are unbiased. However, when the number of control units becomes larger than the number of treatment periods this alternative is no longer an option, and some form of regularization must be accepted. This will introduce bias, but as long as the increased bias does not outweigh the reduction in variance, the countefactual predictions will improve. Indeed, ADH is biased due to its weight restriction, not capturing the fact that the third weight is negative in the data generating process. Lasso allows for negative weights, reducing bias by almost 71\% in this particular case. Taken together, improvement in both bias and variance leads to the situation in Figure \ref{fig_retail_intro_errors}, where each estimator is plotted in a diagram where the root mean squared error (RMSE) of the pretreatment period (horizontal axis) is plotted against the RMSE of the treatment period. OLS has a low pretreatment error due to overfitting, resulting in poor out of sample performance. By allowing a larger in sample error, both ADH and Lasso performs better out of sample, with Lasso performing the best in this case.

\par A third observation is that the identification strategy may be assessed from the bias-variance framework. Observing the situation in the pretreatment period of Figure \ref{fig_retail_intro_ols} must not be interpreted as something close to a ``perfect match'', where the synthetic control closely follows the treated unit. In other words, one may be lead to believe that the synthetic control is capturing the unobserved time-varying confounders and covariates due to the good match, when in reality it is due to a dimensionality problem. Regularization will remove some of the overfitting problem, capturing the data generating process to a larger extent.

\par It is important to note that we can never observe (\ref{eq_retail_framework_est_mse}) in practical applications because $y_{0h}(0)$ is unobserved. The bias-variance decomposition is generally only available when the true data generating process is known, but $L$ may be estimated based on e.g. cross-validation, see Section \ref{sec_retail_methods_reduced}.

\par The simulation study above is conditional in treatment already being assigned. Inference in case studies is not straightforward due to the lack of randomization and that probabilistic sampling is not used to select sample units, see \cite{abadie2015comparative}. Inference is usually conducted by using some form of placebo study. When a high dimensional control pool is available, I find the inference procedure which the authors refer to as ``in-space placebos'' as the most appropriate. The idea is to estimate the counterfactual for each control unit $j>0$ by using the remaining $J-1$ control units as controls. This creates a distribution of treatment effect estimates that can be compared with the original estimate. Concluding that the policy has an effect would require that the treatment effect estimate lies outside or in the tails of the distribution of placebo effects.

\section{Synthetic Control Methods}\label{sec_retail_methods}
In this section I relate both existing and novel methods for synthetic control estimation to the general framework in Section \ref{sec_retail_framework}. I classify a method as a ``reduced form approach'' if a linear model with weight restrictions (i.e. regularization) is used to obtain the synthetic control estimator. If the method explicitly models the structure of the treated unit, the term ``structural approach'' is used. 

\subsection{Reduced form approaches}\label{sec_retail_methods_reduced}
Most synthetic control estimators encountered in the literature assumes a linear model for the relationship between the treated unit and the control units
\begin{equation}\label{eq_retail_methods_linear}
f_{\alpha,\boldsymbol{\omega}}(\y_t) = \alpha + \boldsymbol{\omega}'\y_t
\end{equation}
where the parameters of the model are $(\alpha,\boldsymbol{\omega})=(\alpha,\omega_1,\hdots,\omega_J)$. The first stage of the prediction problem is to estimate the parameters by minimizing squared loss subject to a complexity restriction
\begin{equation}\label{eq_retail_methods_erm}
(\hat{\alpha},\hat{\boldsymbol{\omega}}) = \argmin_{\alpha,\boldsymbol{\omega}\in\mathcal{W}} \left\{\frac{1}{T_0} \sum_{t=1}^{T_0}(y_{0t}-\alpha-\boldsymbol{\omega}'\y_t)^2\right\}
\end{equation}
where $\mathcal{W}$ is the weight set. Below I show that all reduced form approaches differ only by the way $\mathcal{W}$ is specified. Each method is characterised by some penalty parameter $\lambda$ associated with $\mathcal{W}$. For the machine learning methods $\lambda$ is chosen based on the data in order to minimize the tradeoff between bias and variance in (\ref{eq_retail_framework_est_mse}), thereby in theory optimizing predictive performance. In contrast, $\lambda$ is fixed in the traditional approaches. 

\par The data-driven choice of $\lambda$ is determined using cross-validation. Divide the set of preintervention periods into $k$ subsets (``folds'') by random. Typical choices are $k=5$ and $k=10$ for a sufficiently large number of preintervention periods, but for smaller datasets, the leave-one-out alternative, $k=T_0$, is often preferred. For a given $\lambda$, solve (\ref{eq_retail_methods_erm}) by using all but the first fold and compute the mean squared error on the left out fold. Repeat this procedure $k$ times, each time leaving out a new fold, so that all folds are left out once each. The optimal value $\lambda^*$ is chosen to minimize the average of the cross-validated errors, $\hat{L}(\lambda)$, and (\ref{eq_retail_methods_erm}) is solved with the value $\lambda^*$ embedded in $\mathcal{W}$.\\

\noindent \textbf{Principal Component Regression}. In Principal Component Regression (PCR), only the first $\lambda$ principal components of the control units are used as independent variables in a standard regression. Including only the first $\lambda$ principal components captures the directions of largest variance in the data, at the same time as avoiding the dimensionality problem. Let $\Y$ denote the $T_0\times J$ matrix of outcomes for the control units in the pretreatment period and let $\mathbf{V}$ denote the $J\times J$ matrix where the principal components (eigenvectors) of $\Y$ are stored in each column. Furthermore, let $\mathbf{V}_\lambda$ denote the $J\times \lambda$ matrix where only the first $\lambda$ columns (principal components) are included. The control units are projected onto the lower dimensional space using $\Y_\lambda=\Y\mathbf{V}_\lambda$, and a $\lambda\times 1$ vector of parameters is obtained from the regression $\boldsymbol{\omega}_\lambda = (\Y_\lambda'\Y_\lambda)^{-1}\Y_\lambda'\y$, where $\y$ is the vector of pretreatment outcomes for the treated unit. The PCR estimator transfers the low dimensional parameter vector back to weights for each control unit using $\hat{\boldsymbol{\omega}}^\text{PCR} = \mathbf{V}_\lambda\boldsymbol{\omega}_\lambda$. To see how this procedure actually performs regularization, note that restricting the solution to the top $\lambda$ principal components implies that the solution will be orthogonal to the bottom $J-\lambda$ principal components. Thus, we may relate PCR to the general model (\ref{eq_retail_methods_erm}) by letting $\alpha=0$ and defining $\mathcal{W}^\text{PCR}=\{\boldsymbol{\omega}\in\R^J : \mathbf{V}_{-\lambda}'\boldsymbol{\omega}=\zero \}$, where $\mathbf{V}_{-\lambda}$ denotes the bottom $J-\lambda$ eigenvectors. \cite{jolliffe1982note} argues in a general context that there is no guarantee that the chosen principal components provide a good match with the treated unit. The optimal number of principal components $\lambda^*$ can be chosen based on cross-validation.\\

\noindent \textbf{Lasso}. The Least Absolute Shrinkage and Selection Operator (LASSO) performs both variable selection and estimation, see e.g. \cite{hastie2011elements}. A generalisation known as Elastic Net was applied to synthetic control estimation by \cite{doudchenko2016balancing}.\footnote{The elastic net penalty can be obtained from problem (\ref{eq_retail_methods_erm}) by defining the weight set as $\mathcal{W}^\text{EN} = \{\boldsymbol{\omega}\in\R^J: \gamma \sum_{j=1}^J |\omega_j| + (1-\gamma)\sum_{j=1}^J \omega_j^2 \leq c\}$, with Lasso ($\gamma=1$) by \cite{tibshirani1996regression} and Ridge regression ($\gamma=0$) by \cite{hoerl1970ridge} as special cases.} However, since obtaining a sparse set of control units is important in the high dimensional setting, I restrict the discussion in this paper to Lasso. In terms of the general model (\ref{eq_retail_methods_erm}), Lasso may be obtained by defining the parameter set as $\mathcal{W}^\text{LASSO} = \{\boldsymbol{\omega}\in\R^J: \sum_{j=1}^J |\omega_j|\leq c\}$, where $c$ is some constant. In that case (\ref{eq_retail_methods_erm}) has an alternative formulation
\begin{equation}\label{eq_retail_methods_lasso}
(\hat{\alpha}^\text{LASSO},\hat{\boldsymbol{\omega}}^\text{LASSO}) = \argmin_{\alpha,\boldsymbol{\omega}}\text{ }\left\{ \frac{1}{T_0}\sum_{t=1}^{T_0} (y_{0t}-\alpha-\boldsymbol{\omega}'\y_t)^2 + \lambda \sum_{j=1}^J|\omega_j|\right\}
\end{equation}
where $\lambda$ denote the Lagrange multiplier associated with the parameter restriction $\mathcal{W}^\text{LASSO}$. Like regression, Lasso aims at minimizing the sum of squared residuals, but unlike regression, it impose a penalty on the coefficients, effectively shrinking several of the control unit weight estimates to zero. In practice, the Lasso synthetic control estimator often reduces to a function of only a few control units even when $J$ is large. Solving (\ref{eq_retail_methods_lasso}) with the cross-validated penalty $\lambda^*$ gives the Lasso estimator $(\hat{\alpha}^\text{LASSO},\hat{\boldsymbol{\omega}}^\text{LASSO})$.\\

\noindent \textbf{The ADH synthetic control}. The ADH synthetic control method was first applied by \cite{abadie2003economic} to estimate the economic costs of conflicts by using the Basque Country as a case study. The method was later formalised in \cite{abadie2010synthetic}. 

\par ADH is related to the general problem (\ref{eq_retail_methods_erm}) by restricting weights to be positive and sum to one, $\mathcal{W}^\text{ADH}=\{\boldsymbol{\omega}\in\R^J:\boldsymbol{\omega}\geq 0, \mathbf{1}'\boldsymbol{\omega} = 1\}$, where $\mathbf{1}$ is an $J\times 1$ vector of ones.\footnote{The definition used of the ADH estimator in this paper is a simplified version of the original ADH estimator, which in general includes covariates and no intercept.} This restriction makes ADH intuitively appealing because it is as simple weighted combination of the control units. Furthermore, \cite{athey2017state} point out that ADH tend to assign positive weights to only a subset of units, thereby performing variable selection. I show below that the ADH estimator is a restricted version of Lasso, and therefore in many cases work well for control unit selection. However, it is not obvious that the restrictions imposed by ADH are necessary. 

\par To see the connection between ADH and Lasso, define a restricted version of Lasso as $\mathcal{W}^\text{RLASSO} = \{\boldsymbol{\omega}\in\R^J:\sum_{j=1}^J |\omega_j| \leq 1, \mathbf{1}'\boldsymbol{\omega}=1\}$, where the constant of the penalty is fixed, $c=1$, and an additional summation constraint is imposed. Since $\mathcal{W}^\text{ADH}=\mathcal{W}^\text{RLASSO}$, the ADH estimator is equal to a restricted version of Lasso, where coefficients sum to one and $c=1$ implies a fixed value for $\lambda$. ADH thus perform regularization, but there is no garantee that the fixed value of the penalty minimize the bias-variance tradeoff. In line with the above argument, \cite{athey2017state} argue that the ADH restrictions might hurt predictive performance. \cite{abadie2010synthetic} emphasise another shortcoming of the ADH restrictions, namely that the method should be avoided in cases where the treated unit lies outside the distribution of control units. In that case, no convex combination of control units can construct the treated unit and the counterfactual estimates could be severely biased.\\

\noindent \textbf{Difference-in-differences with matching}. The difference-in-differences with matching (MDD) approach is a well known method for estimating the average treatment effect of the treated, applied by empirical researchers in several forms. In the setting here, the difference between the treated unit outcome and an average of the outcomes for a set of $\lambda<J$ control units is adjusted by the difference between the treated unit and the same group of controls prior to the treatment. Matching esitmators may be used to select the subset of $\lambda$ control units based on the distance between the treated unit outcome and the control unit outcomes in the pretreatment period, see e.g. \cite{o2016estimating}. The idea is that the matching procedure may remove biases from unit-specific, time-varying confounders, while the use of difference-in-differences may remove any remaining bias in time-invariant confounders. 

\par Let $\mathcal{M}$ denote the index set of the $\lambda$ chosen control units based on any matching procedure. In terms of the general problem (\ref{eq_retail_methods_erm}), the weight set is $\mathcal{W}^\text{MDD} = \{\boldsymbol{\omega}\in\R^J : \omega_j = 1/\lambda \text{ for all } j\in \mathcal{M} \text{ and } \omega_j=0 \text{ for all } j\not\in\mathcal{M}\}$. In other words, an equal weight is distributed to all included (matched) control units, while the remaining units receives weights of zero. Solving (\ref{eq_retail_methods_erm}) with respect to $\alpha$ gives the intercept $\hat{\alpha}^\text{MDD} = \frac{1}{T_0}\sum_{t=1}^{T_0}\left(y_{0t} - \frac{1}{J}\sum_{j=1}^J y_{jt}\right)$. In the case of a single control and a single time period both before and after the treatment, it is straigthforward to show that $\hat{\tau}_2^\text{MDD} = (y_{02}-y_{01})-(y_{12}-y_{11})$, which is the formula discussed in e.g. \cite{angrist2008mostly}.

\par MDD is a special case of another variable selection method known as best subset selection. To see this connection, define the best subset $\mathcal{W}^\text{BS} = \{\boldsymbol{\omega}\in\R^J : \sum_{j=1}^J I(\omega_j\neq 0) \leq \lambda\}$, where $I(\omega_j\neq 0)$ is the indicator function taking the value 1 if unit $j$ is included and zero otherwise. In this case, both the parameters $\boldsymbol{\omega}$ and the number of chosen control units $c$ can be determined from the data. In contrast, MDD is more restrictive, as it fixes the threshold to some prespecified number of matches and assigns the same weight to all of the included control units. Even if the matching algorithm chooses $\lambda$ based on cross-validation, it will still be the case that all of the included units receive the same weight. To summarise, MDD is not in general regularizing the coefficients in a way that is consistent with the bias-variance tradeoff. Best subset selection can be used to illustrate this problem, but I will not apply it to synthetic control estimation due to its close connection to Lasso.\footnote{\cite{james2013introduction} argues that solving the best subset specification is infeasible when $J$ is large, but that Lasso is a related, computationally feasible alternative.} \\

\noindent \textbf{Spatial econometrics}. Another way of handling the dimensionality problem is by explicitly computing the weights based on distance measures. Typically such measures are based on geographical proximity, but in principle any variable may be used for computing the distance. This is somewhat similar to matching, but a key difference is that there is no pre-defined number of matches and weights are not constant and equal for the included units. Specifically, weights could be determined based on the distance $a_j$ between the treated unit and control unit $j$. A common specification is $\hat{\omega}_j^\text{S} = \frac{\exp(-\zeta a_j)}{\sum_{j=1}^J \exp(-\zeta a_j)}$ if $a_j < c$, and $\hat{\omega}_j^\text{S}=0$ otherwise, where $\zeta>0$ is the distance decay parameter and $c$ is some distance threshold. In terms of the general framework, let $\alpha=0$ and use the above weights directly in the linear model (\ref{eq_retail_methods_linear}). This specification let weights exponentially decline as control units are further away from the treated unit in terms of the distance $a_j$. It is difficult to see how the conditional independence assumption is satisfied in this case, as there is no attempt to balance the treated unit with the control units. However, it might serve as a benchmark, and give intuition for why some weights are considered important by other estimation methods.

\subsection{Structural approaches}\label{sec_retail_methods_structural}
The structural approach allows components of the treated unit to be modelled explicitly and use control units used to capture any remaining variation. Formally, let $f_{\boldsymbol{\theta}}(\y_t)$ in (\ref{eq_retail_framework_est_model}) be given by
\begin{equation}\label{eq_retail_methods_structural}
f_{\boldsymbol{\theta}}(\y_t) = g(\xi_t,\iota_t,\psi_t) + \boldsymbol{\omega}'\y_t
\end{equation}
where $g(.)$ is some function, $\xi_t$ denotes the trend of the treated unit, $\iota_t$ is the cycle and $\psi_t$ is the seasonality. Such systems fit the state space framework, see e.g. \cite{durbin2012time}, where (\ref{eq_retail_methods_structural}) is the measurement equation and suitable transition equations are defined for the structural components. Consider e.g. the case where only the trend is modelled, $g(\xi_t,\iota_t,\psi_t)=\xi_t$. It is implicitly assumed that the control units $\y_t$ will capture the remaining cyclical and seasonality patterns. In any case, if a component is modelled explicitly, this component should be removed from the control units before estimation. The flexibility of this approach can be beneficial in high dimensional problems. Intuitively, when components of the treated unit are modelled explicitly, we are ``narrowing down'' what we are looking for in the large pool of control units.\\

\noindent \textbf{Bayesian Structural Time Series}. The Bayesian Structural Time Series (BSTS) approach was first proposed by \cite{varian2014big} and applied to a synthetic control setting by \cite{brodersen2015inferring}. BSTS includes a variety of different models where $g(.)$ is linear and the particular version discussed here focus on the trend. Define the measurement equation of the state space model using (\ref{eq_retail_framework_est_model}) and (\ref{eq_retail_methods_structural}) with $g(\xi_t,\iota_t,\psi_t)=\xi_t$. The state space model is
\begin{equation}\label{eq_retail_methods_bsts_model}
\begin{aligned}
y_{0t} &= \xi_t + \boldsymbol{\omega}'\y_t + \varepsilon_t \\
\xi_t &= \xi_{t-1} + \nu_{t-1} + \eta_{1t}\\
\nu_t &= \nu_{t-1} + \eta_{2t}
\end{aligned}
\end{equation} 
where $\eta_{1t}$ and $\eta_{2t}$ are mutually uncorrelated white noise disturbances with mean zero and variances $\sigma_1^2$ and $\sigma_2^2$, respectively. The trend is assumed to be linear with stochastic level and slope parameters.\footnote{This is described in \cite{harvey1990forecasting}. Define the linear trend, $\xi_t=\mu + \nu t$. Let $\mu$ and $\nu$ be stochastic. Then we may write $\xi_t=\xi_{t-1}+\nu$ initialised at $\xi_0=\mu$.} The noise $\eta_{1t}$ allow the level of the trend to shift stochastically, while $\eta_{2t}$ allow the slope of the trend to change over time. In the special case that $\sigma_1^2=\sigma_2^2=0$ we are left with the deterministic trend case. The model is estimated on the pretreatment period in a Bayesian framework using Gibbs sampling. Once converged, each Gibbs sampling trajectory may be iterated forward using the estimated state variables and parameters to construct $\hat{y}_{0h}(0)$ for $h>T_0$.

\par It is convenient to work with a more compact version of (\ref{eq_retail_methods_bsts_model}). Specifically, define the state vector as $\x_t = \begin{bmatrix} \xi_t & \nu_t & 1 \end{bmatrix}'$, and gather the model components in the following set of matrices and vectors
$$\mathbf{Z}_t = \begin{bmatrix}
1 & 0 & \boldsymbol{\omega}'\y_t
\end{bmatrix} \text{, } \mathbf{T} = \begin{bmatrix}
1 & 1 & 0\\
0 & 1 & 0\\
0 & 0 & 1
\end{bmatrix} \text{, } \mathbf{R} = \begin{bmatrix}
1 & 0 \\
0 & 1 \\
0 & 0
\end{bmatrix} \text{, } \boldsymbol{\eta}_t = \begin{bmatrix}
\eta_{1t} \\ \eta_{2t} \end{bmatrix} \text{, } \mathbf{Q} = \begin{bmatrix}
\sigma_1^2 & 0 \\0 & \sigma_2^2
\end{bmatrix}$$
The linear state space model
\begin{eqnarray}\label{eq_retail_methods_bsts_ssm}
\begin{aligned}
y_{0t}&=\mathbf{Z}_t\x_t+\varepsilon_t \text{ where } \varepsilon_t \sim \N(0,\sigma^2)\\ 
\x_{t+1}&=\mathbf{T}\x_{t}+\mathbf{R}\boldsymbol{\eta}_t  \text{ where }\boldsymbol{\eta}_t\sim \N(\zero,\mathbf{Q})
\end{aligned}
\end{eqnarray}
is referred to as the local linear trend model by \cite{durbin2012time}, but with one notable exception; the first equation includes a set of contemperanous covariates. Estimation of the system in (\ref{eq_retail_methods_bsts_ssm}) requires obtaining an estimate of the state variables $\x=\{\x_1,\hdots,\x_{T_0}\}$ and the set of parameters collectively denoted by $\boldsymbol{\theta}=(\boldsymbol{\omega},\sigma^2,\sigma_1^2,\sigma_2^2)$. If not for the regression component $\boldsymbol{\omega}'\y_t$ of the measurement equation, estimation of both the states and parameters would proceed in a straightforward fashion by using the Kalman filter and smoother. However, in the presence of the high dimensional regression component estimation can be carried out in a Bayesian framework combining a ``spike and slab'' prior with a simulation smoother as descried in \cite{brodersen2015inferring}. Estimation will in general depend on what kind of structural model that is chosen, and I provide the estimation details for the specific model (\ref{eq_retail_methods_bsts_model}) below. 

\par Denote the observed data in the preintervention period by $\y=\{y_{0t}\}_{t=1}^{T_0}$. First, a \emph{simulation smoother} is constructed to draw from the posterior of state variables given parameters and data, $p(\x|\boldsymbol{\theta},\y)$. Second, draws of all parameters in $\boldsymbol{\theta}$ are made conditional on the states and data, $p(\boldsymbol{\theta}|\x,\y)$ by explicitly incorporating priors $p(\boldsymbol{\theta})$. By using Gibbs sampling, draws from the first and second step converge to the posterior of interest $p(\boldsymbol{\theta},\x|\y)$. Third, predictions of the counterfactual are drawn from $p(\boldsymbol{\theta},\x|\y)$. Each of these steps are discussed below.\\

\noindent \emph{1. State posterior $p(\x|\boldsymbol{\theta},\y)$}. The draw from the posterior state distribution can be made using several different algorithms. I apply the simulation smoother discussed in \cite{durbin2012time}. First, run the Kalman filter and smoother given the data $\y$ and the parameters $\boldsymbol{\theta}$ to obtain the smoothed states $\hat{\x}_t$. Second, simulate a new dataset $\y^*$ and $\x^*$ from (\ref{eq_retail_methods_bsts_ssm}) and compute the Kalman smoothed states as $\hat{\x}_t^*$ based on $\y^*$. A draw from the posterior of interest is now characterised by $\tilde{\x}_t = \x_t^* - \hat{\x}_t^* + \hat{\x}_t$.\\

\noindent \emph{2. Parameter posterior $p(\boldsymbol{\theta}|\x,\y)$}. Given the state estimates from Step 1, two of the parameters in $\boldsymbol{\theta}$ are independent and easy to draw given a prior. Denote the likelihood of the model by $p(\y,\x|\boldsymbol{\theta})$ and the posterior by $p(\boldsymbol{\theta}|\x,\y) \propto p(\y,\x|\boldsymbol{\theta})p(\boldsymbol{\theta})$, where the prior may be decomposed as $p(\boldsymbol{\theta}) = p(\sigma_1^2)p(\sigma_2^2)p(\boldsymbol{\omega},\sigma^2)$. For the transition variances the draw is straightforward. Since the two transition equations are independent given the state $\x$, we may write their corresponding distributions as $p(\y,\x|\sigma_1^2)\sim\N(\xi_t+\nu_t,\sigma_1^2)$ and $p(\y,\x|\sigma_2^2)\sim\N(\nu_t,\sigma_2^2)$. By assuming indepenent inverse Gamma priors $p(\sigma_1^2)\sim\mathcal{G}^{-1} (a_0,c_0)$ and $p(\sigma_2^2)\sim\mathcal{G}^{-1}(d_0,e_0)$ the posteriors are also inverse Gamma
\begin{equation}
p(\sigma_1^2|\x,\y) \sim \mathcal{G}^{-1}(a_1,c_1) \text{ and } p(\sigma_2^2|\x,\y) \sim \mathcal{G}^{-1}(d_1,e_1)
\end{equation}
with parameters given by $a_1=a_0+\frac{1}{2}T_0$ and $c_1 = c_0 + \frac{1}{2}\sum_{t=2}^{T_0}(\xi_t - \xi_{t-1} - \nu_{t-1})^2$ for the level component and $d_1=d_0+\frac{1}{2}T_0$ and $e_1 = e_0 +\frac{1}{2}\sum_{t=2}^{T_0}(\nu_t-\nu_{t-1})^2$ for the slope component. These expressions are easily computed since $\xi_t$ and $\nu_t$ are available from Step 1. The remaining parameters are the coefficients $\boldsymbol{\omega}$ and the observation noise $\sigma^2$. Define $\tilde{y}_t=y_t-\xi_t$ so that $\tilde{y}_t=\boldsymbol{\omega}'\y_t+\varepsilon_t$. Obtaining $\boldsymbol{\omega}$ and $\sigma^2$ is a problem of Bayesian variable selection in linear regression with a closed form expression for the posterior given a specific choice of priors.\footnote{Further details on Bayesian variable selection can be found in e.g. \cite{george1993variable}, \cite{george1997approaches}, \cite{brown1998multivariate} and \cite{murphy2012machine}.} To introduce the variable selection element define a $J\times 1$ vector $\boldsymbol{\kappa}$, where $\kappa_j=1$ if control unit $j$ is relevant for predicting the treated unit and $\kappa_j=0$ otherwise. The likelihood and choice of priors are 
\begin{equation}\label{eq_retail_methods_bsts_spikeslab}
\begin{aligned}
\tilde{\y}|\boldsymbol{\omega}_\kappa,\sigma^2 & \sim  \N(\Y_\kappa\boldsymbol{\omega}_\kappa,\sigma^2\I)\\
\boldsymbol{\omega}_\kappa|\sigma^2,\boldsymbol{\kappa} & \sim  \N(\boldsymbol{\omega}_{0\kappa},\sigma^2 \mathbf{V}_{0\kappa})\\
\sigma^2|\boldsymbol{\kappa} & \sim  \mathcal{G}^{-1}(s_0,r_0)\\
\boldsymbol{\kappa} & \sim  \mathcal{B}(\mathbf{q})
\end{aligned}
\end{equation}
The first distribution is the likelihood where the parameters $\boldsymbol{\omega}$ and $\sigma^2$ are unknown and $\Y$ is the $T_0\times J$ matrix of control unit information. The subscript $\kappa$ indicate the dependence between the variable selection element and the priors and data. Any variable with this subscript is reduced to the dimensionality of the included units, i.e. units with $\kappa_j=1$. To simplify exposition, I will omit this from the notation in the rest of the paper. Define a joint prior for $\boldsymbol{\omega}$ and $\sigma^2$ as $p(\boldsymbol{\omega},\sigma^2) = p(\boldsymbol{\omega}|\sigma^2)p(\sigma^2)$. By using Bayes' rule we may obtain the following joint prior for the parameters, the noise and the selection variable  
\begin{equation}\label{eq_retail_methods_bsts_prior}
p(\boldsymbol{\omega},\sigma^2,\boldsymbol{\kappa}) = p(\boldsymbol{\omega},\sigma^2|\boldsymbol{\kappa})p(\boldsymbol{\kappa})=p(\boldsymbol{\omega}|\sigma^2,\boldsymbol{\kappa})p(\sigma^2|\boldsymbol{\kappa})p(\boldsymbol{\kappa})
\end{equation}
The posterior describing what control units that are relevant for predicting the treated unit, $p(\boldsymbol{\kappa}|\tilde{\y}) \propto p(\tilde{\y}|\boldsymbol{\kappa})p(\boldsymbol{\kappa})$, has a closed form solution given the prior specifications above. The derivation involves integrating out $\boldsymbol{\omega}$ and $\sigma^2$, which gives the following posterior distributions for the weights and the observation noise
\begin{equation}\label{eq_retail_methods_bsts_posterior_pars}
\boldsymbol{\omega}|\sigma^2,\boldsymbol{\kappa},\tilde{\y} \sim \N(\boldsymbol{\omega}_1,\sigma^2\mathbf{V}_1) \text{ and } \sigma^2|\boldsymbol{\kappa},\tilde{\y} \sim \mathcal{G}^{-1}(r_1,s_1)
\end{equation}
where
\begin{eqnarray}
\mathbf{V}_1 = (\Y'\Y + \mathbf{V}_0^{-1})^{-1} &\text{  }& \boldsymbol{\omega}_1 = \mathbf{V}_1(\Y'\tilde{\y} + \mathbf{V}_0^{-1} \boldsymbol{\omega}_0) \\
s_1 = s_0 + \frac{1}{2}T_0 &\text{  }& r_1 = r_0 + \frac{1}{2}(\tilde{\y}'\tilde{\y} + \boldsymbol{\omega}_0' \mathbf{V}_0^{-1}\boldsymbol{\omega}_0 - \boldsymbol{\omega}_1'\mathbf{V}_1^{-1}\boldsymbol{\omega}_1)
\end{eqnarray} 
Finally, the control unit inclusion posterior is
\begin{equation}\label{eq_retail_methods_bsts_posterior_idx}
p(\boldsymbol{\kappa}|\tilde{\y}) \propto \frac{1}{(2\pi)^{T_0/2}} \frac{|\mathbf{V}_1|^{1/2}}{|\mathbf{V}_0|^{1/2}} \frac{\Gamma(s_1)}{\Gamma(s_0)} \frac{r_0^{s_0}}{r_1^{s_1}} \prod_{j=1}^J q_j^{\kappa_j} (1-q_j)^{1-\kappa_j}
\end{equation}
See Appendix \ref{sec_retail_appendix_bsts} for derivation details.\\
\begin{algorithm}
\caption{BSTS}
\begin{algorithmic}[1]
\item Set starting values $\boldsymbol{\theta}^{(0)}=((\sigma_1^2)^{(0)},(\sigma_2^2)^{(0)},\boldsymbol{\kappa}^{(0)},(\sigma^2)^{(0)},\boldsymbol{\omega}^{(0)})$.
\item \textsc{Estimation}. For Gibbs sampling iteration $j=1,\hdots,M$
\subitem Draw $\x^{(j)}$ from $p(\x|\boldsymbol{\theta}^{(j-1)},\y)$ by using a simulation smoother.
\subitem Draw $(\sigma_1^2)^{(j)}$ from $p(\sigma_1^2|\x^{(j)},\y)$ and $(\sigma_2^2)^{(j)}$ from $p(\sigma_2^2|\x^{(j)},\y)$.
\subitem Draw $\boldsymbol{\kappa}^{(j)}$ from $p(\boldsymbol{\kappa}|\tilde{\y}^{(j)})$ where $\tilde{\y}^{(j)}$ is the vector with elements $\tilde{y}_t^{(j)}=y_t-\xi_t^{(j)}$.
\subitem Draw $(\sigma^2)^{(j)}$ from $p(\sigma^2|\boldsymbol{\kappa}^{(j)},\tilde{\y}^{(j)})$.
\subitem Draw $\boldsymbol{\omega}^{(j)}$ from $p(\boldsymbol{\omega}|(\sigma^2)^{(j)},\boldsymbol{\kappa}^{(j)},\tilde{\y}^{(j)})$.
\subitem Update vector $\boldsymbol{\theta}^{(j)}=((\sigma_1^2)^{(j)},(\sigma_2^2)^{(j)},\boldsymbol{\kappa}^{(j)},(\sigma^2)^{(j)},\boldsymbol{\omega}^{(j)})$.
\item \textsc{Prediction}. For Gibbs sampling iteration $j=1,\hdots,M$ 
\subitem For time period $h=T_0+1,\hdots,T$
\subsubitem Draw $\hat{\boldsymbol{\eta}}_h^{(j)}\sim\N(\zero,\mathbf{Q}^{(j)})$, generate $\hat{\x}_{h}^{(j)} = \mathbf{T}\hat{\x}_{h-1}^{(j)} + \mathbf{R}\hat{\boldsymbol{\eta}}_h^{(j)}$, with $\hat{\x}_{T_0}^{(j)}=\x_{T_0}^{(j)}$.
\subsubitem Draw $\hat{\varepsilon}_h^{(j)}\sim\N(0,(\sigma^2)^{(j)})$, generate $\hat{y}_{0h}(0) = \mathbf{Z}_h\hat{\x}_h^{(j)} + \hat{\varepsilon}_h^{(j)}$.
\end{algorithmic}
\label{tab_retail_methods_bsts_algorithm}
\end{algorithm}

\noindent \emph{3. Predictions}. The estimated states and parameters may be used to forecast trajectories of the treated unit $\hat{y}_{0h}(0)$ for $h=T_0+1,\hdots,T$. Starting from $T_0$, the treated unit is forecasted using the estimated states and parameters from each Gibbs sampling iteration together with the observed control unit outcomes $\y_h$. An outline of the estimation and prediction procedure is given in Algorithm \ref{tab_retail_methods_bsts_algorithm}.\\

\par BSTS offers a flexible way of modelling components of the treated unit combined with information from the control units. In terms of the bias-variance discussion in Section \ref{sec_retail_framework_est}, BSTS prevents overfitting by regularizing priors on the control unit weights using the ``spike and slab'' prior defined in (\ref{eq_retail_methods_bsts_spikeslab}). Furthermore, BSTS does not commit to one specific choice of weights, but rather forecast different trajectories based on weights that are allowed to change during the Gibbs sampling procedure. Thus, in contrast to the reduced form approaches that all rely on point forecasts, BSTS takes model uncertainty into account when forecasting the counterfactual.\\

\noindent \textbf{Nonlinear BSTS}. BSTS is restricted to linear state space models where the treated unit outcomes are linearly dependent on the structural components and a linear model is assumed for the evolution of these components. However, many macroeconomic time series show some form of interaction between trend, cycle and seasonality components. For instance, \cite{koopman2009seasonality} find increasing seasonal variation of US unemployment rates. They suggest to model this interaction using the specification $g(\xi_t,\iota_t,\psi_t) = \xi_t + \exp(\beta \xi_t) \psi_t$, where $\beta$ is the parameter determining the size and direction of the change in the seasonality $\psi_t$ following a change in the trend $\xi_t$. I outline an approach for extending the BSTS framework to allow for such nonlinear specifications below. 

\par Combining (\ref{eq_retail_framework_est_model}) and (\ref{eq_retail_methods_structural}) with the nonlinear function for $g$, and adding trend and seasonal transition equations gives the state space model
\begin{equation}
\begin{aligned}
y_{0t} &= \xi_t + \exp(\beta \xi_t) \psi_t + \boldsymbol{\omega}'\y_t + \varepsilon_t\\
\xi_t &= \xi_{t-1} + \nu_{t-1} + \eta_{1t}\\
\nu_t &= \nu_{t-1} + \eta_{2t}\\
\psi_t &= - \sum_{s=1}^{S-1}\nolimits\psi_{t-s} + \eta_{3t}
\end{aligned}
\end{equation}
where the trend evolution is identical to (\ref{eq_retail_methods_bsts_model}) and the last transition equation is a common seasonality specification, where $S$ denotes the number of seasons. In this model, the trend and trend-dependent seasonality pattern are modelled explicitly for the treated unit, while the role of the control units becomes to pick up specific events or cyclical patterns.

\par Estimation of the states in the BSTS model was done using a simulation smoother based on the linear Kalman filter and smoother. This approach is not feasible in the nonlinear setting, and I propose to substitute the simulation smoother with draws from the posterior of the state variables, obtained from e.g. a Bootstrap particle filter, \cite{gordon1993novel}. By conditioning on the state variables, draws can be made for each of the parameters in a similar setup to Algorithm \ref{tab_retail_methods_bsts_algorithm}.

\section{Simulation study}\label{sec_retail_sim}
The purpose of this simulation study is to compare the synthetic control methods discussed in Section \ref{sec_retail_methods} in different scenarios. In particular, I consider both parallel and heterogenous trend cases, a case where a convex combination of controls is insufficient to recreate the treated unit, a case where trends are not easily detected due to noise and a case with a short pretreatment period and high noise. I assume the following data generating process
\begin{equation}\label{eq_retail_sim_model}
\begin{aligned}
y_{jt}(0) &= \xi_{jt} + \psi_{jt} + \varepsilon_{jt} \text{ for } j=1,\hdots,J \\
y_{0t}(0) &= \sum_{j=1}^J \omega_j(\xi_{jt}+\psi_{jt}) + \varepsilon_{0t}\\
y_{0t}(1) &= y_{0t}(0) + d_{0t}\tau_t
\end{aligned}
\end{equation} 
for $t=1,\hdots,T$ where $\varepsilon_{jt}\sim\N(0,\sigma^2)$ and where $d_{0t}$ is an indicator variable taking the value 1 in the treatment period and zero otherwise. The pretreatment period is initially set to $T_0=100$ with $T-T_0=10$ treatment periods and $J=150$ control units. The trend component $\xi_{jt}$ consist of two parts; a common trend and a unit-specific time-varying confounding factor. Furthermore, $\psi_{jt}$ represents the seasonality of the time series. I set $\omega_1=0.7$ and $\omega_2=0.3$ with $\omega_j=0$ for $j=3,\hdots,150$. In other words, of all the 150 control units, only the first two are relevant for forecasting values of the counterfactual of the treated unit. All results provided below are based on the assumption that data draws are made conditional on the treatment assignment. Each set of potential outcomes drawn from (\ref{eq_retail_sim_model}) is made conditional on fixed values for the treatment indicator and the weights $\omega_j$, where $\xi_j$, $\psi_{jt}$ and $\varepsilon_{jt}$ are drawn using the specifications outlined below.
\begin{figure}
\centering
\begin{minipage}{1\textwidth}
\begin{subfigure}{0.32\textwidth}
\includegraphics[width=\linewidth]{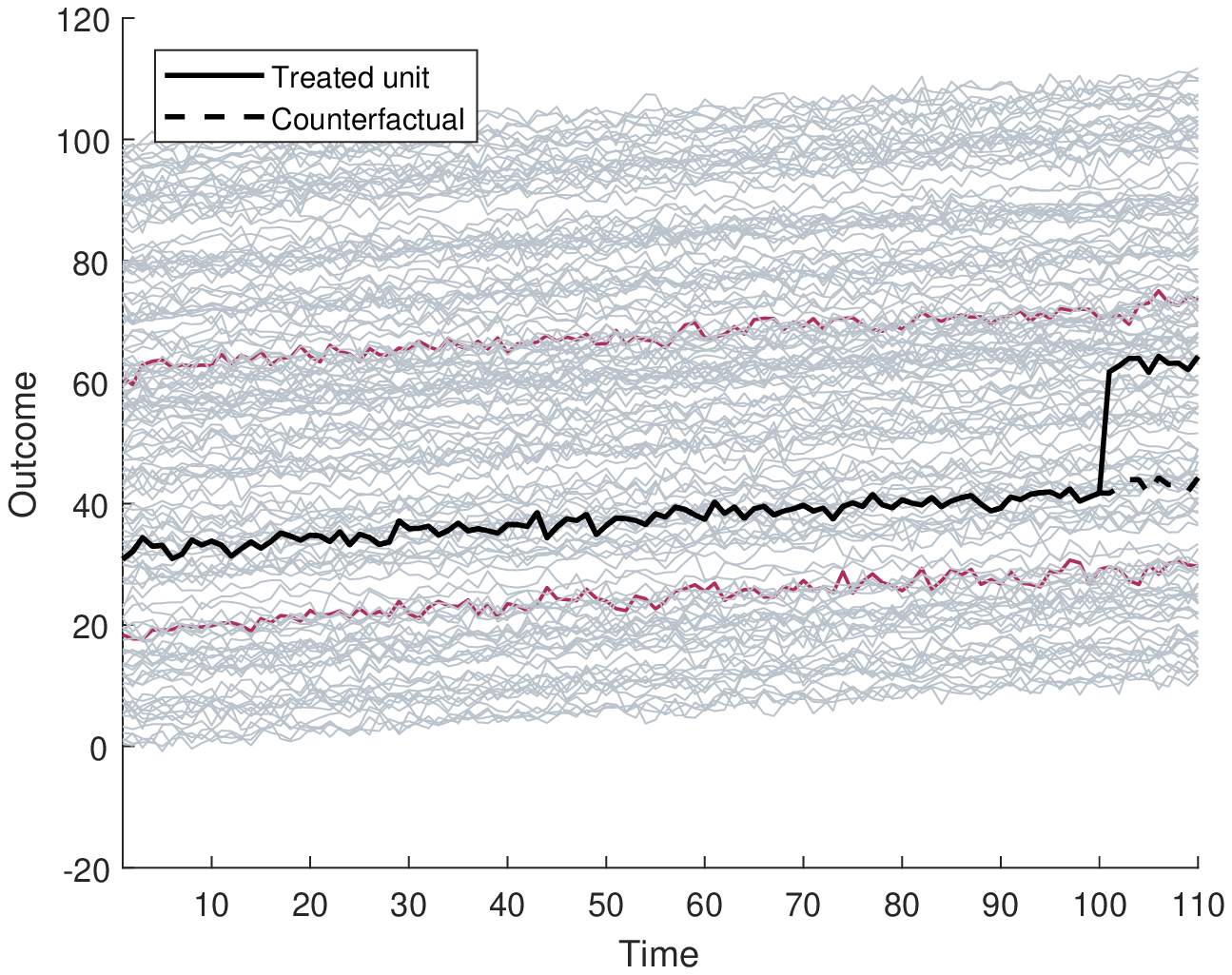}
\caption*{(A) Parallel trends}
\end{subfigure}\hspace*{\fill}
\begin{subfigure}{0.32\textwidth}
\includegraphics[width=\linewidth]{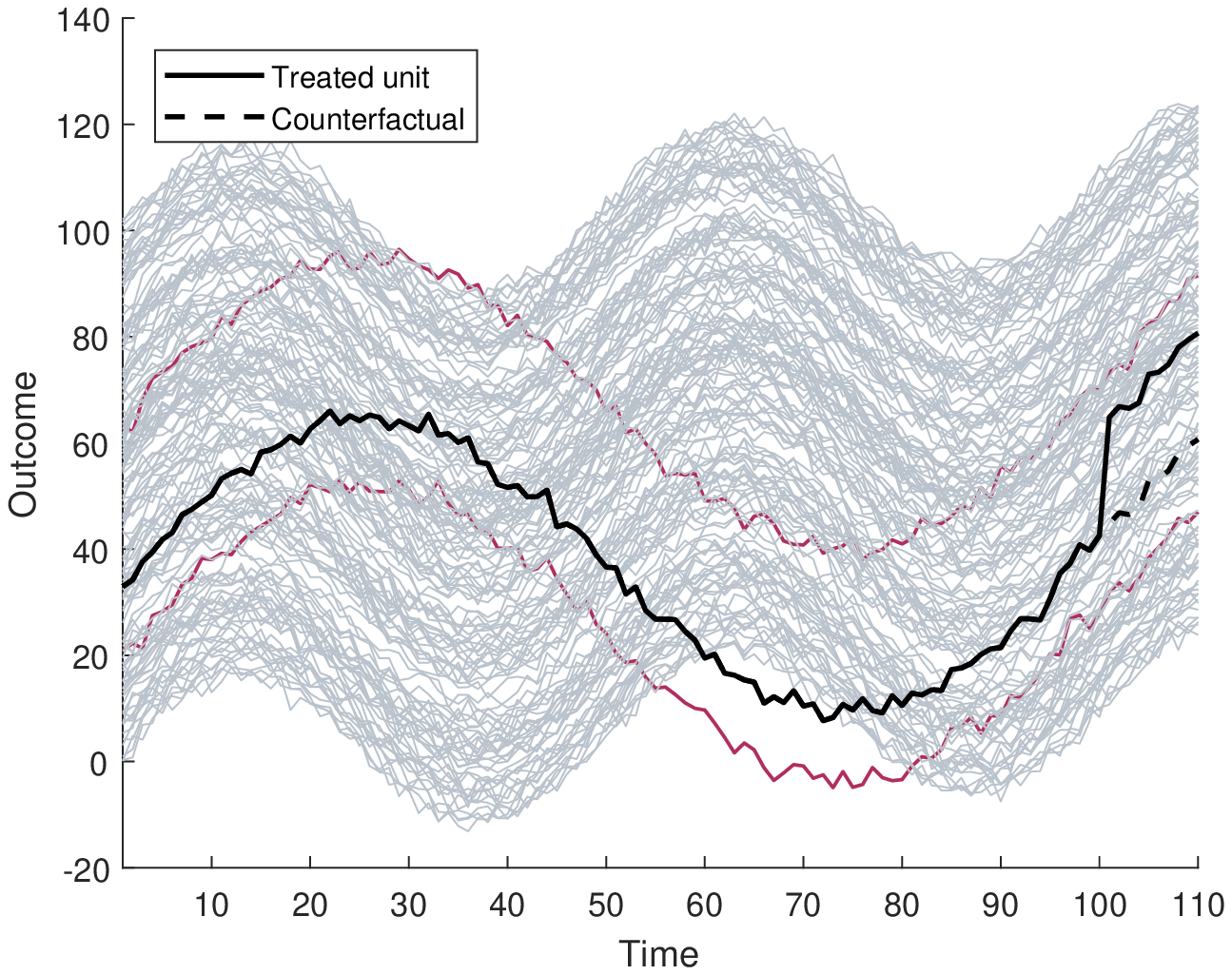}
\caption*{(B) Seasonality}
\end{subfigure}\hspace*{\fill}
\begin{subfigure}{0.32\textwidth}
\includegraphics[width=\linewidth]{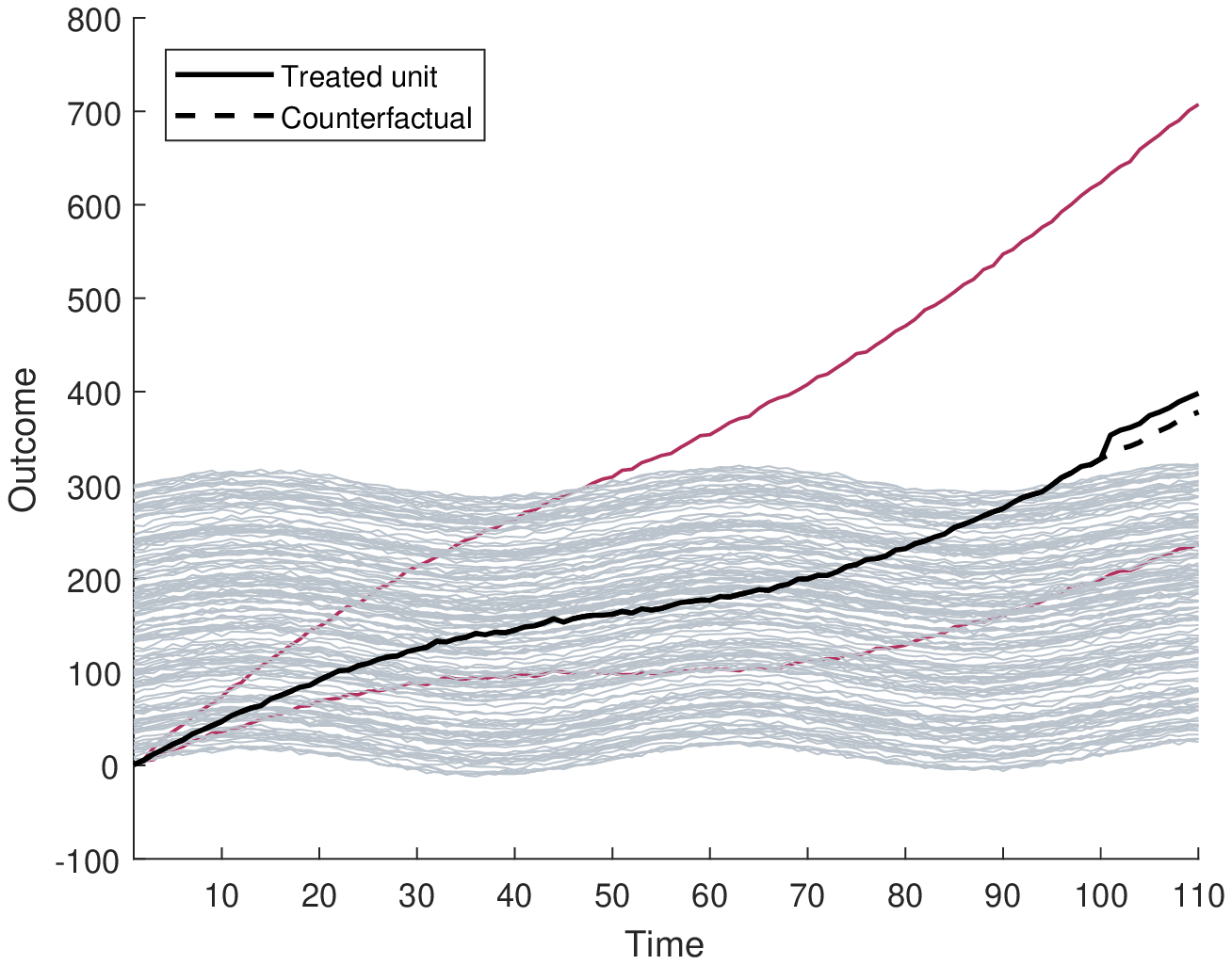}
\caption*{(C) Heterogeneous trends}
\end{subfigure}\\
\medskip
\begin{subfigure}{0.32\textwidth}
\includegraphics[width=\linewidth]{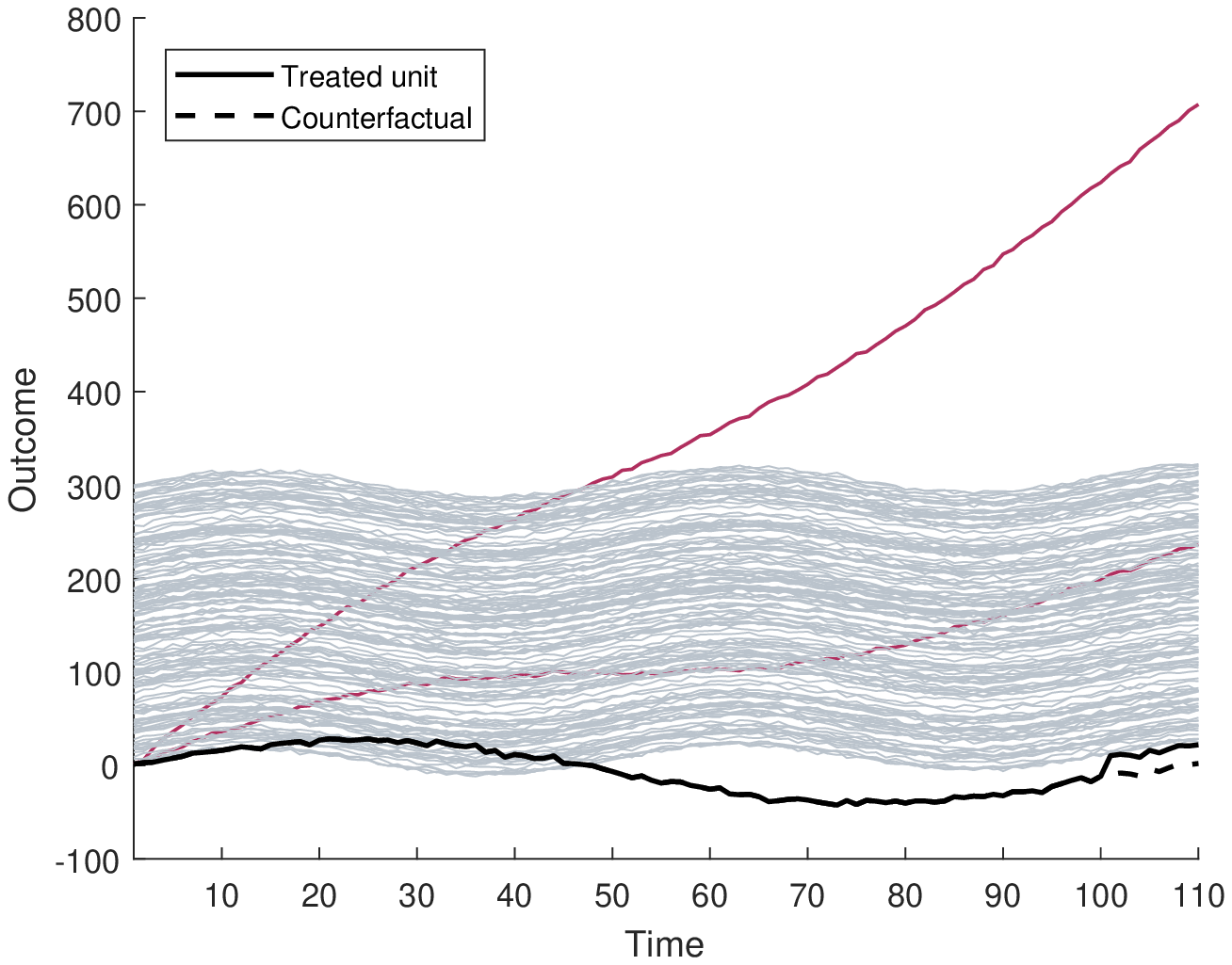}
\caption*{(D) Extreme end}
\end{subfigure}\hspace*{\fill}
\begin{subfigure}{0.32\textwidth}
\includegraphics[width=\linewidth]{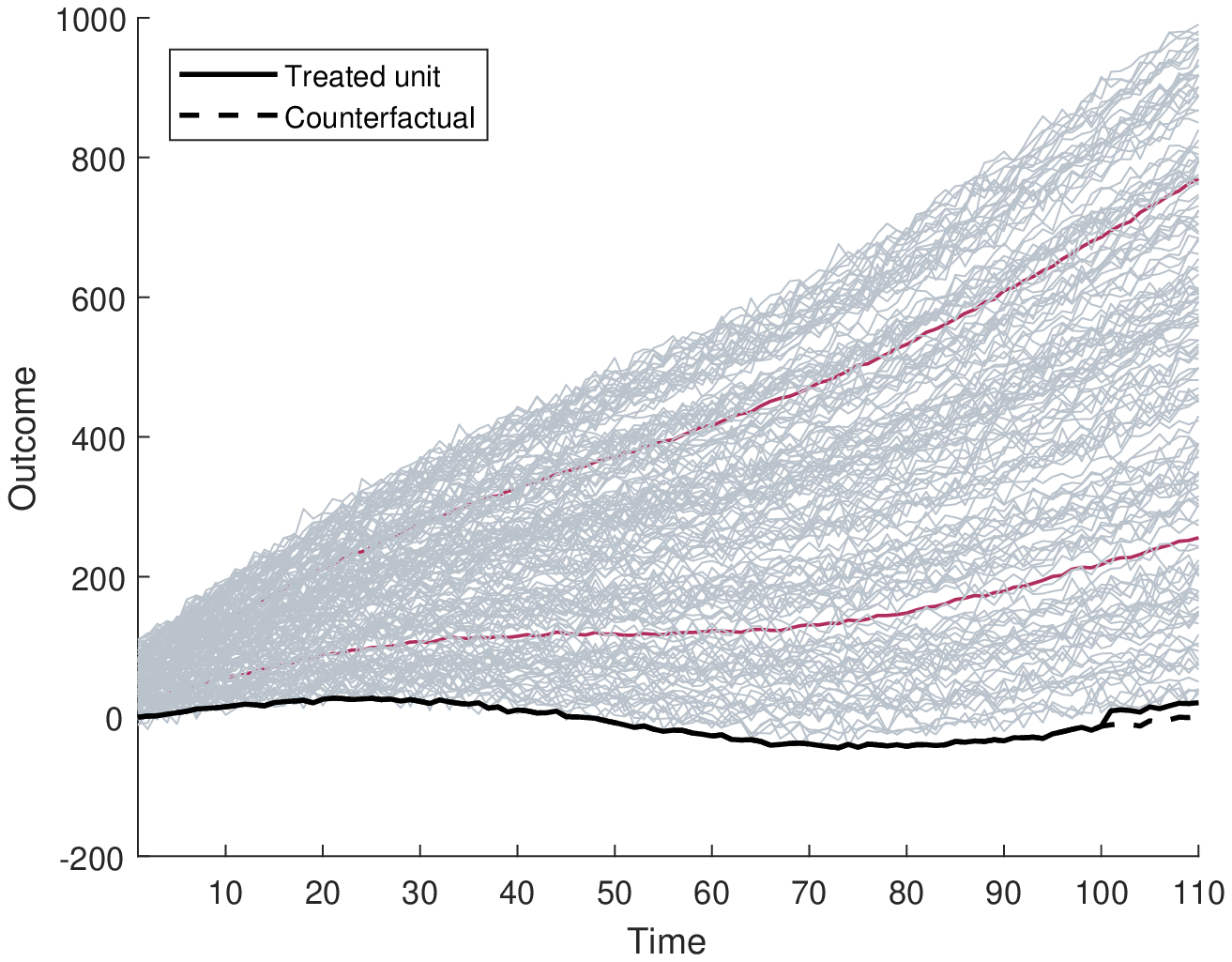}
\caption*{(E) Hidden trends}
\end{subfigure}\hspace*{\fill}
\begin{subfigure}{0.32\textwidth}
\includegraphics[width=\linewidth]{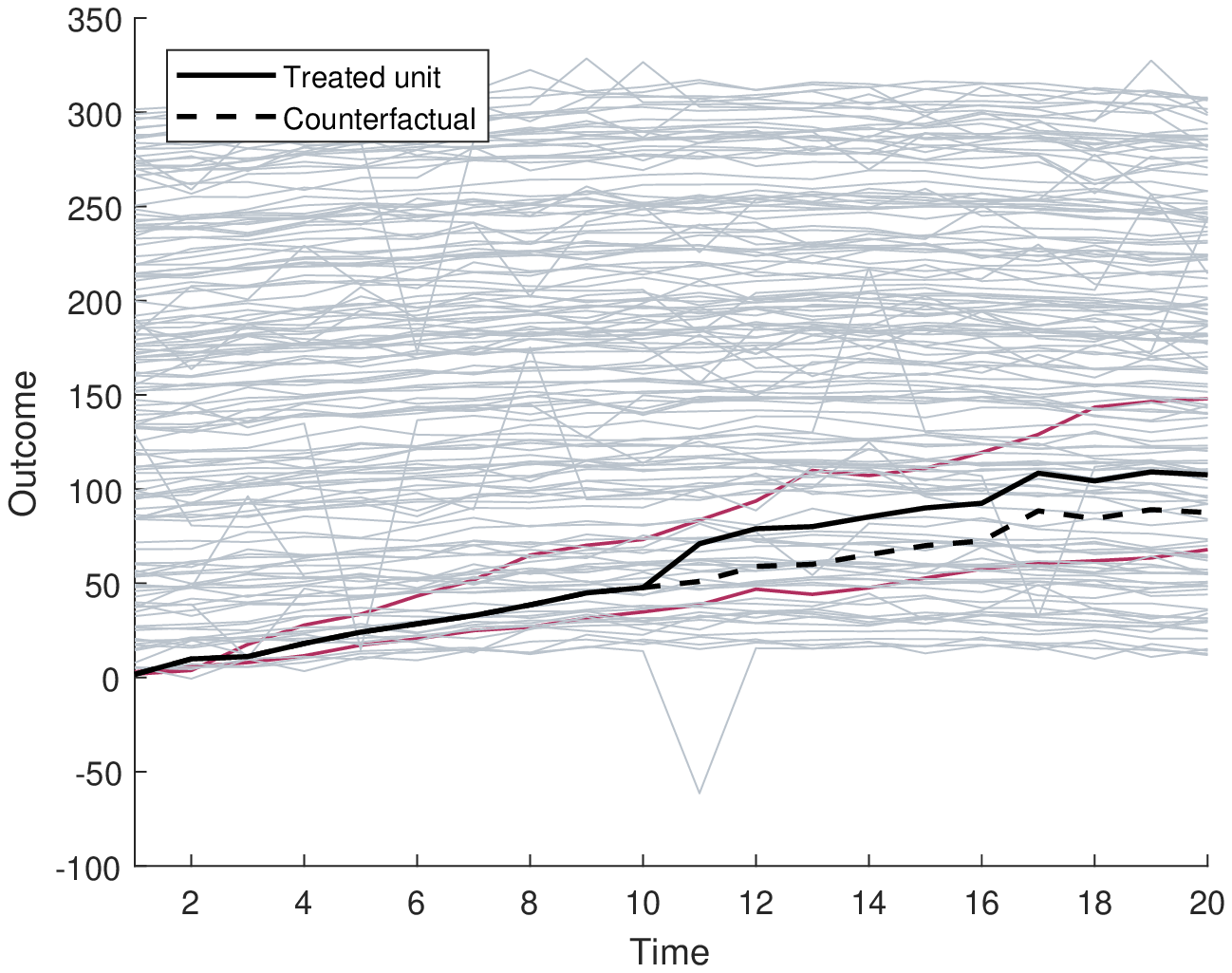}
\caption*{(F) Short and noisy}
\end{subfigure}\\
\caption{\footnotesize The figures show one draw of the potential outcomes model defined in (\ref{eq_retail_sim_model}) for the treated unit $y_{0t}(0)$ (dotted line) and $J=150$ control units $y_{jt}(0)$ (red and gray lines) for each scenario. For each $h>T_0$, the true average treatment effect is constant and equal to $\tau_h=20$. The solid line is the observed value for the treated unit, $y_{0t}(1)=y_{0t}(0)+d_{0t}\tau_t$. The pretreatment period is set to $T_0=100$ with $T=110$. The following assumptions describes Scenario (C), and all other scenarios follow by altering some of these specifications. Only the two first control units are relevant, $\omega_1=0.7$ and $\omega_2=0.3$ and $\omega_j=0$ for $j=3,\hdots,150$. The common trend is assumed to be linear deterministic trend with slope 0.1. The unit-specific component is a fixed number between 1 and 100 for each $j$ multiplied with a trend. This trend is deterministic with slope 0.1 for the two first units and flat and equal to 3 for the remaining units. The seasonality component $\phi_{jt}$ is modelled as a sinus curve plus noise, where the seasonality pattern is more frequent for the irrelevant units. Scenario (A): flat specific trends and no seasonality, $\psi_{jt}=0$ for all $j$ and $t$. Scenario (B): flat specific trends $\xi_{jt}$ for all $j$ and $t$. Scenario (D): Weights are $\omega_1=1.5$ and $\omega_2=-0.5$. Scenario (E): same weights as for Scenario (D). T specific trend of the two first units is $1+0.1t$ and $1+0.08t$ for the remaining units. All seasonality patterns are equal to that of the first two units. Scenario (F): $\varepsilon_{jt}$ t-distributed with degrees of freedom 0.99 and $T_0=10$.}
\label{fig_retail_sim_dgp}
\end{minipage}
\end{figure}

\par I have designed six different scenarios that are summarised in Figure \ref{fig_retail_sim_dgp}. Scenario (A) assumes parallel trends and no seasonality. A positive common trend is combined with an unobserved confounding factor that is allowed to vary between units, but is constant across the time dimension. Scenario (B) is identical to (A), but adds the seasonality component $\psi_{jt}$. I assume that the first two controls have the same low-frequency seasonality pattern while the remaining units follow a seasonality pattern with a higher frequency. Scenario (C) extends scenario (B) by additionally allowing trends $\xi_{jt}$ to be heterogeneous. In particular, the first two units follow a positive trend while the remaining units have constant unit-specific trends. Scenario (D) is identical to (C), but I impose that the treated unit lie outside of the distribution of the control units by altering the weights to $\omega_1=1.5$ and $\omega_2=-0.5$. In Scenario (E) all units have increasing unit-specific confounding factors, but the slope of the two first units are larger than for the remaining units. In addition, the seasonality pattern is equal for all units and the noise is large. Finally, Scenario (F) is identical to Scenario (C), with an additional t-distributed noise and a shorter pretreatment period. A more detailed description of the data generating process for each scenario can be found in the text below Figure \ref{fig_retail_sim_dgp}. 

\par Each method was implemented with the following properties. Matching prior to DD analysis (``MDD'') was implemented by minimizing the Mahalonobis distance between the treated unit and each control unit in the pretreatment period. The five units closest to the treated unit was chosen to represent the control group, each given a weight of 0.2 in the DD analysis. The original synthetic control method (``ADH'') was implemented with all control units as outcome variables and no covariates. Principal Component Regression (``PCR'') was implemented based on the first $k=5$ principal components of the control unit data. Lasso (``LASSO'') was implemented using five-fold cross validation on the pretreatment period. Bayesian Structural Time Series (``BSTS'') was implemented with the following specification for the priors. The inverse Gamma priors of the transition equations was chosen with shape parameters $a_0=c_0=0.01$ and scale parameters $b_0=e_0=0.1$. Following \cite{brodersen2015inferring}, the normal prior for the coefficient vector was specified with mean $\boldsymbol{\omega}_0=\zero$ and the Zellner's g-prior for the covariance $\boldsymbol{V}_0=\frac{1}{T_0}(w\Y'\Y + (1-w)\text{diag}(\Y'\Y))$ with $w=0.5$. The observation noise prior was specified as inverse Gamma with shape $s_0=0.1$ and scale $r_0=s_0(1-R^2)s_y^2$ where $R^2$ is the expected R-squared set equal to $R^2=0.5$ and $s_y^2=\frac{1}{T_0-1}\sum_{t=1}^{T_0}(y_{0t}-\bar{y}_0)^2$ is the sample variance of the treated unit. Last, the variable inclusion prior is uninformative, $q_j=0.5$ for each $j$.

\begin{table}
\caption{\textsc{Simulation Study Results}}
\label{tab_retail_sim_results}
\scriptsize
\centering
\begin{threeparttable}
\begin{tabular}{l rrrr|rrr}
\midrule
& \multicolumn{4}{c}{\textbf{Estimation}} & \multicolumn{3}{c}{\textbf{Tradeoff}}\\
Description 		& Mean & Std & Sum & Controls & MSE & Bias$^2$ & Variance\\
\midrule
\textbf{Scenario A}\\
MDD			& 19.9 & 1.0 & 200.3 & 2.0  & 1.1 & 0.0 & 1.1 \\
ADH			& 19.8 & 1.0 & 200.4 & 16.8 & 1.0 & 0.0 & 1.0\\
PCR			& 19.9 & 1.0 & 200.8 & 51.3 & 0.9 & 0.0 & 0.9\\
LASSO		& 19.9 & 1.0 & 201.6 & 17.2 & 1.0 & 0.0 & 1.0\\ \\
\textbf{Scenario B}\\
MDD			& 19.6 & 1.3 & 204.2 & 2.0  & 1.8 & 0.2 & 1.6 \\
ADH			& 20.0 & 1.2 & 199.9 & 2.1 & 1.5 & 0.0 & 1.5\\
PCR			& 20.3 & 1.2 & 201.7 & 10.2 & 1.6 & 0.1 & 1.5\\
LASSO		& 20.1 & 1.3 & 201.1 & 4.9 & 1.6 & 0.0 & 1.6\\
BSTS		& 20.1 & 1.4 & 201.2 & 7.1 & 2.1 & 0.0 & 2.1\\ \\
\textbf{Scenario C}\\
MDD			& 133.1 & 1.3 & 1440.9 & 2.0  & 12802.6 & 12800.9 & 1.7 \\
ADH			& 19.9 & 1.4 & 201.2 & 2.1 & 1.8 & 0.0 & 1.8\\
PCR			& 20.8 & 1.4 & 213.3 & 47.5 & 2.6 & 0.6 & 2.0\\
LASSO		& 24.9 & 1.3 & 265.7 & 2.0 & 25.7 & 24.0 & 1.7\\
BSTS		& 20.1 & 1.4 & 199.2 & 9.1 & 1.9 & 0.0 & 1.9\\ \\
\textbf{Scenario D}\\
MDD			& 9.1 & 1.4 & 73.2 & 2.0  & 119.8 & 117.9 & 2.2 \\
ADH			& 9.0 & 1.5 & 73.3 & 2.3 & 122.8 & 120.6 & 2.2\\
PCR			& 22.1 & 2.0 & 232.1 & 104.8 & 8.6 & 4.6 & 4.1\\
LASSO		& 20.8 & 2.1 & 213.3 & 4.4 & 4.9 & 0.6 & 4.4\\
BSTS		& 21.1 & 1.6 & 192.1 & 63.5 & 3.7 & 1.2 & 2.6\\ \\
\textbf{Scenario E}\\
MDD			& 6.2 & 7.1 & 30.1 & 2.0  & 240.3 & 190.3 & 50.4 \\
ADH			& 4.9 & 5.6 & 24.9 & 3.9 & 257.9 & 226.7 & 31.5\\
PCR			& 20.8 & 1.8 & 191.1 & 77.8 & 3.7 & 0.6 & 3.1\\
LASSO		& 21.4 & 2.1 & 203.5 & 38.7 & 6.6 & 2.1 & 4.6\\
BSTS		& 21.3 & 1.4 & 193.2 & 1.9 & 3.5 & 1.7 & 1.8\\ \\
\textbf{Scenario F}\\
MDD			& 35.0 & 3.8 & 583.8 & 2.0 & 238.3 & 225.3 & 14.4 \\
ADH			& 22.1 & 4.4 & 250.7 & 6.1 & 21.7 & 4.5 & 19.2\\
PCR			& 30.7 & 5.9 & 473.0 & 90.7 & 145.7 & 114.5 & 34.7\\
LASSO		& 20.7 & 3.7 & 218.7 & 3.7 & 13.0 & 0.5 & 13.9\\
BSTS		& 19.7 & 6.4 & 250.8 & 134.8 & 37.2 & 0.1 & 41.2\\
\bottomrule
\end{tabular}
\begin{tablenotes}
\item The table reports estimation results from each scenario of the simulation study. Results are based on 100 repeated draws of the potential outcomes in (\ref{eq_retail_sim_model}). Methods employed are Difference in Differences with Matching (MDD), the original Synthetic Control (ADH), Principal Component Regression (PCR) Lasso regularization (LASSO) and Bayesian Structural Time Series (BSTS). Column ``Mean'' reports the mean average treatment effect in repeated samples at time period $T_0+1$ and column ``Std'' reports its standard deviation. The column ``Sum'' reports the sum over all treatment periods, which according to the data generating process is $\tau_h\times (T-T_0)=200$. The average number of controls with weights above 1\% is reported in column ``Controls''. The mean squared error (MSE) is decomposed int o squared bias and variance by using (\ref{eq_retail_framework_notation_att}) together with (\ref{eq_retail_framework_est_mse}) to obtain $L = (\tau_h-\E[\hat{\tau}_h])^2 + \V[\hat{\tau}_h]$.
\end{tablenotes}
\end{threeparttable}
\end{table}

\par Table \ref{tab_retail_sim_results} presents the estimation results from applying each synthetic control method to 100 repeated samples of data for each scenario. The table contains the mean treatment effect on the treated, its standard deviation, the total treatment effect, the average number of controls and a bias-variance decomposition of the treatment effect estimates. The results show that all methods work well for estimating the treatment effect in Scenario (A). In fact, when potential outcomes are parallel without seasonality, it is straightforward to construct a weighted combination of controls that resembles the treated unit. In this case any combination of controls will work, since confounding factors are constant in time. The average number of controls and misclassification is therefore irrelevant in this scenario.

\par
\begin{figure}
\centering
\begin{minipage}{1\textwidth}
\begin{subfigure}{0.20\textwidth}
\includegraphics[width=\linewidth]{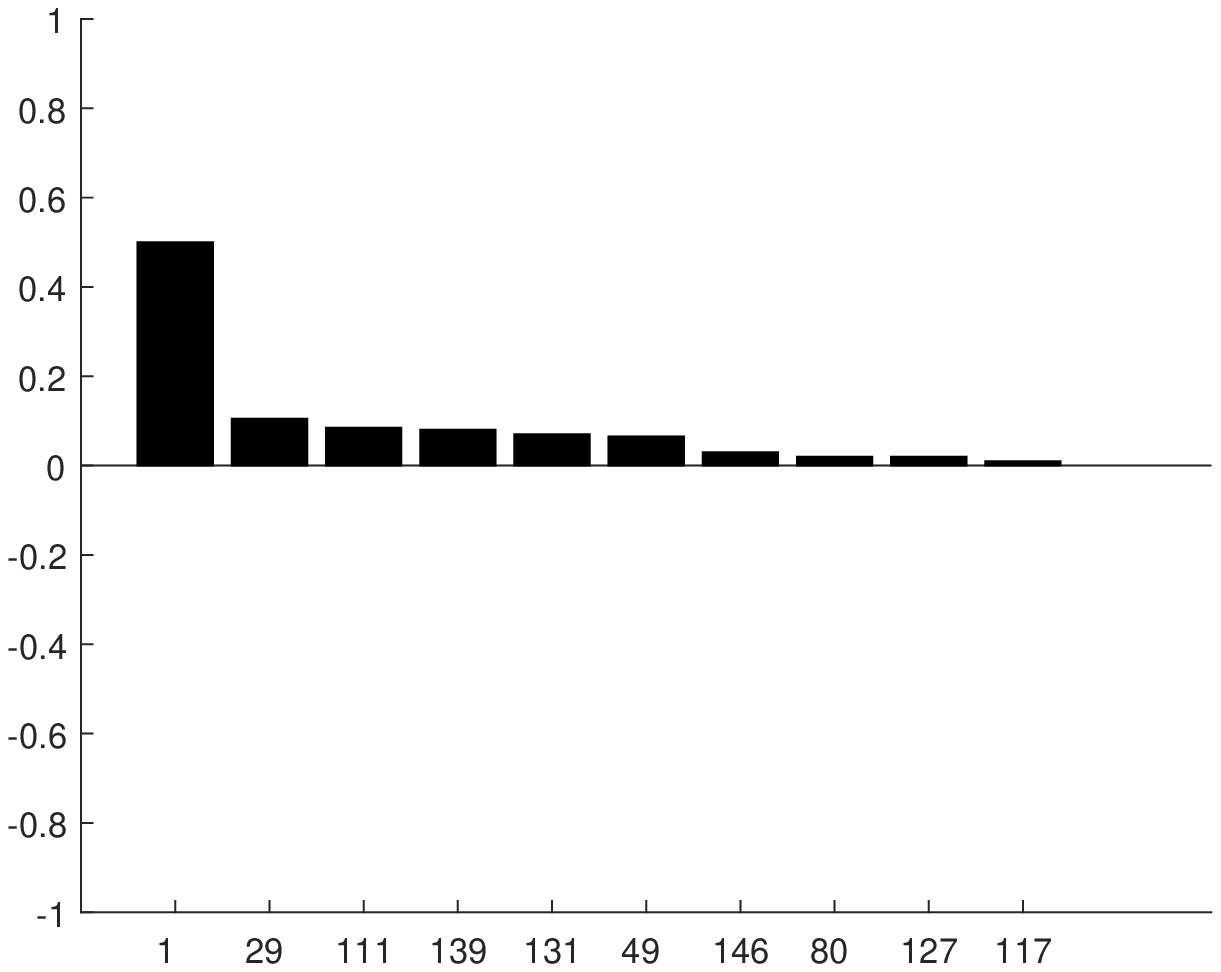}
\caption*{(B) MDD}
\end{subfigure}\hspace*{\fill}
\begin{subfigure}{0.20\textwidth}
\includegraphics[width=\linewidth]{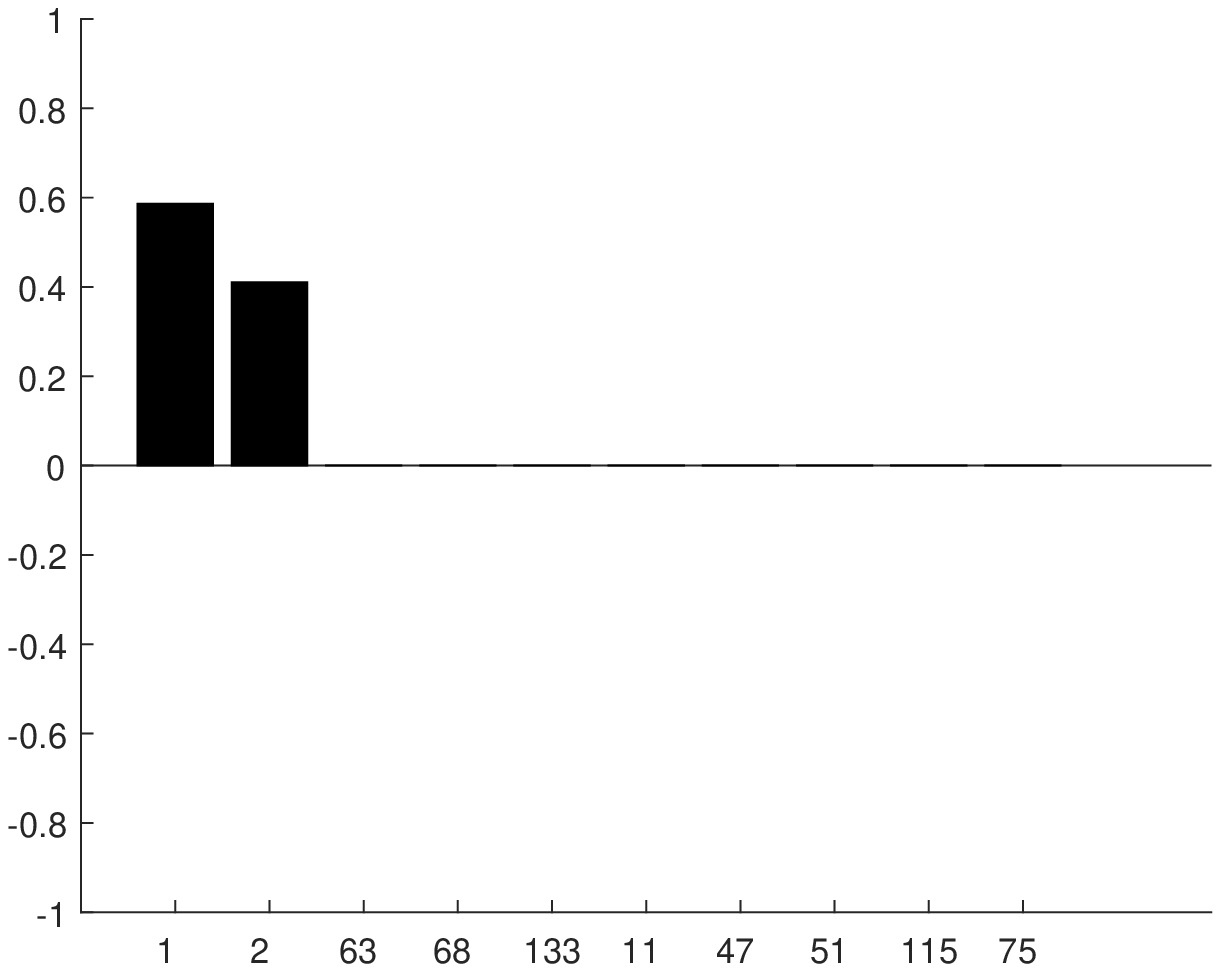}
\caption*{(B) ADH}
\end{subfigure}\hspace*{\fill}
\begin{subfigure}{0.20\textwidth}
\includegraphics[width=\linewidth]{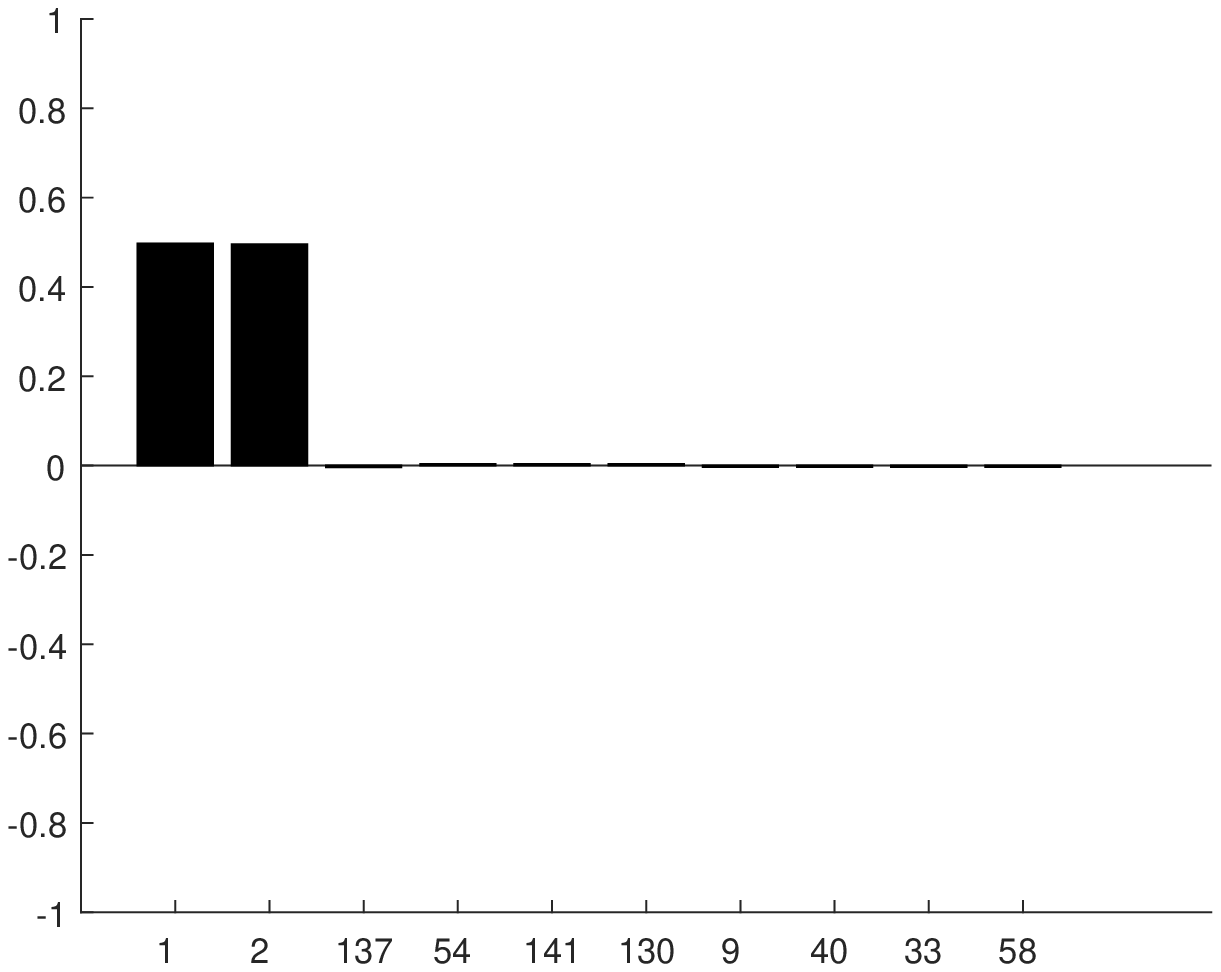}
\caption*{(B) PCR}
\end{subfigure}\hspace*{\fill}
\begin{subfigure}{0.20\textwidth}
\includegraphics[width=\linewidth]{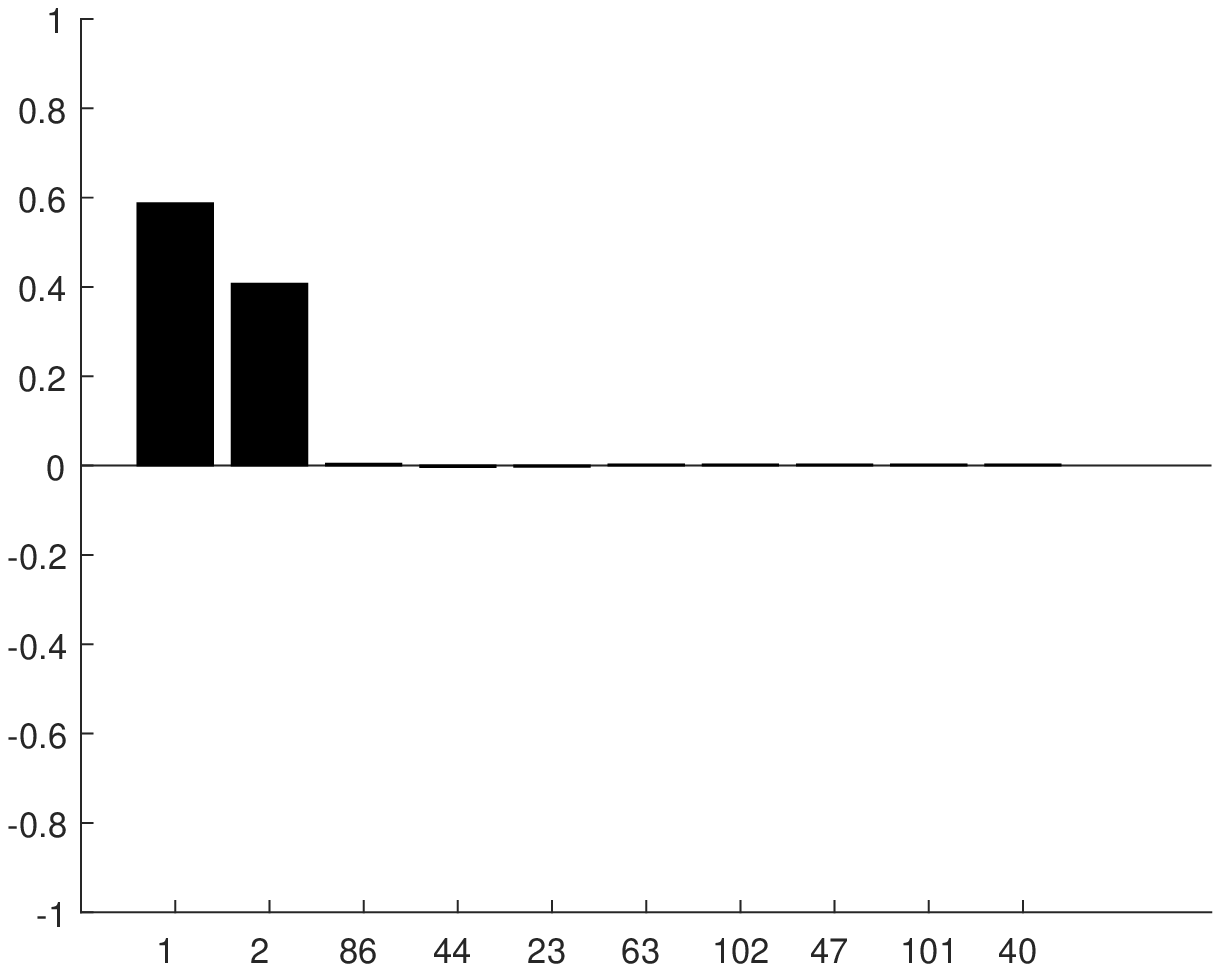}
\caption*{(B) LASSO}
\end{subfigure}\hspace*{\fill}
\begin{subfigure}{0.20\textwidth}
\includegraphics[width=\linewidth]{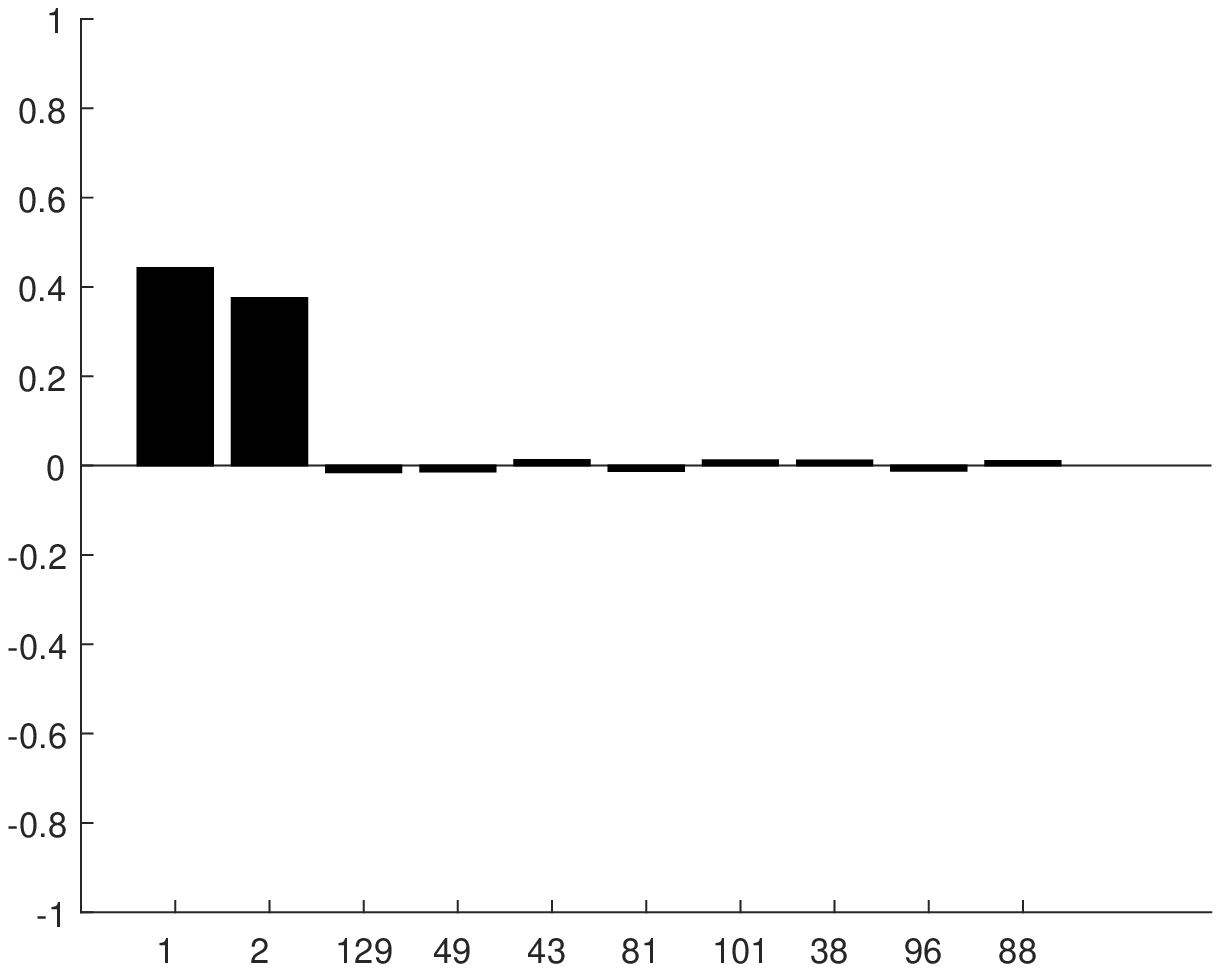}
\caption*{(B) BSTS}
\end{subfigure}
\medskip
\begin{subfigure}{0.20\textwidth}
\includegraphics[width=\linewidth]{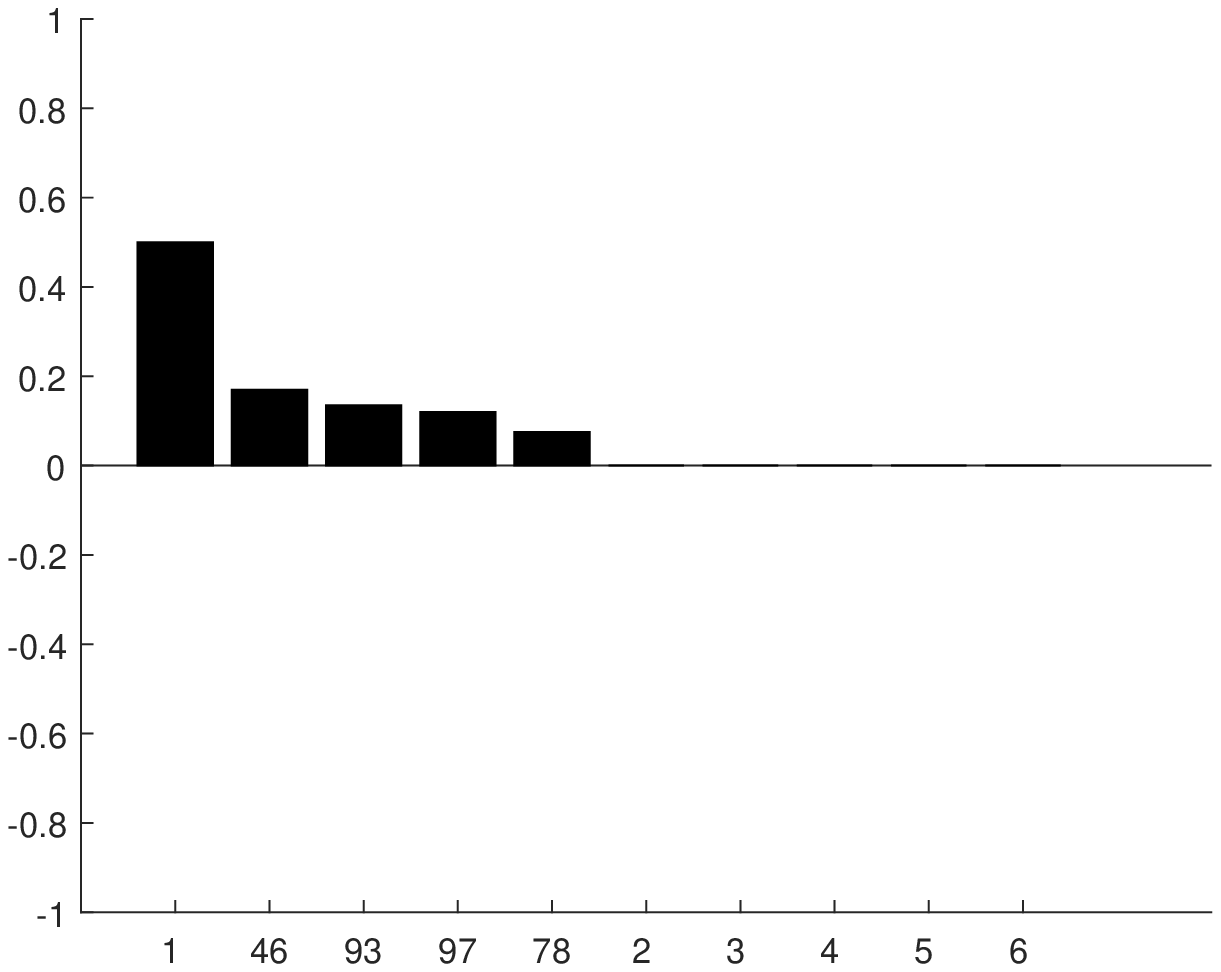}
\caption*{(C) MDD}
\end{subfigure}\hspace*{\fill}
\begin{subfigure}{0.20\textwidth}
\includegraphics[width=\linewidth]{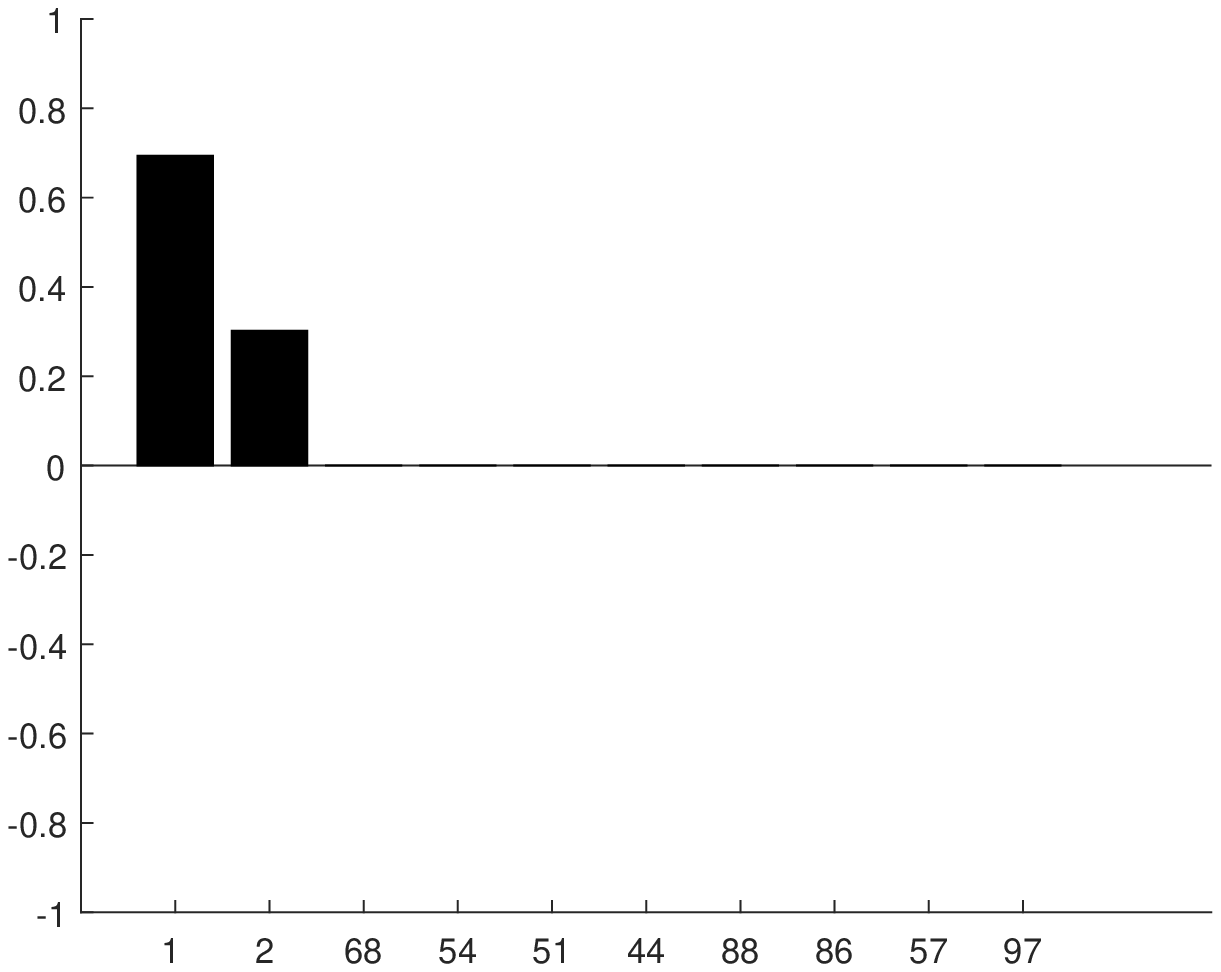}
\caption*{(C) ADH}
\end{subfigure}\hspace*{\fill}
\begin{subfigure}{0.20\textwidth}
\includegraphics[width=\linewidth]{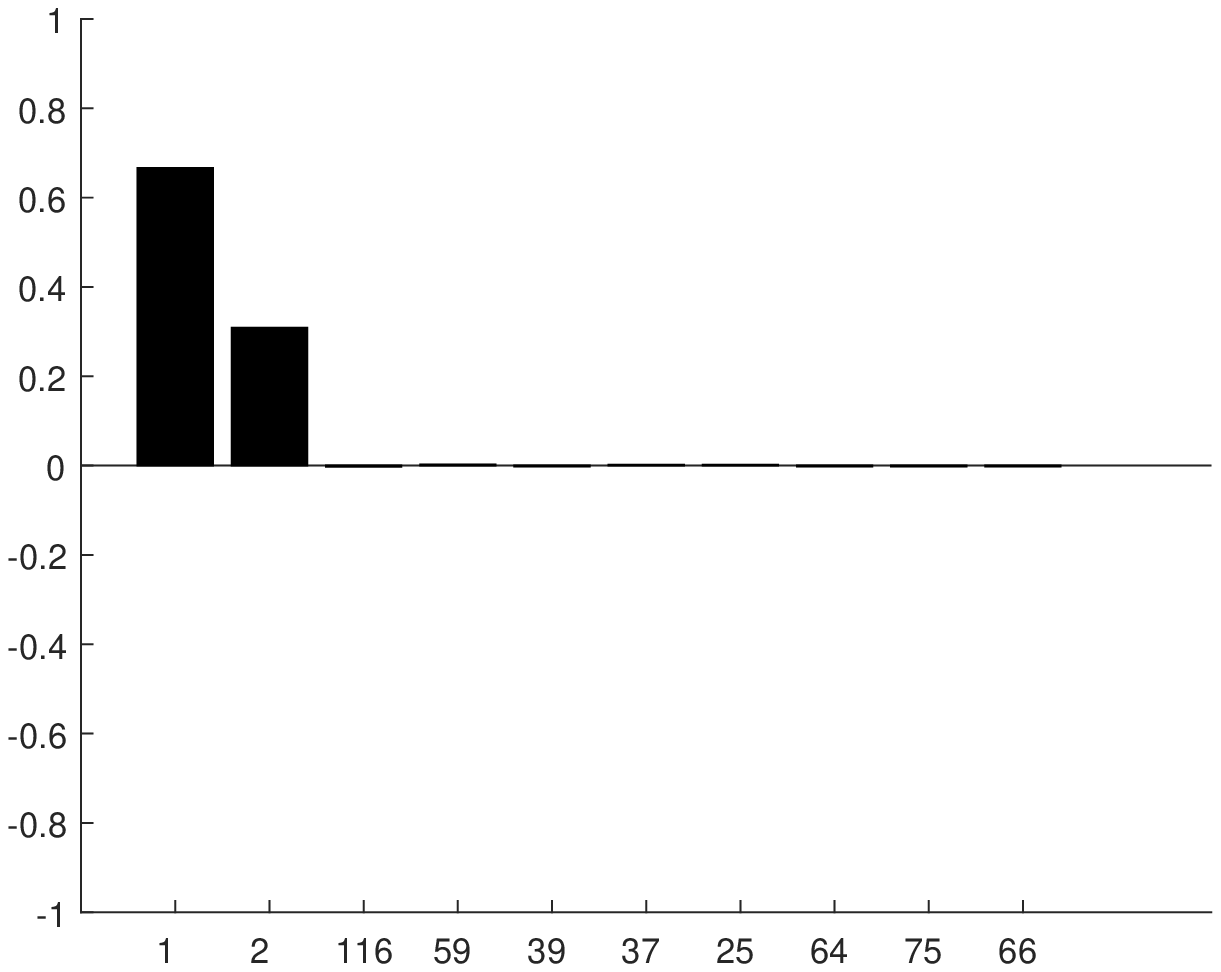}
\caption*{(C) PCR}
\end{subfigure}\hspace*{\fill}
\begin{subfigure}{0.20\textwidth}
\includegraphics[width=\linewidth]{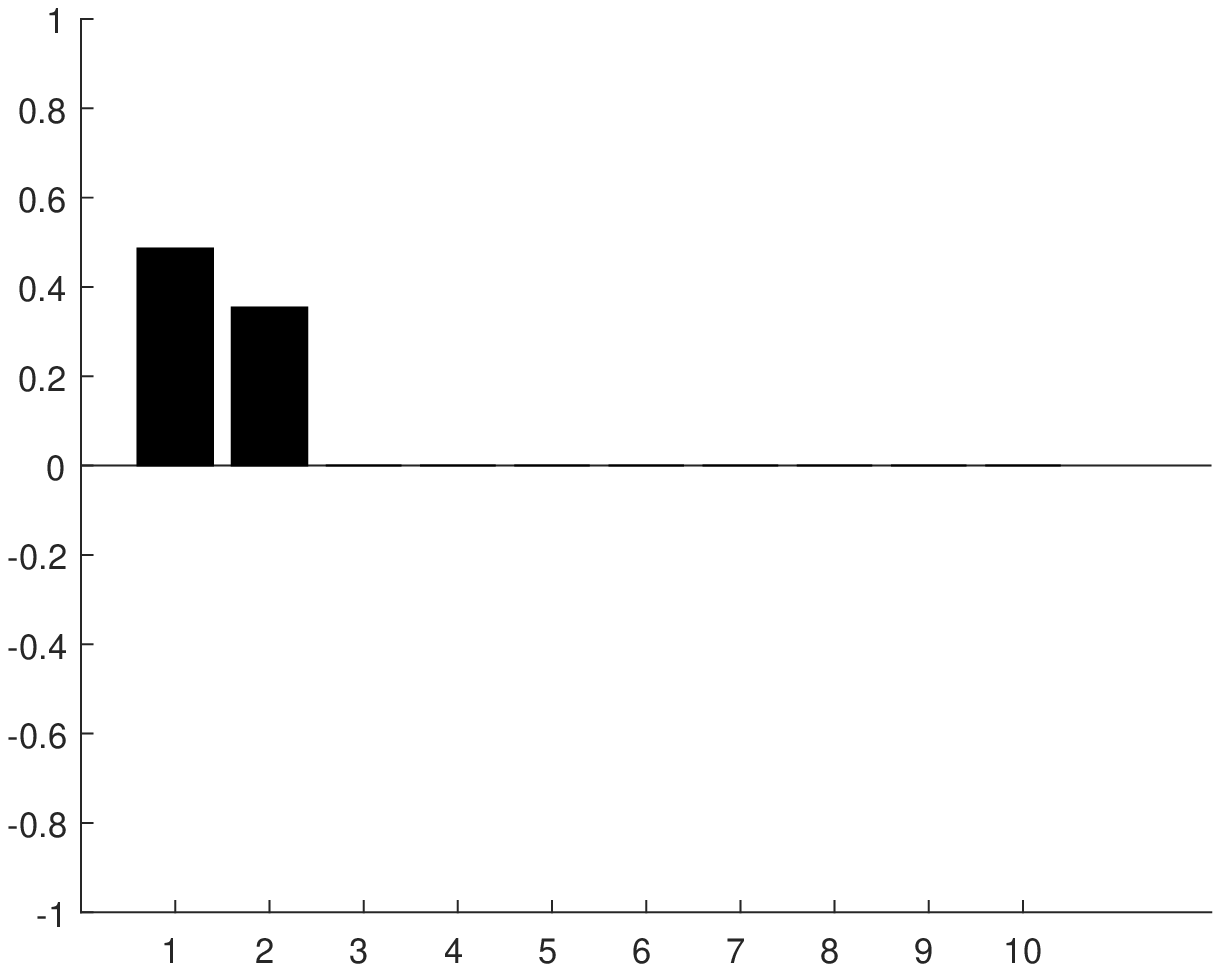}
\caption*{(C) LASSO}
\end{subfigure}\hspace*{\fill}
\begin{subfigure}{0.20\textwidth}
\includegraphics[width=\linewidth]{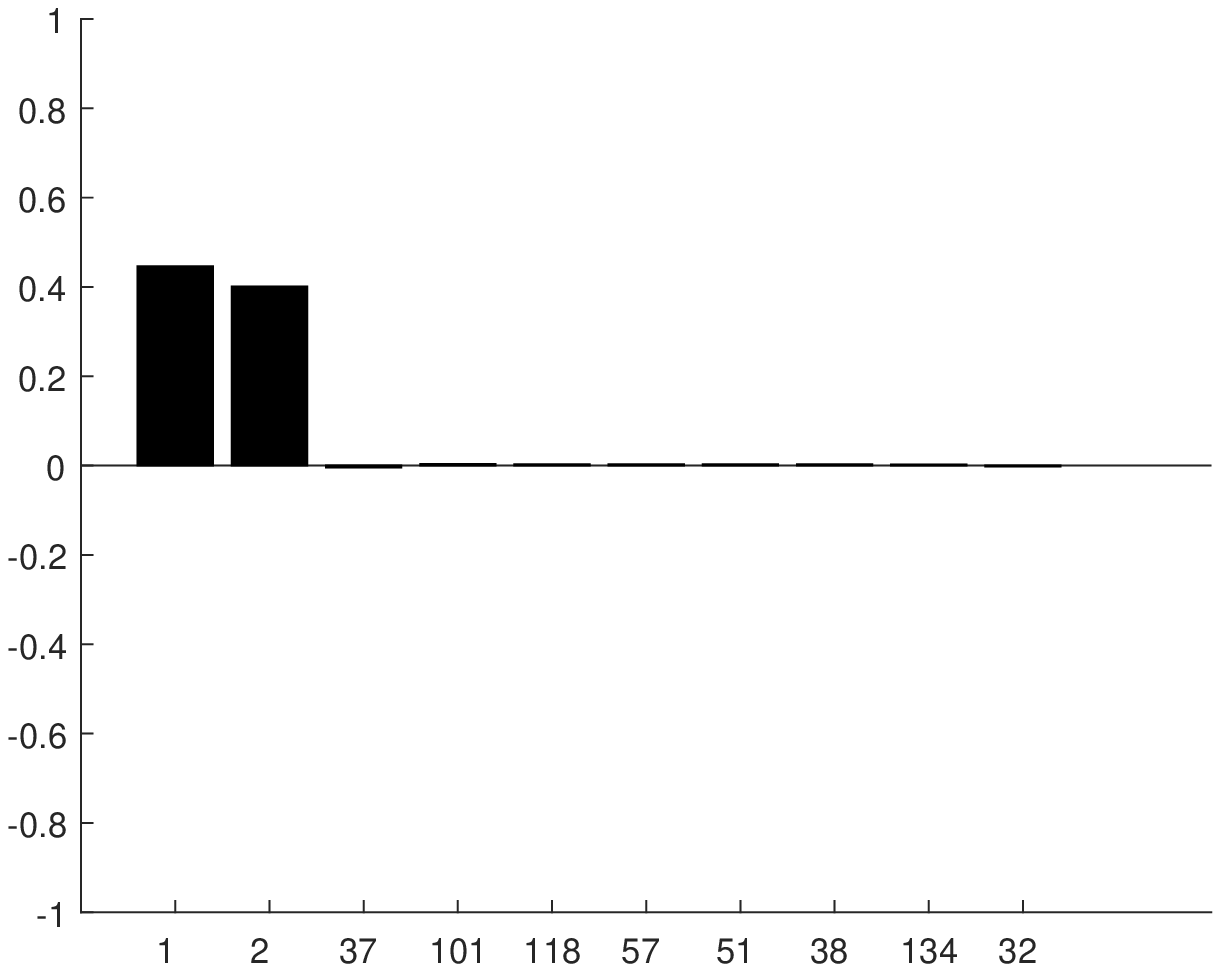}
\caption*{(C) BSTS}
\end{subfigure}
\medskip
\begin{subfigure}{0.20\textwidth}
\includegraphics[width=\linewidth]{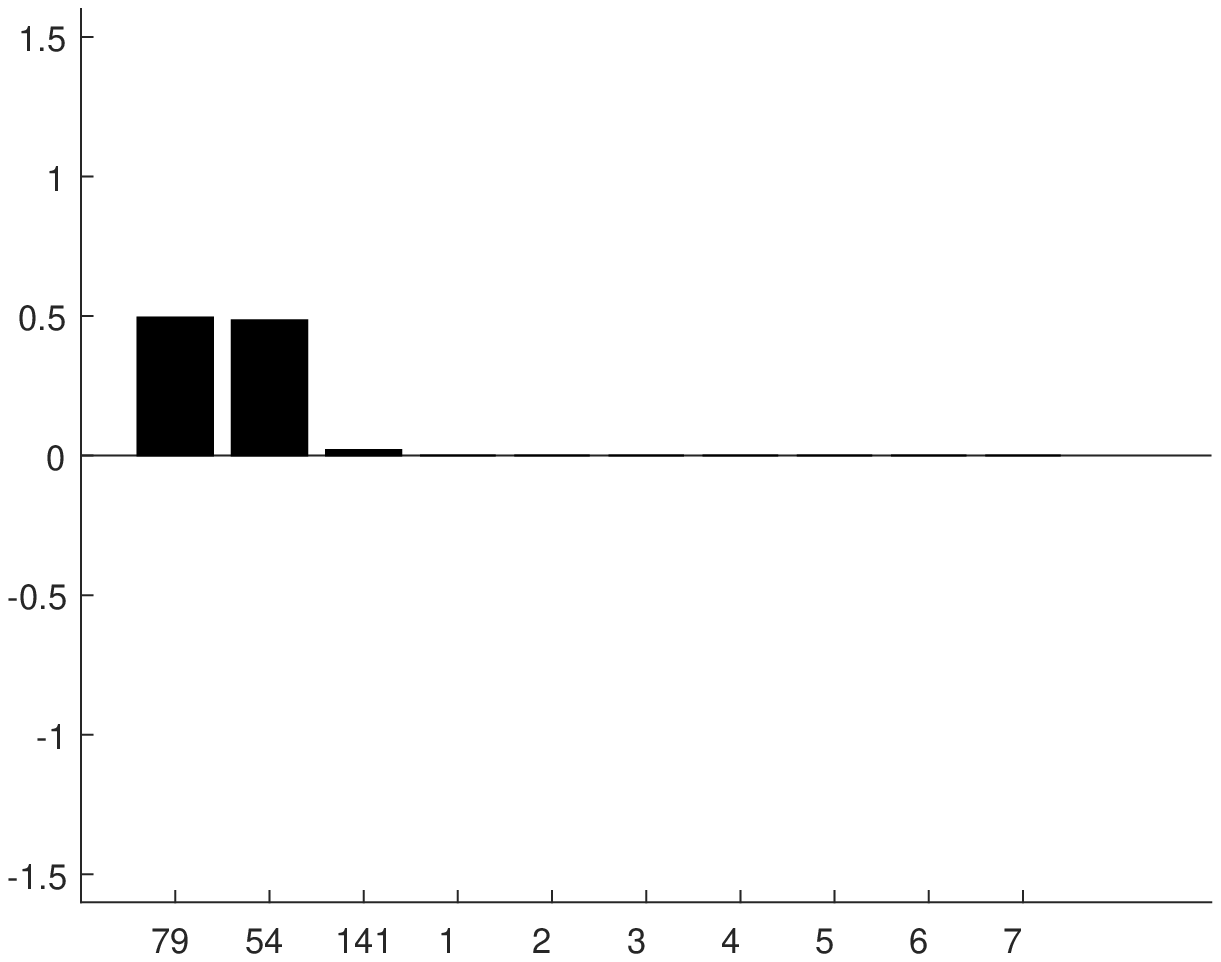}
\caption*{(D) MDD}
\end{subfigure}\hspace*{\fill}
\begin{subfigure}{0.20\textwidth}
\includegraphics[width=\linewidth]{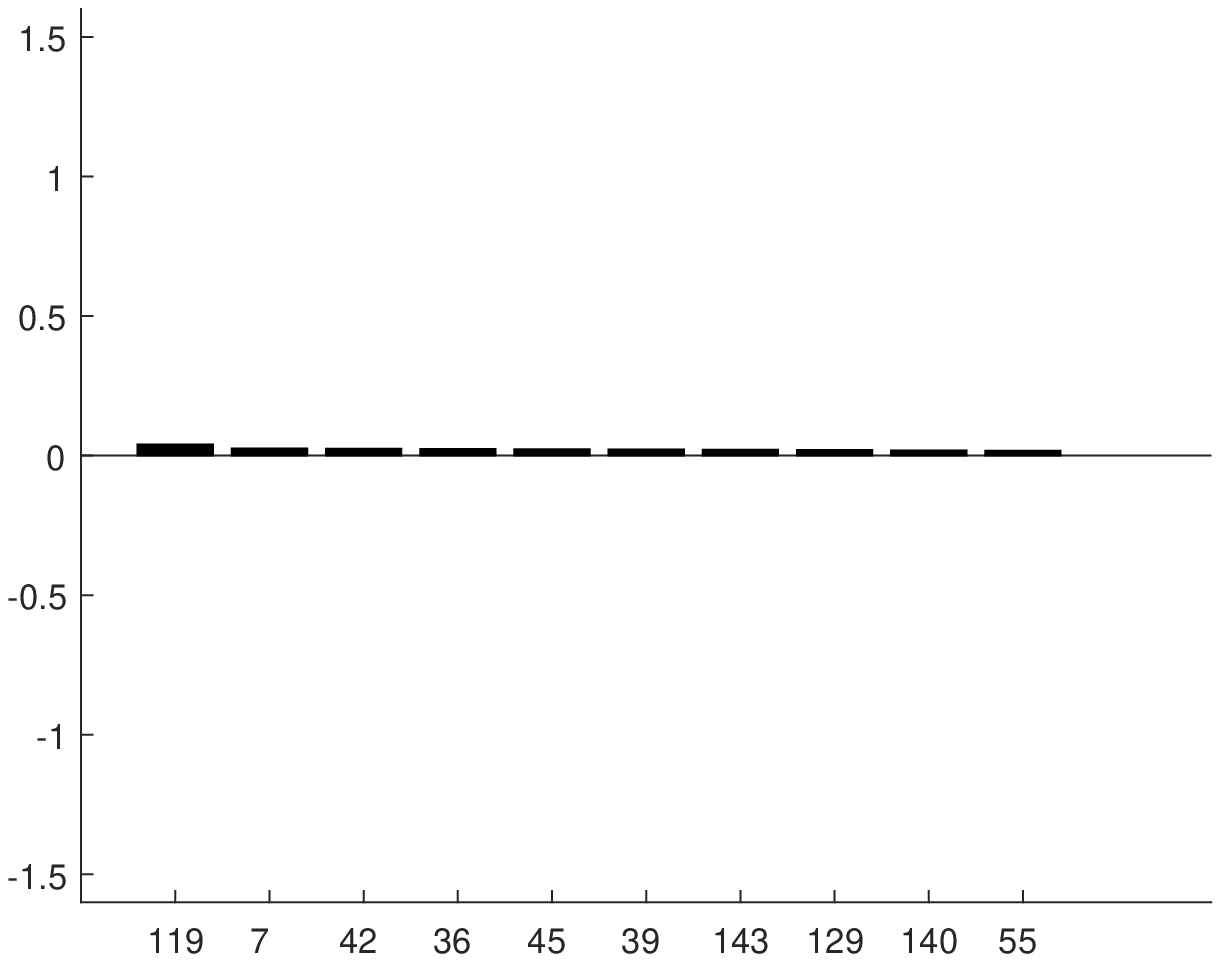}
\caption*{(D) ADH}
\end{subfigure}\hspace*{\fill}
\begin{subfigure}{0.20\textwidth}
\includegraphics[width=\linewidth]{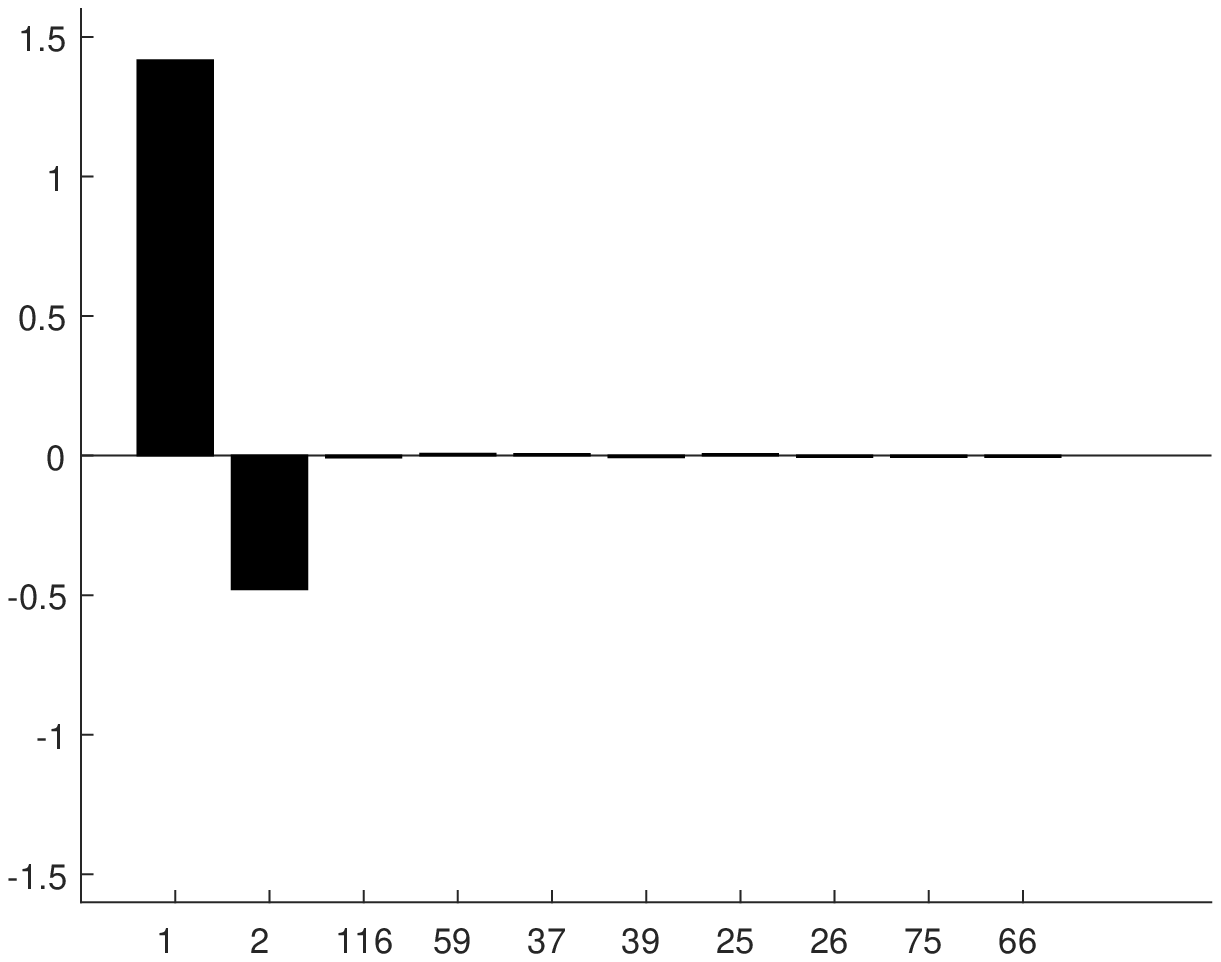}
\caption*{(D) PCR}
\end{subfigure}\hspace*{\fill}
\begin{subfigure}{0.20\textwidth}
\includegraphics[width=\linewidth]{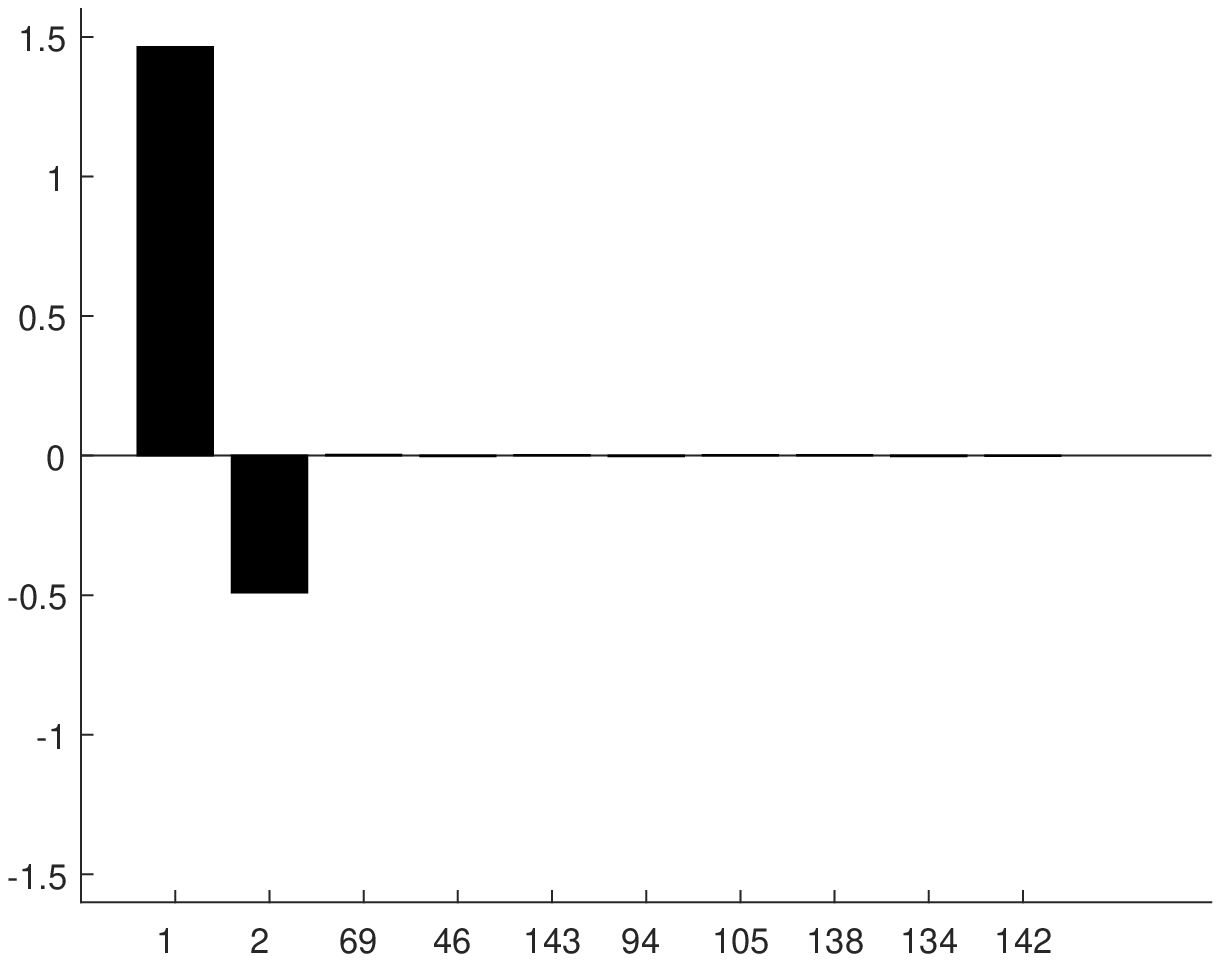}
\caption*{(D) LASSO}
\end{subfigure}\hspace*{\fill}
\begin{subfigure}{0.20\textwidth}
\includegraphics[width=\linewidth]{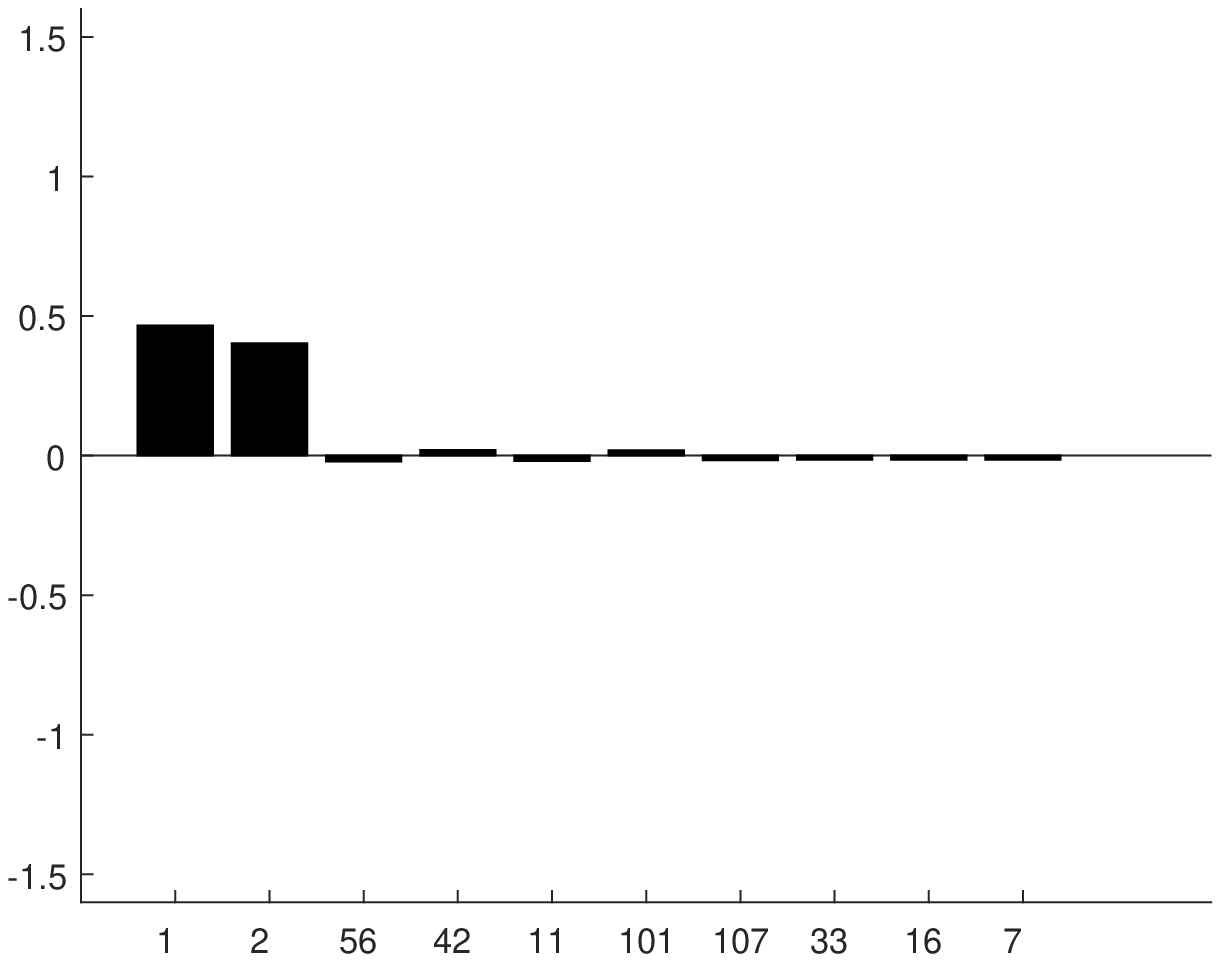}
\caption*{(D) BSTS}
\end{subfigure}
\medskip
\begin{subfigure}{0.20\textwidth}
\includegraphics[width=\linewidth]{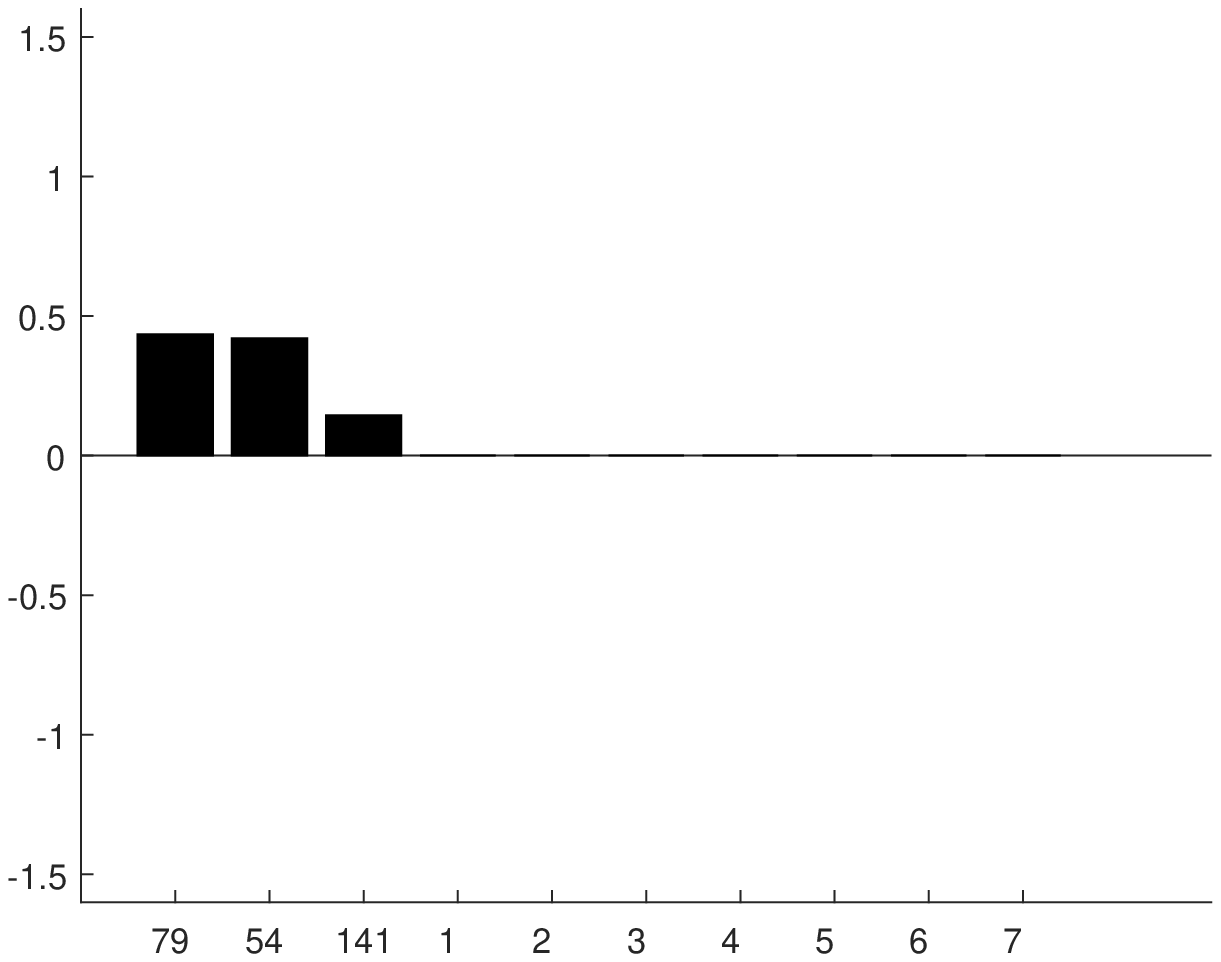}
\caption*{(E) MDD}
\end{subfigure}\hspace*{\fill}
\begin{subfigure}{0.20\textwidth}
\includegraphics[width=\linewidth]{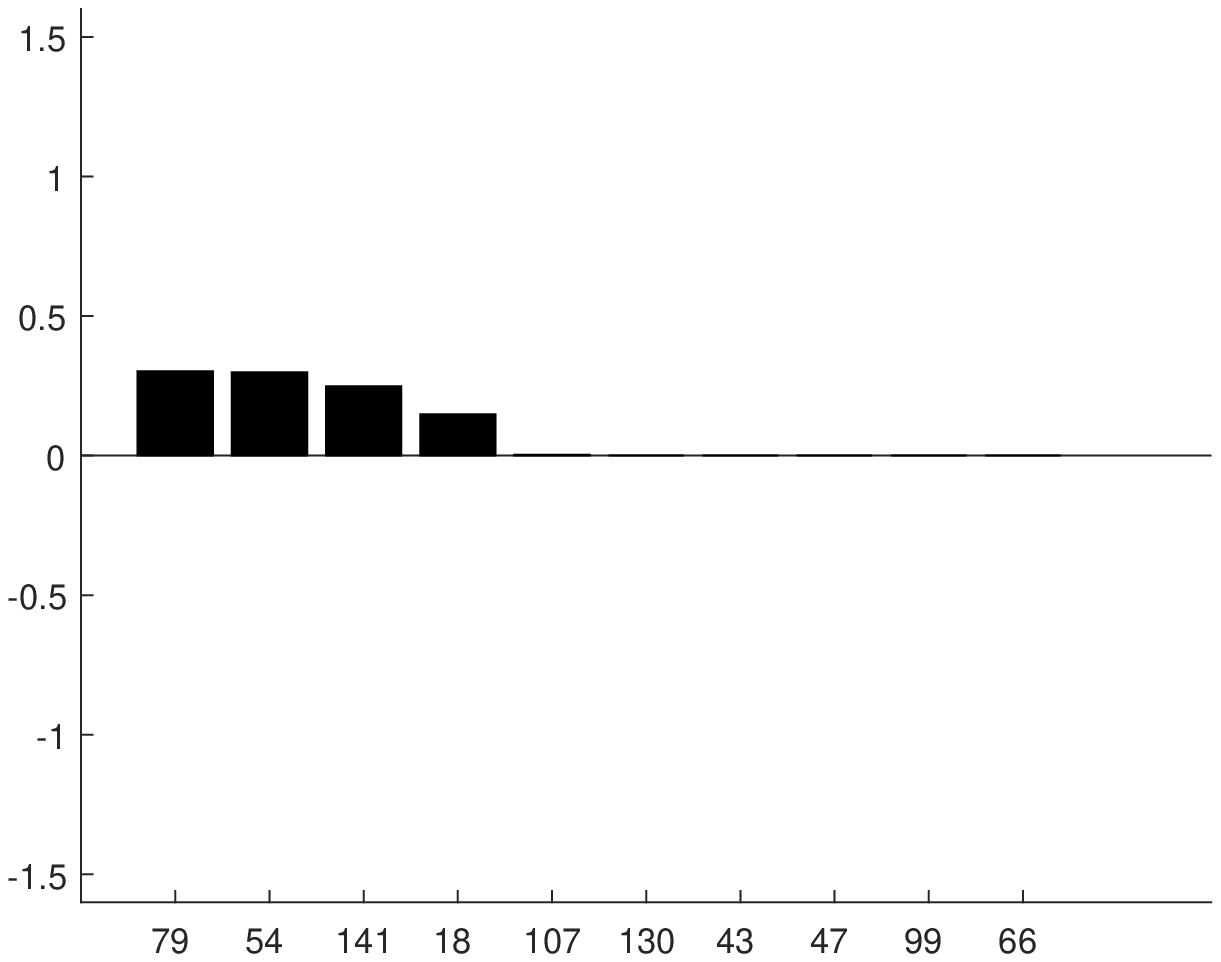}
\caption*{(E) ADH}
\end{subfigure}\hspace*{\fill}
\begin{subfigure}{0.20\textwidth}
\includegraphics[width=\linewidth]{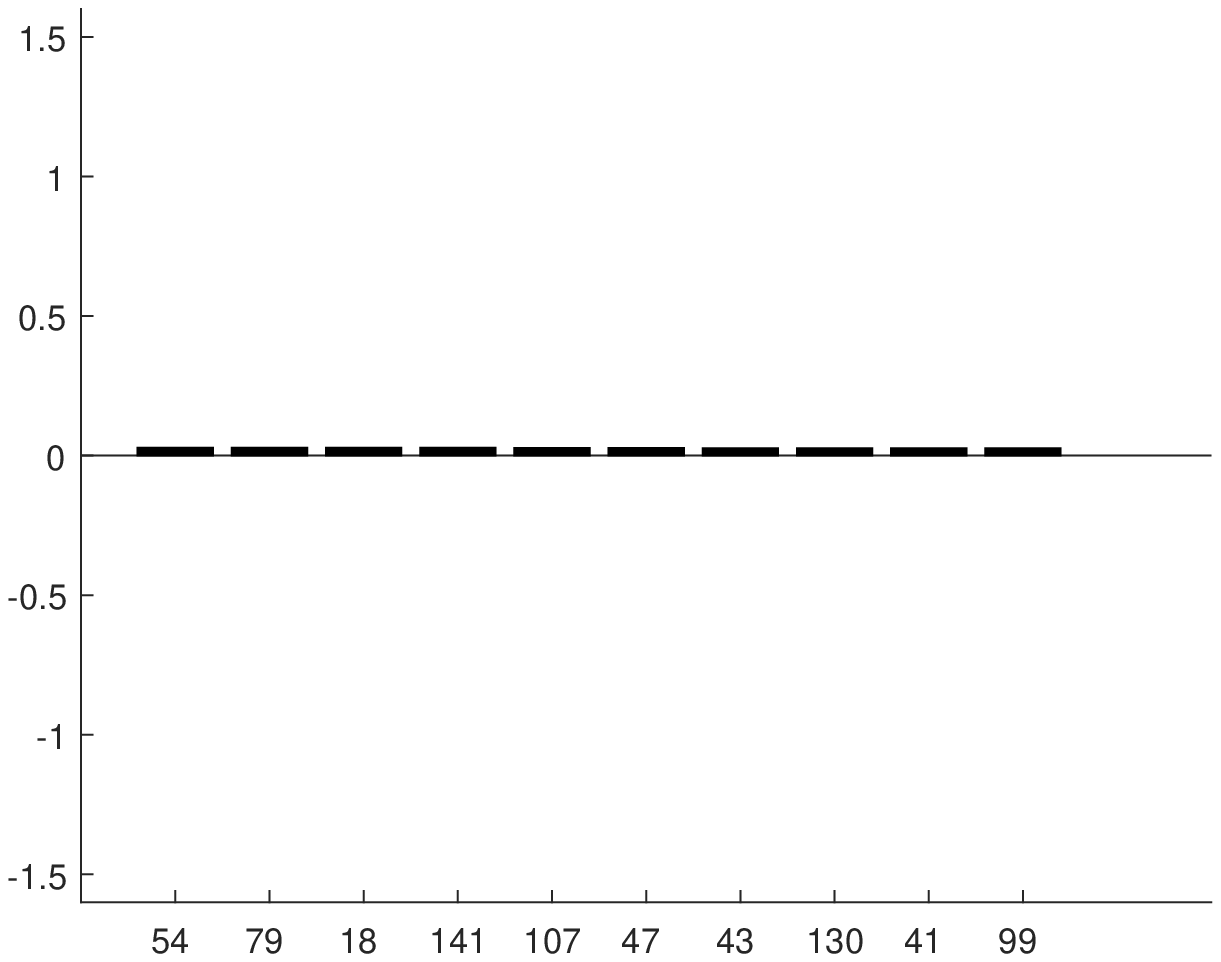}
\caption*{(E) PCR}
\end{subfigure}\hspace*{\fill}
\begin{subfigure}{0.20\textwidth}
\includegraphics[width=\linewidth]{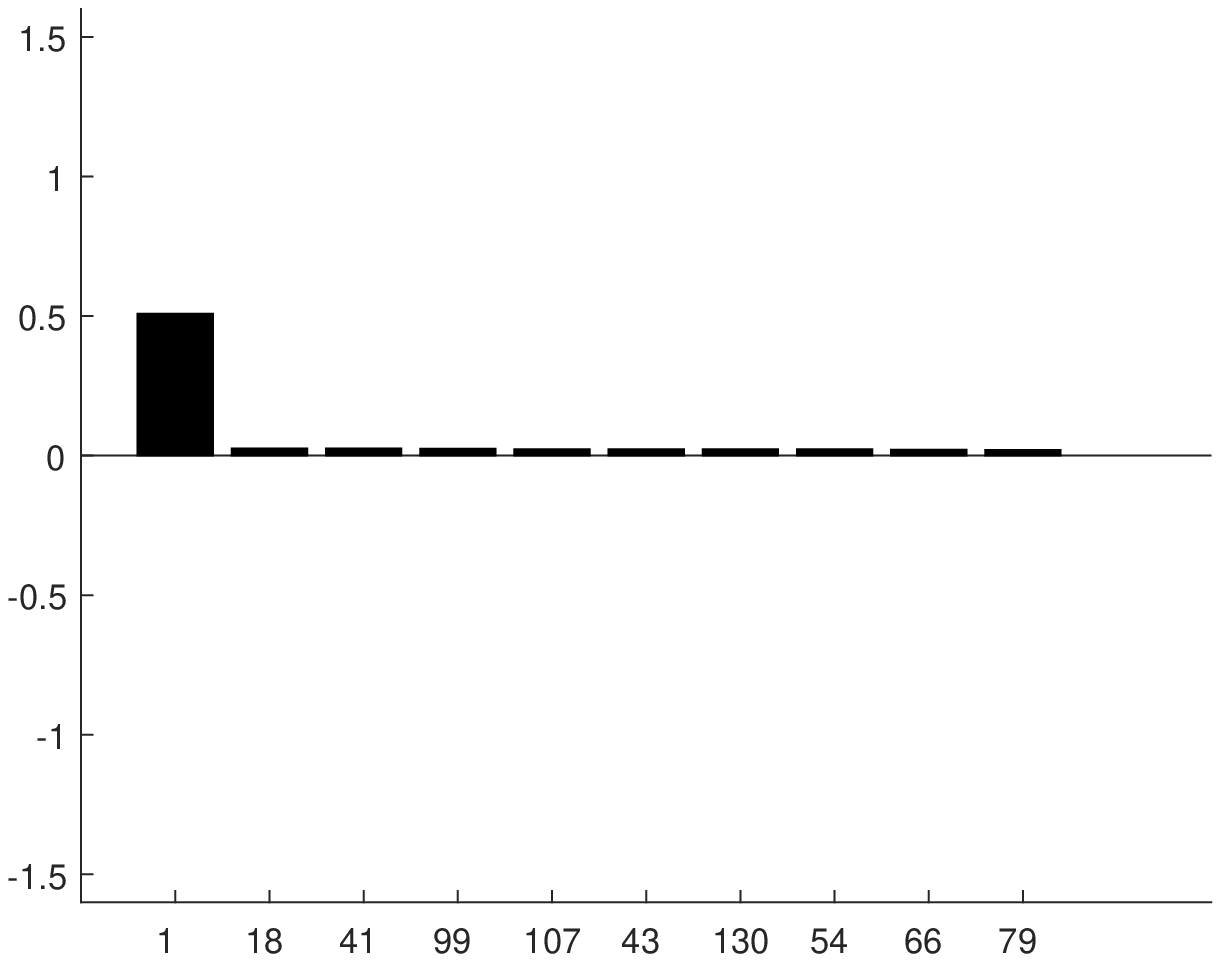}
\caption*{(E) LASSO}
\end{subfigure}\hspace*{\fill}
\begin{subfigure}{0.20\textwidth}
\includegraphics[width=\linewidth]{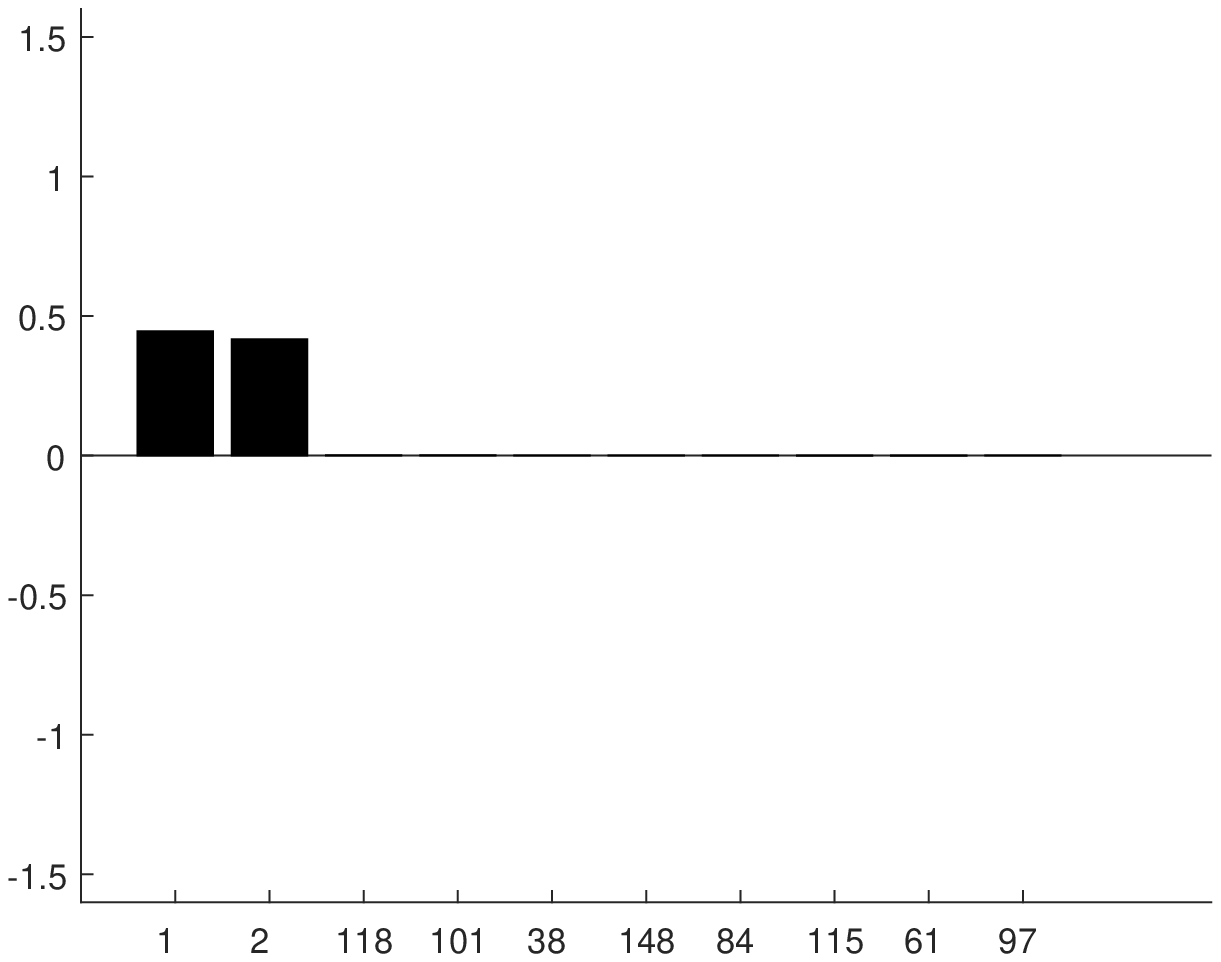}
\caption*{(E) BSTS}
\end{subfigure}
\medskip
\begin{subfigure}{0.20\textwidth}
\includegraphics[width=\linewidth]{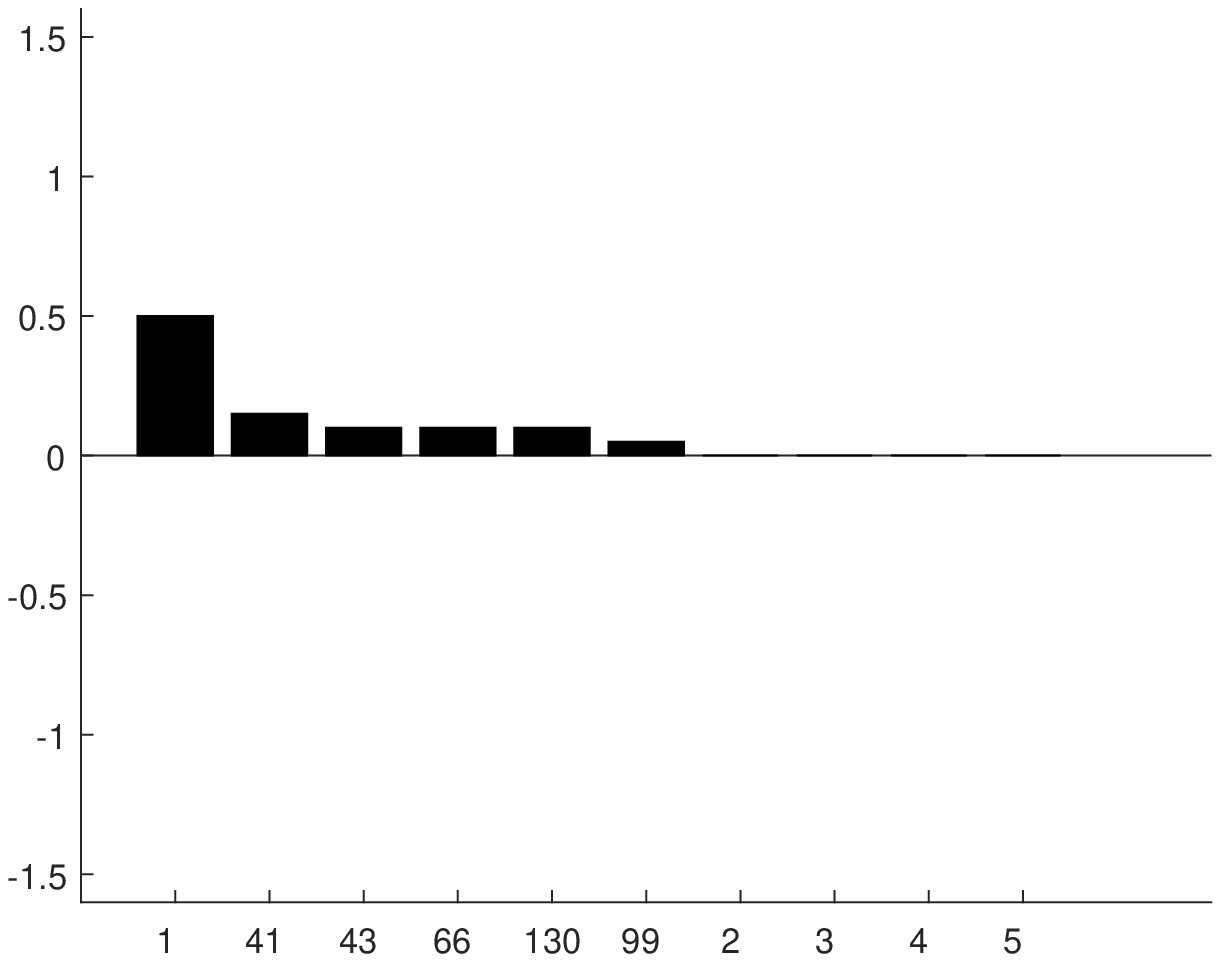}
\caption*{(F) MDD}
\end{subfigure}\hspace*{\fill}
\begin{subfigure}{0.20\textwidth}
\includegraphics[width=\linewidth]{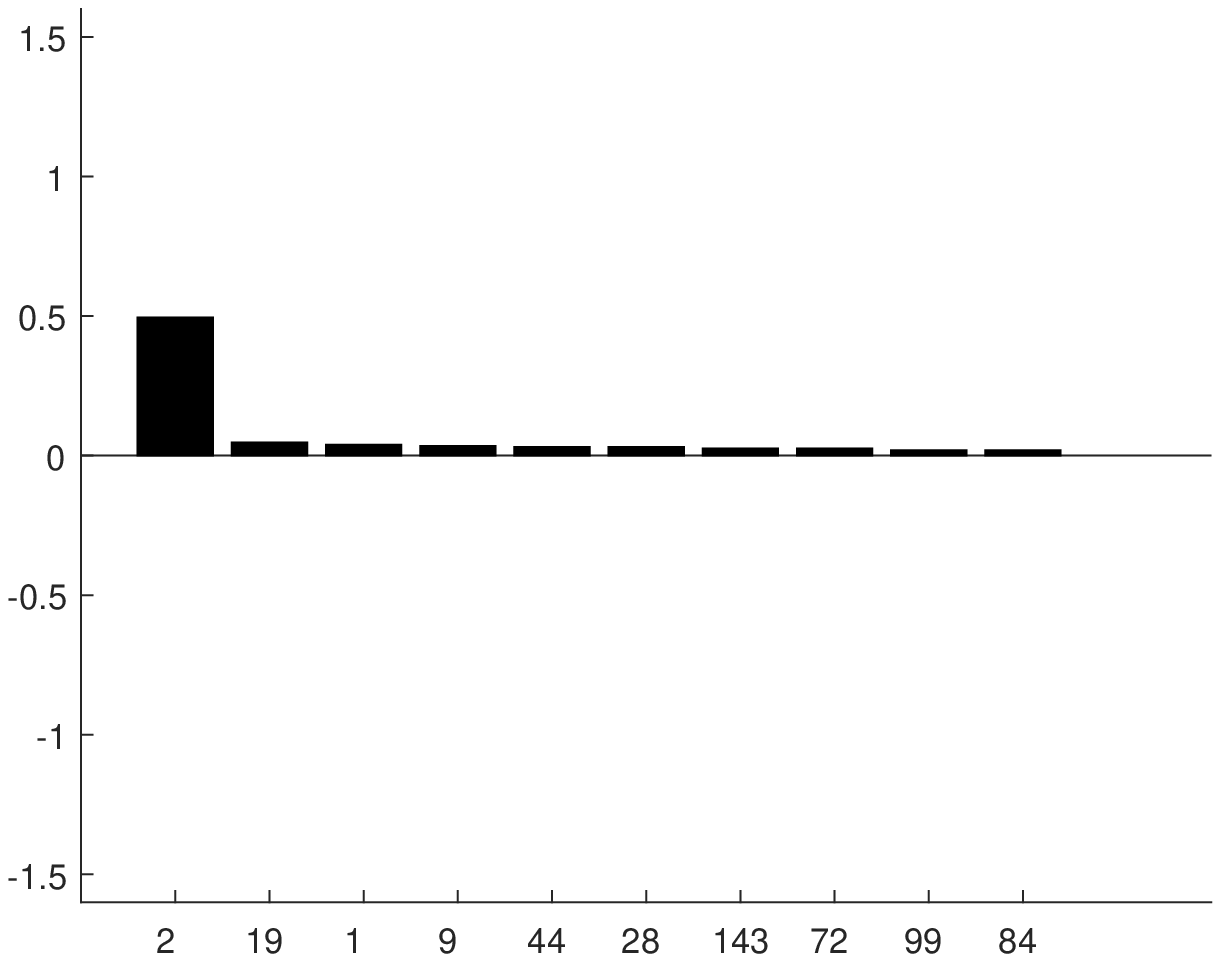}
\caption*{(F) ADH}
\end{subfigure}\hspace*{\fill}
\begin{subfigure}{0.20\textwidth}
\includegraphics[width=\linewidth]{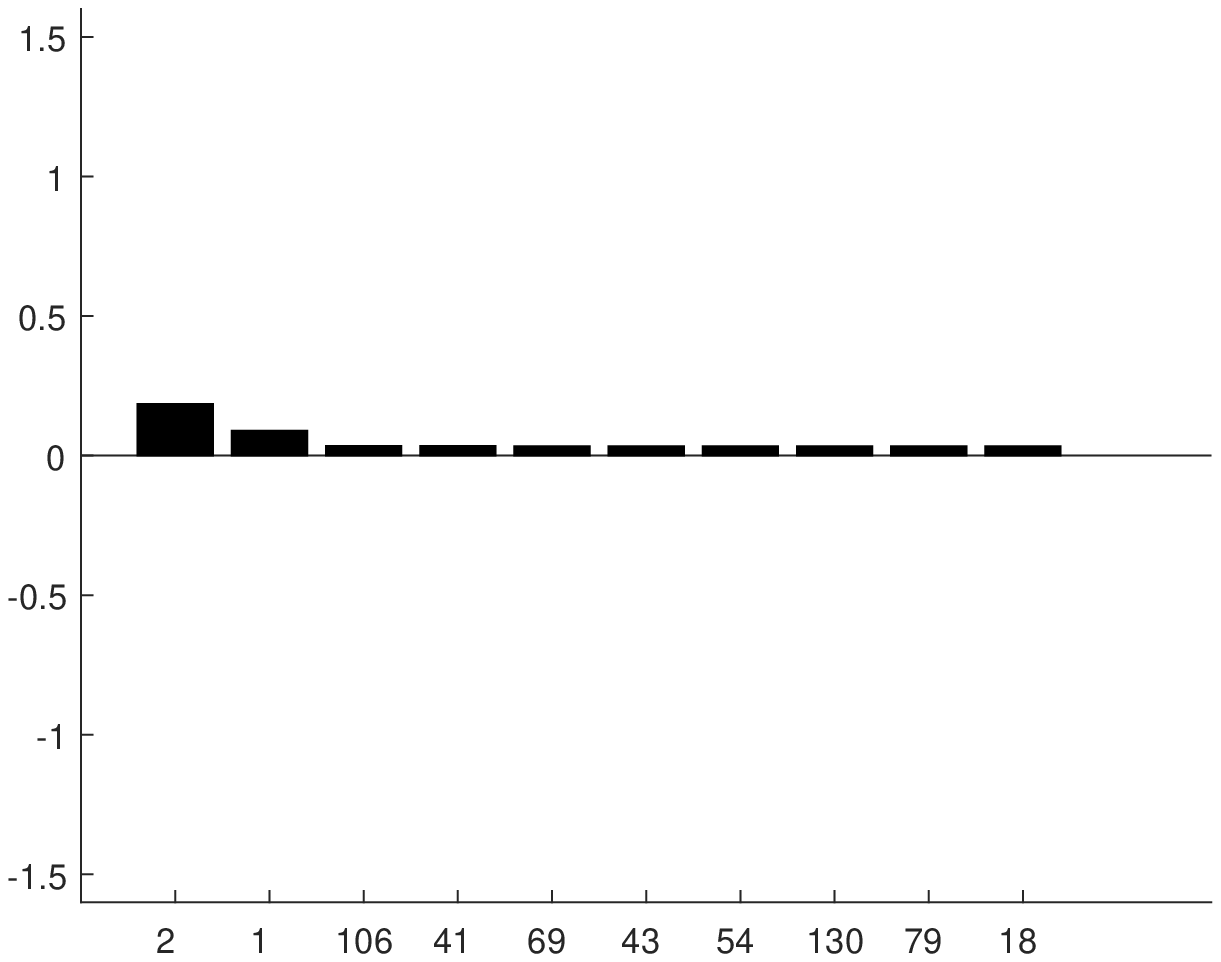}
\caption*{(F) PCR}
\end{subfigure}\hspace*{\fill}
\begin{subfigure}{0.20\textwidth}
\includegraphics[width=\linewidth]{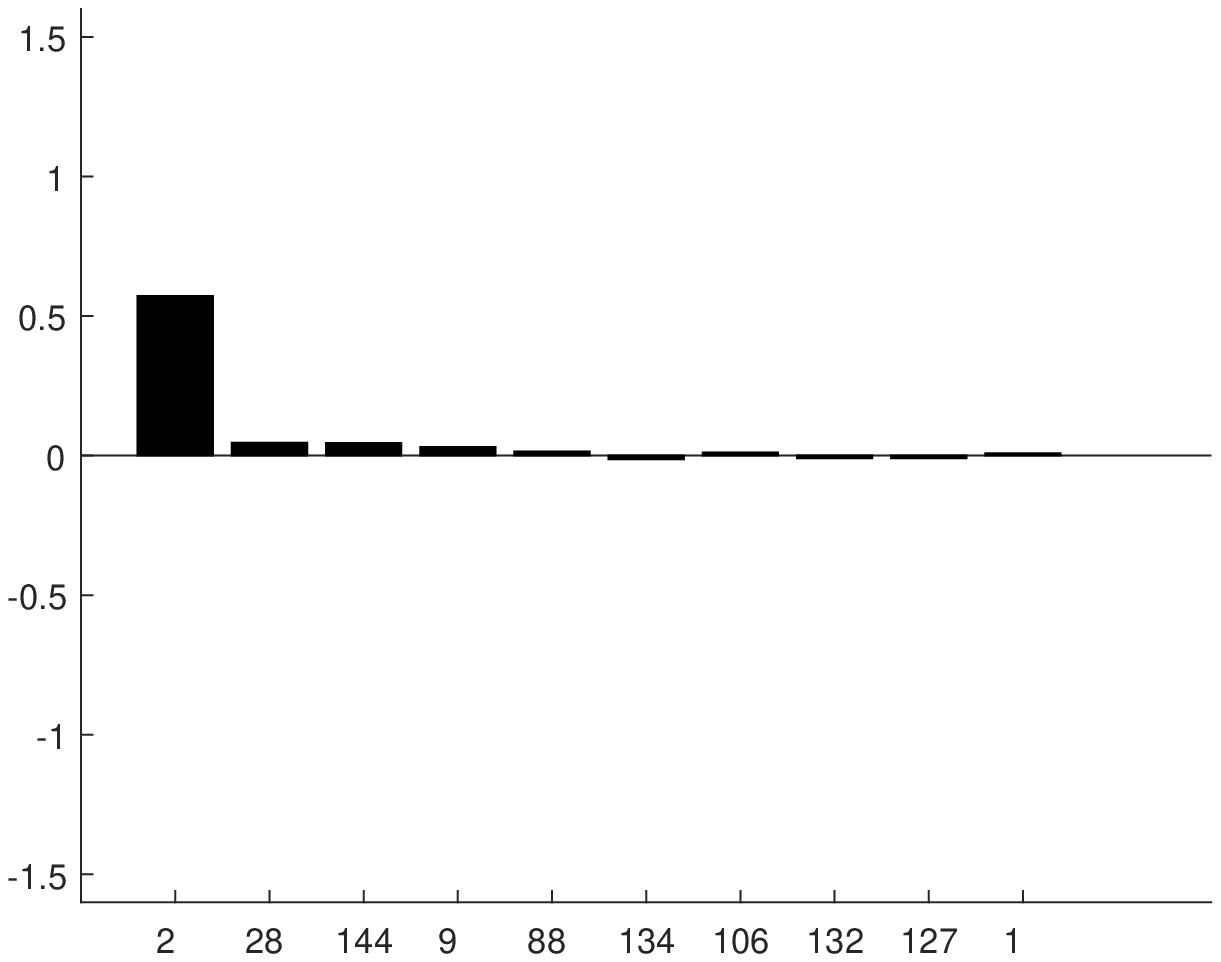}
\caption*{(F) LASSO}
\end{subfigure}\hspace*{\fill}
\begin{subfigure}{0.20\textwidth}
\includegraphics[width=\linewidth]{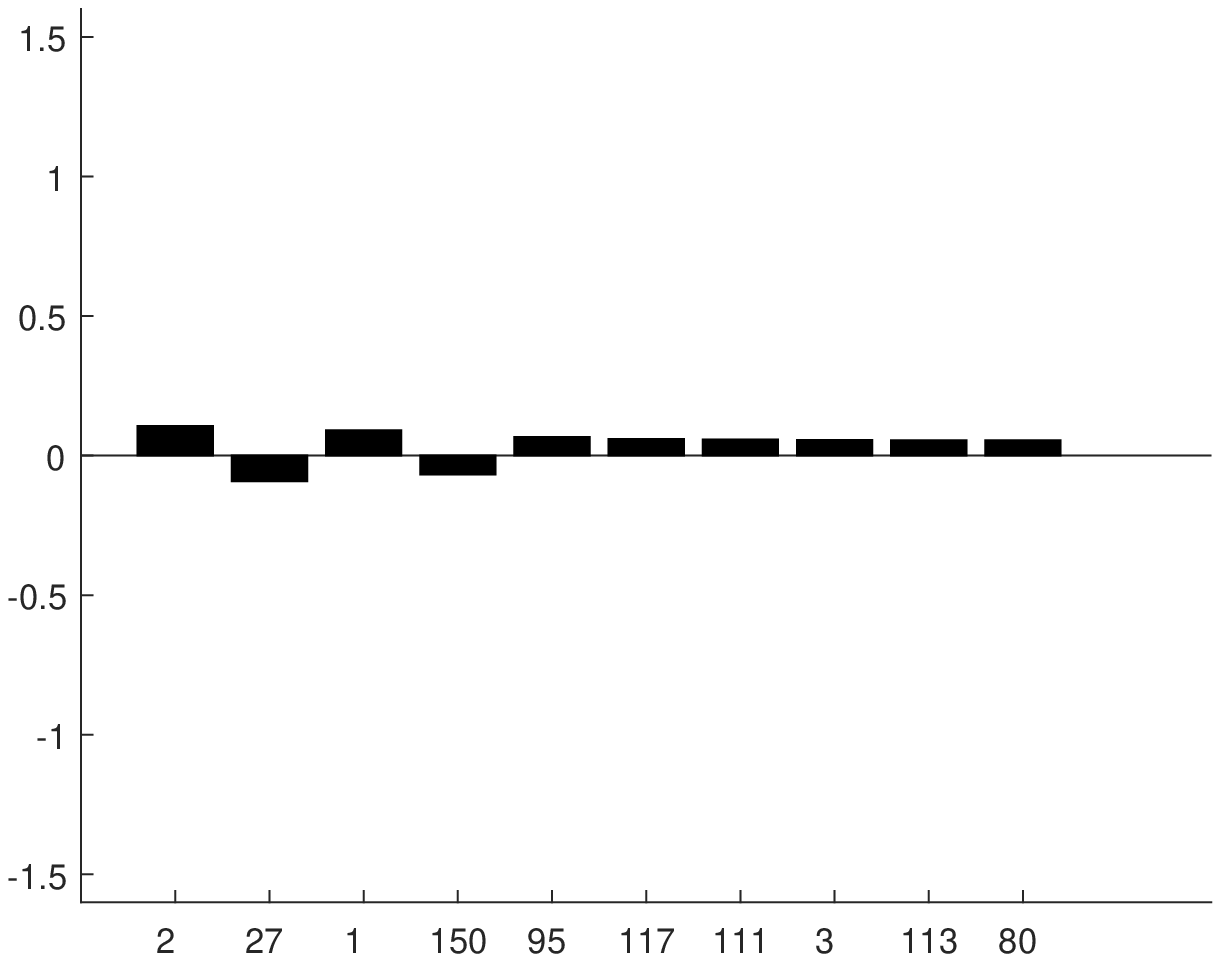}
\caption*{(F) BSTS}
\end{subfigure}
\caption{\footnotesize The 10 largest average weights for Difference in Differences post Matching (MDD), the original Synthetic Control (ADH), Principal Component Regression (PCR), Lasso regularization (LASSO) and Bayesian Structural Time Series (BSTS) based on 100 samples of potential outcomes from the model in (\ref{eq_retail_sim_model}) for scenarios (B)-(F).} 
\label{fig_retail_sim_weights}
\end{minipage}
\end{figure}

 Scenario (B) introduces the seasonality component. Choosing the two first control units is now important, as this will balance the the trend and seasonality component of the treated unit with the combination of the control units. Figure \ref{fig_retail_sim_weights} shows histograms of the 10 largest weights in terms of average values in repeated samples. All methods are able to select sparse sets of control units, in large containing the relevant units. Furthermore, average treatment effect estimates are precise and comparable across methods. The scenario highlights the importance of distinguishing between parallel trends and seasonality. Variable selection is important in the presence of seasonality, even when the trends of the potential outcomes are parallel. Figure \ref{fig_retail_sim_bsts} shows results from the Gibbs sampling procedure for BSTS for a specific draw of the data for Scenario (B). As argued by \cite{brodersen2015inferring}, an advantage with BSTS is the possibility to construct an uncertainty band around the counterfactual prediction that takes into account both parameter uncertainty and model selection. The figure show that the Gibbs sampling procedure has converged and parameter estimates are in line with the true values. 

\par Scenario (C) introduce time-varying confounding factors. In this setting it is cruicial that the synthetic control picks the correct subset of control units in order to balance the confounding factor. In theory, a good match removes the time-varying confounding factors and subsequently applying DD to the matched data may remove any remaining imbalances in time-invariant confounding factors, see e.g. \cite{o2016estimating}. The results in Table \ref{tab_retail_sim_results} suggest that matching combined with DD does not succeed in removing the bias. Indeed, Figure \ref{fig_retail_sim_weights} show that control unit 2 is almost never chosen in this scenario. This has to do with the particular matching algorithm applied here and it is likely that a more sophisticated algorithm could improve the results. ADH obtain an accurate estimate of the average treatment effect in this scenario. This result is consistent with \cite{abadie2010synthetic}, who argue that ADH allows the effects of confounding unobserved characteristics to vary in time. Interestingly, BSTS perform similar to ADH, but is based on a larger pool of controls in most cases. However, the weights assigned to the irrelevant controls are negligible, in large due to the fact that control unit selection is chosen in each Gibbs sampling iteration. The results show similar variances in both cases, indicating no signs of overfitting. Figure \ref{fig_retail_sim_weights} reveals that Lasso, BSTS and even control unit selection based on the leading principal components tend to assign high weights to the relevant control units. The key takeway from this scenario is that machine learning methods produce comparable results to ADH when the unobserved confounding factor is assumed to vary in time. In summary, these methods tend to pick relevant control units which in turn yields credible estimates of the average treatment effect.

\par Scenario (D) highlights the situation where the treated unit lies outside the distribution of relevant control units. It is well known in the literature that ADH fails in this setting due to the restrictive weight assumption, see e.g. \cite{abadie2010synthetic} and \cite{athey2017state}. The shortcomings of ADH is easily seen from Figure \ref{fig_retail_sim_weights}, showing that the method fails in choosing a sparse set of control units. Intuitively, for each draw of the data ADH will select the unit that lies closest to the treated unit and assign a value of 1.0 to this control unit. This results in low variance, but high bias. The machine learning approaches, however, use negative weights and thus provide more flexibility to reconstruct the treated unit. Indeed, PCR, Lasso and BSTS obtain weights in line with the data generating process, leading to improved estimates of the treatment effect. 

\par In Scenario (E), the similarity between units and the large noise makes it hard do detect the underlying trend. BSTS should have an edge in this setting. Intuitively, since the trend is modelled without consideration for control units, the variable selection exercise reduce to choosing a combination of seasonality patterns. In contrast, any reduced form approach chooses control units based on both trend and seasonality. The results in Table \ref{tab_retail_sim_results} and Figure \ref{fig_retail_sim_weights} show that BSTS yield the lowest mean squared error and is the only approach that on average selects both relevant control units. Lasso and PCR obtain similar results for the mean treatment effect estimate, but with less precision.

\par In Scenario (F) the pretreatment period is short and the noise is large with abrupt jumps. Identification based on the conditional independence assumption is likely to fail in this setting, since the synthetic control is unlikely to capture the correct units in a noisy, short pretreatment period. Table \ref{tab_retail_sim_results} show that the mean treatment effect estimates are acceptable for Lasso and BSTS. However, there are clear signs of overfitting, resulting in a high out of sample mean squared error for all methods. In other words, if the pretreatment period is short, the number of controls must be relatively small for the methods to work.

\section{Applications} \label{sec_retail_applications}

\subsection{Liberalisation episodes and economic growth}
Economic theory suggest a positive relationship between economic liberalisation and welfare. \cite{billmeier2013assessing} (``BN'') propose to investigate this relationship empirically by using a large set of country case studies, for each country assessing the effect of economic liberalization on development in income per capita. Several of the case studies in BN do not indicate that the economic liberalisation had an effect on income. By using the machine learning tools presented in this paper I find positive effects of liberalisation for several countries in Africa, Asia and the Middle East. 

\par I use the same data as in BN, covering a total of $J=180$ countries over the period 1963 to 2000. The liberalisation indicator $d_{jt}$ is equal to one if country $j$ is open at time $t$ and zero otherwise. A country is considered closed if (a) average tariffs exceed 40\%, (b) nontariff barrieers cover more than 40\% of imports, (c) it has a socialist economic system, (d) the black market premium on the exchange rate exceds 20\% and (e) many of its exports are controlled by a sate monopoly.\footnote{BN argue that the identification strategy could suffer from reverse causation in some cases. If the timing of liberalisation is set due to expectations of future growth, estimates of the liberalisation effect could be biased.}

\par I revisit each of the 30 case studies in the original paper using ADH, PCR, LASSO and BSTS. Consistent with the identification strategy outlined in Section \ref{sec_retail_framework_id}, I focus on the outcome variable ``GDP per capita'' and ignore covariates. In contrast, BN only consider one method, ADH, but chooses controls based on additional variables such as schooling, population growth, investment share, inflation and a democracy indicator. Despite this difference, my ADH results are in large close to the original study; compare column 6 and 7 of Table \ref{tab_retail_app_countries}. 

\par In Table \ref{tab_retail_app_countries} I report that a majority of my conclusions are consistent with BN. I divide the case studies into three groups. The first group contain cases where BN find a robust or somewhat robust effect on welfare 10 years after the economic liberalisation. This group include Singapore, South Korea, Indonesia, Colombia, Costa Rica, Botswana, Mauritius, Ghana, Guinea, Benin and Morocco. Using the machine learning algorithms outlined above I obtain the same conclusion for all of these countries.

\par The second group consist of countries where BN do not find any effect 10 years post treatment. This group includes Chile, Gambia, Cape Verde, Guinea-Bissau, Zambia, Kenya, Cameroon, Niger, Mauritania and Egypt. For Cape Verde and Guinea-Bissau the conclusion in BN seem to be unclear, as outperformance of the counterfactual only slowly emerges after the economic liberalisation. I arrive at the same conclusion for all of the countries in the second group. The results suggest that the machine learning approaches broadly agree with the ADH for a large set of country liberalisation episodes, even when the additional covariates used by BN is discarded.

\par The third group gather all countries where I obtain a different conclusion from BN. This group include Barbados, Mexico, Nepal, Mali, Uganda, Philippines, South Africa, Ivory Coast and Tunisia. To assess these differences in some detail, I further divide the case studies into three subcategories.

\par The first category include Barbados and Mexico, where BN find a strong, robust effect that is rejected by the machine learning approaches. The results for Mexico are reported in Figure \ref{fig_retail_app_countries_mexico}. BN find a strong positive effect on welfare due to a large decline in the counterfactual. I find no such decline based on the machine learning approaches, suggesting at best that the effect of liberalisation is not as pronounced as in the original study. The Barbados study is particularly hard due to the short pretreatment period, $T_0=3$. This is not a good scenario for the conditional independence assumption, and my results should be interpreted with caution in this case.

\par The second subcategory consist of Nepal, Mali and Uganda, where the conclusion from BN is somewhat unclear. In each of these cases, an effect is obtained 10 years after the liberalisation. However, it is hard to link this effect directly to the treatment, as GDP is only gradually outperforming the counterfactual. In all of these cases, the flexibility of the machine learning approaches improves the match to the treated unit, leading to more pronounced effects. The case of Uganda is reported in Figure \ref{fig_retail_app_countries_uganda}. The results show some robustness, but are in large less robust than the cases in group 1. 

\par Philippines, South Africa, Ivory Coast and Tunisia defines the final subcategory. In each of these studies, BN do not find an effect of the economic liberalisation. In contrast, I find positive effects on welfare based on several of the machine learning approaches. However, placebo studies indicate that the results only show limited robustness. The case of Tunisia is shown in Figure \ref{fig_retail_app_countries_tunisia}. 

\subsection{Effect of bank-specific capital requirements}
A key part of bank regulation is to ensure that banks have enough equity to handle losses. The aim of this regulation is to ensure solvency of banks, in order to protect its customers and avoid large government costs due to bank failure. I revisit the study by \cite{getz2016consequences}, where a Norwegian bank reform is used to study how banks respond to higher capital requirements.
\begin{figure}
\centering
\includegraphics[scale=0.7]{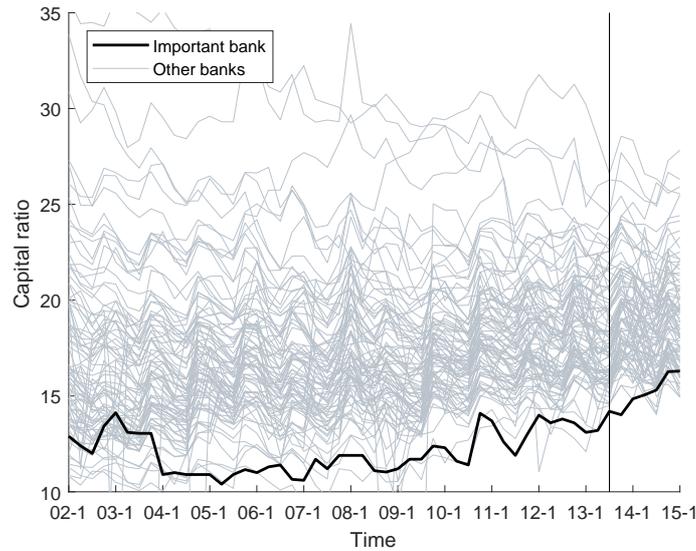}
\caption{\footnotesize Development in capital ratio from Q1 2002 to Q1 2015 for Norwegian banks. The systematically important bank is relatively low-capitalised. The policy reform was introduced in Q3 2013.}\label{fig_retail_app_bank_data} 
\end{figure}
\par Following the Basel III requirements for capital ratios in the aftermath of the financial crisis of 2007/2008, a specific Norwegian bank reform was implemented in 2013. In this reform one systematically important bank was given a higher requirement than other banks.\footnote{Three banks were deemed as systematically important, but the data used in this study only includes one of these banks.} The case study presented here aims at answering if the systematically important bank exposed to higher capital requirements than the other banks increased its capital ratio relative to the other banks. 

\par The data is from The Norwegian Banks' Guarantee Fund covering quarterly data for a sample of 120 banks from January 2002 to September 2015, covering 74\% of the total bank assets in Norway.\footnote{I am grateful to Ragnar Juelsrud for providing this data.} Removing banks with missing and irregular values, I am left with $J=87$ banks. The decision that systematically important banks was given higher requirements than other banks was announced by the Ministry of Finance in July 2013. I thus use Q3 of 2013 as the treatment date. 

\par Figure \ref{fig_retail_app_bank_data} shows the capital ratio for the important bank together with all other banks in the sample. Clearly, the bank under study is low-capitalised, lying well below most banks in terms of capital ratio.\footnote{\cite{getz2016consequences} define the capital ratio as $\text{CR}=E/A$ where $E$ denotes equity and $A = \sum_i \zeta_i A_i$ denotes risk-weighted assets, where $A_i$ is asset class $i$ and $\zeta_i$ is the corresponding risk weight, exogenously given by the regulator.} In fact, the important bank is the lowest capitalised bank during short periods between 2010 and 2012. This makes the case particularly challenging for ADH, which relies on non-negative weights.
\begin{figure}
\centering
\begin{minipage}{1\textwidth}
\begin{subfigure}{0.25\textwidth}
\includegraphics[width=\linewidth]{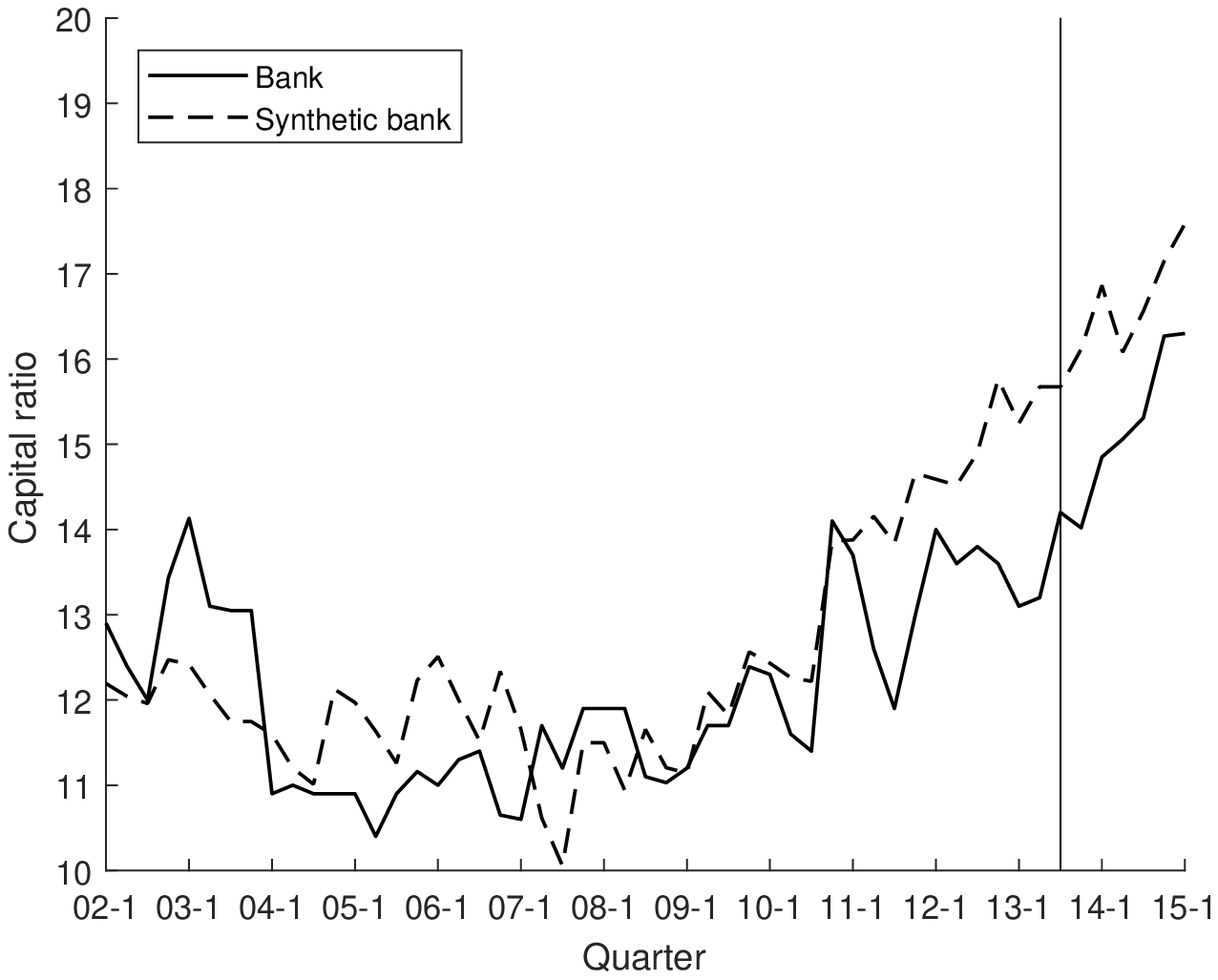}
\end{subfigure}\hspace*{\fill}
\begin{subfigure}{0.25\textwidth}
\includegraphics[width=\linewidth]{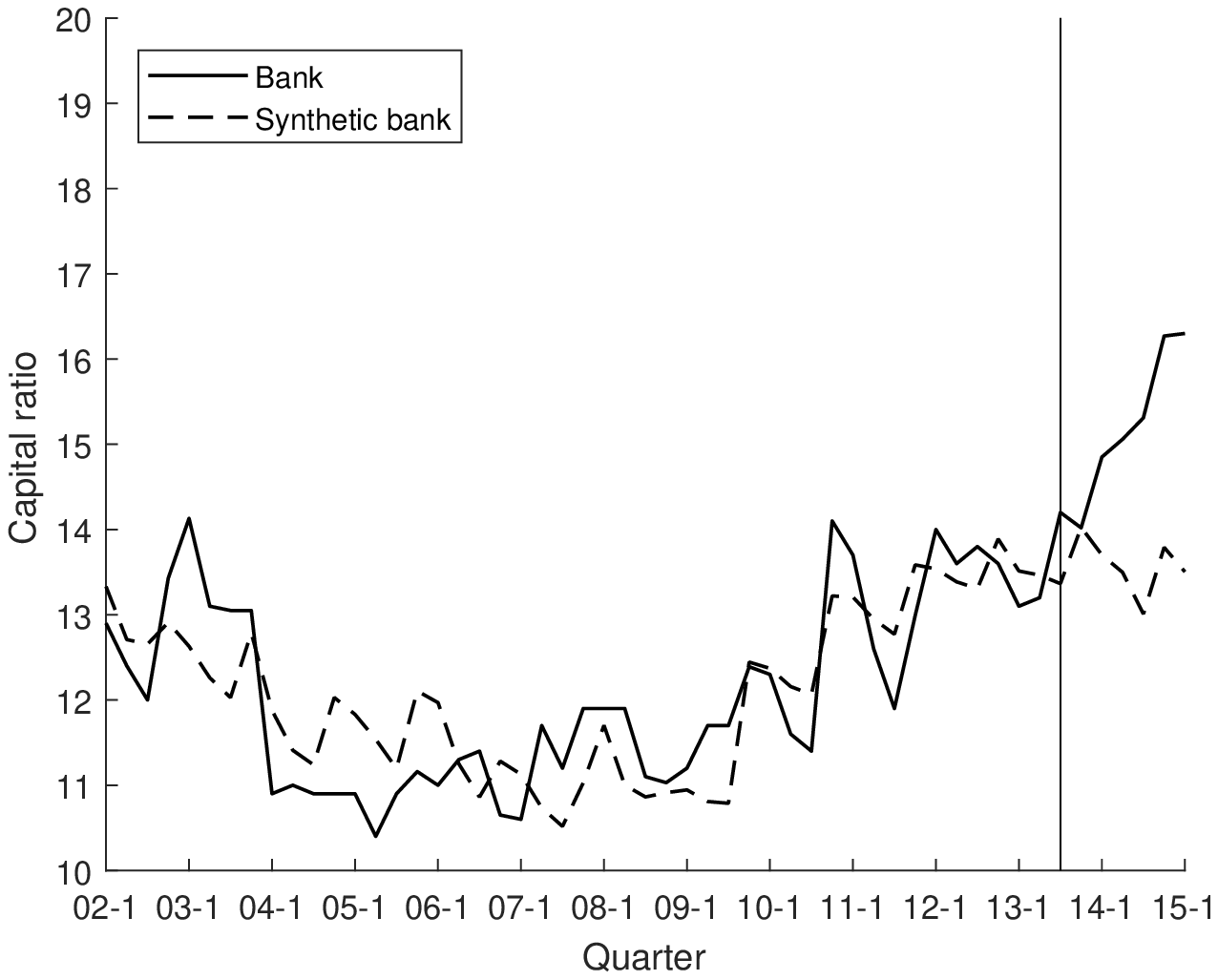}
\end{subfigure}\hspace*{\fill}
\begin{subfigure}{0.25\textwidth}
\includegraphics[width=\linewidth]{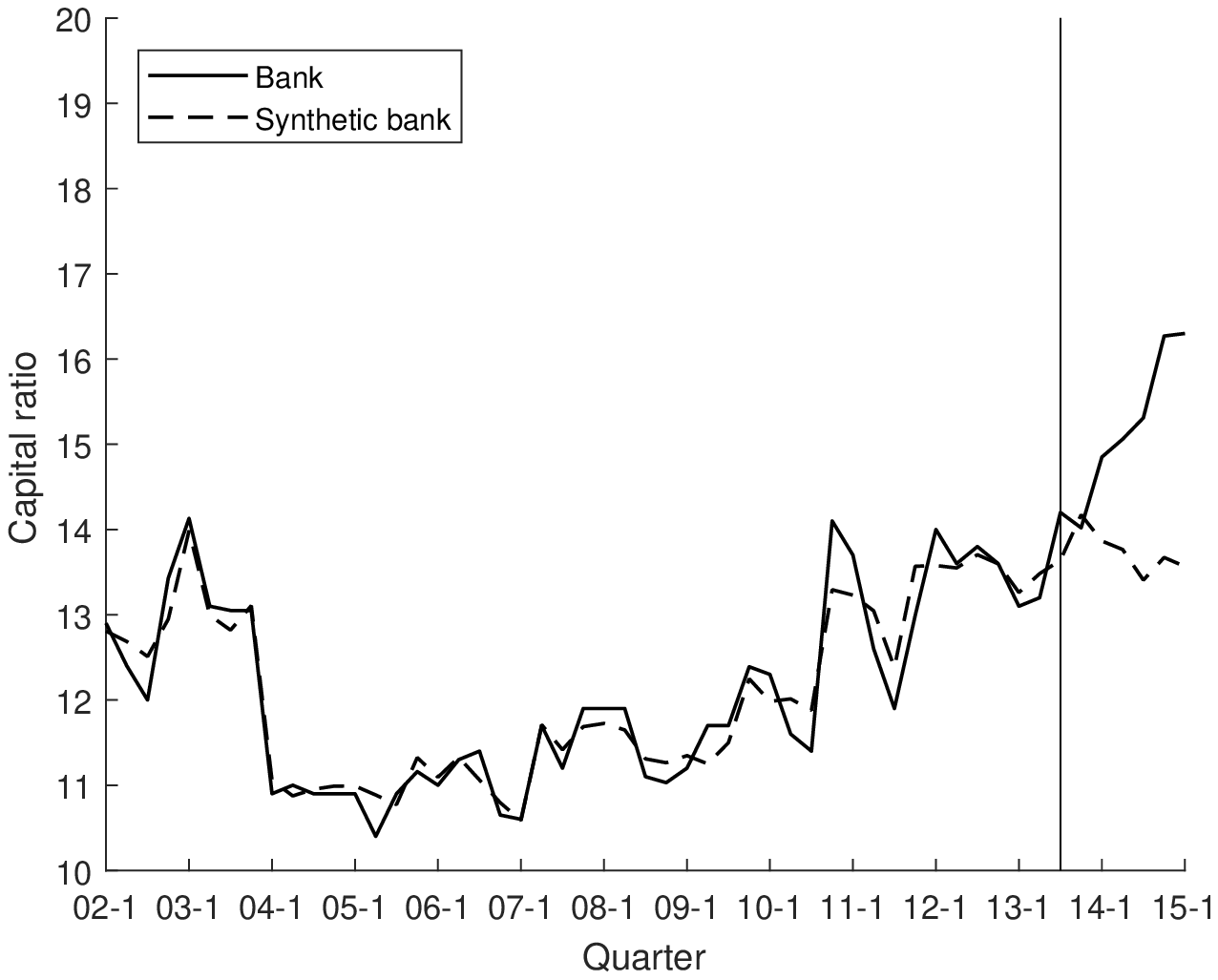}
\end{subfigure}\hspace*{\fill}
\begin{subfigure}{0.25\textwidth}
\includegraphics[width=\linewidth]{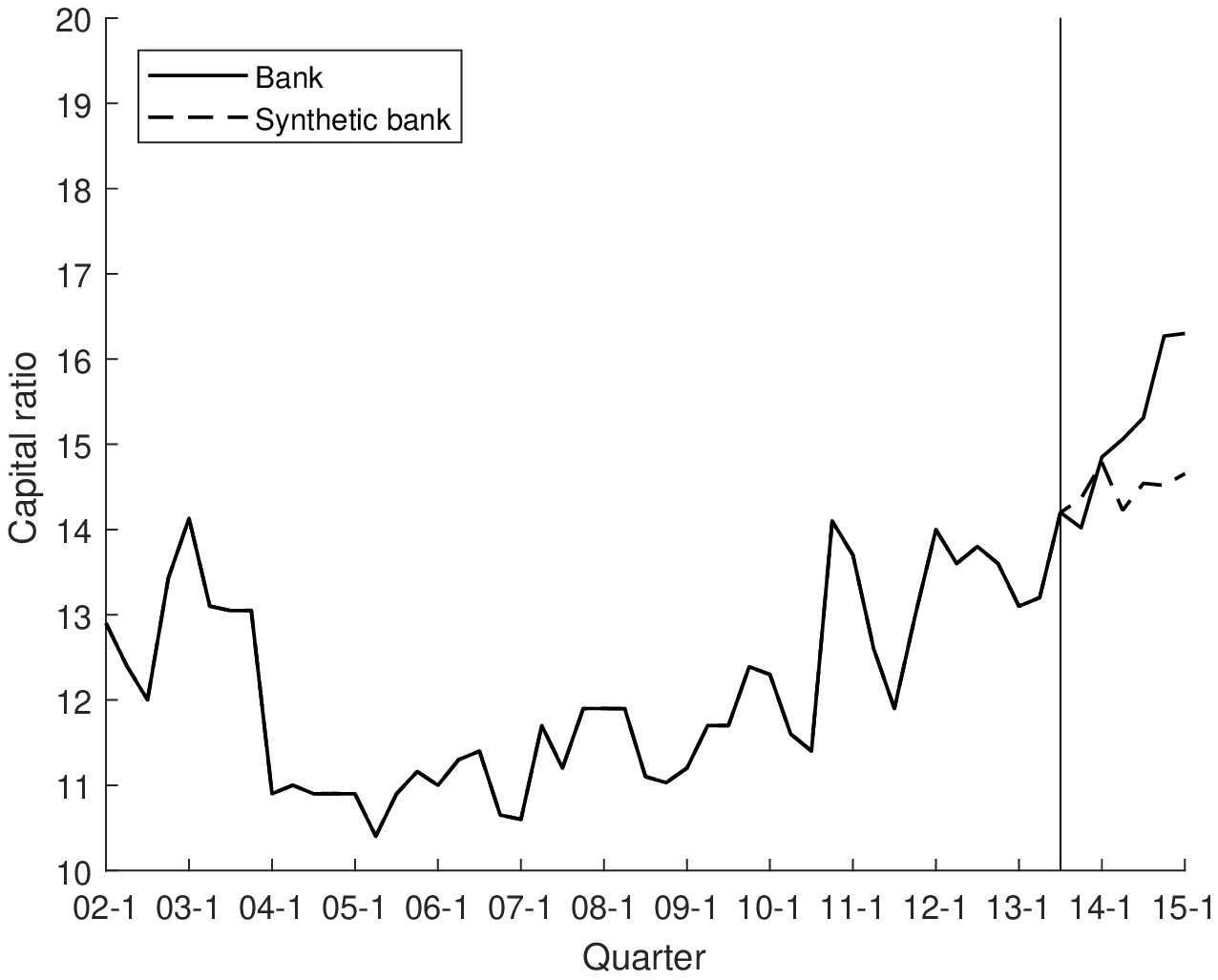}
\end{subfigure}
\medskip
\begin{subfigure}{0.25\textwidth}
\includegraphics[width=\linewidth]{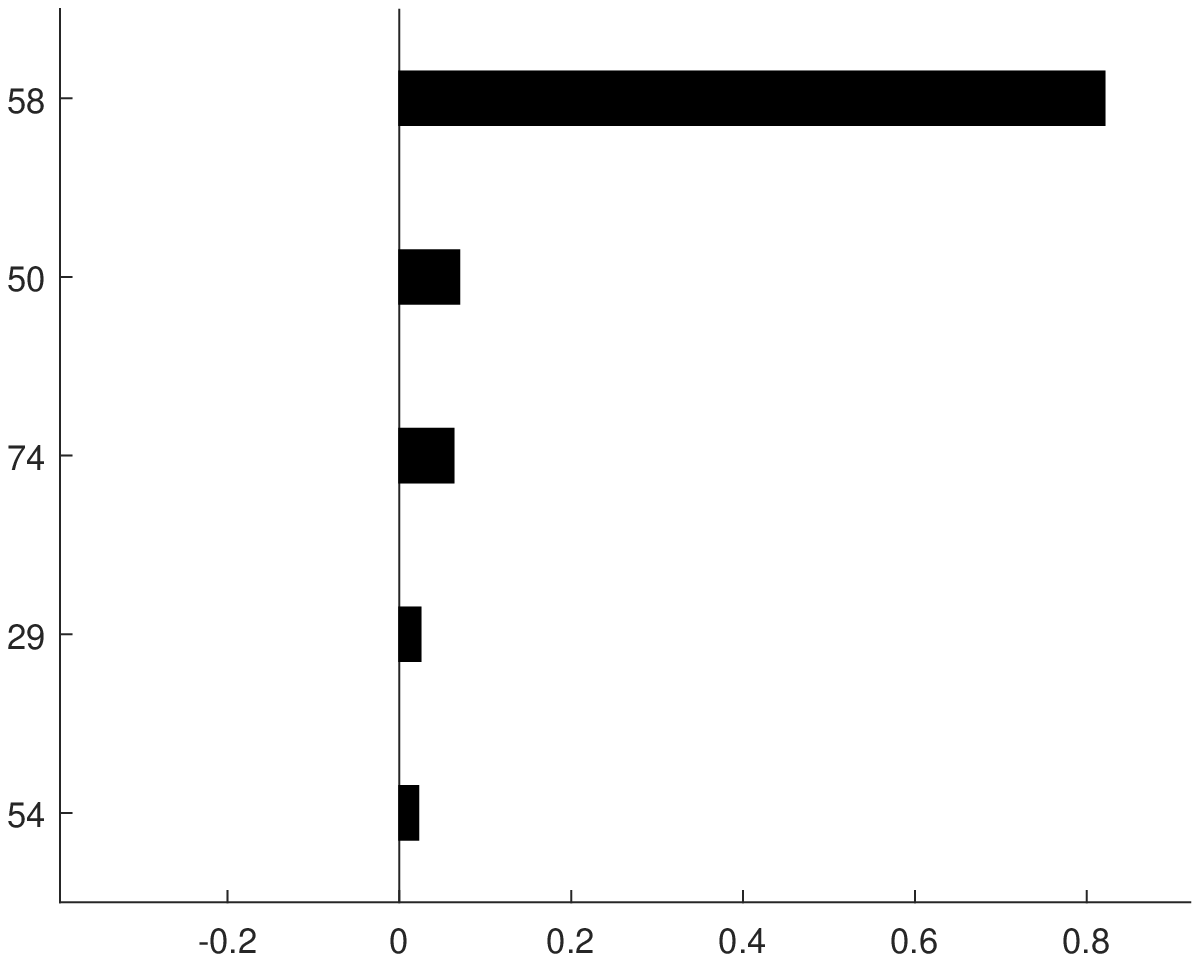}
\caption*{ADH}
\end{subfigure}\hspace*{\fill}
\begin{subfigure}{0.25\textwidth}
\includegraphics[width=\linewidth]{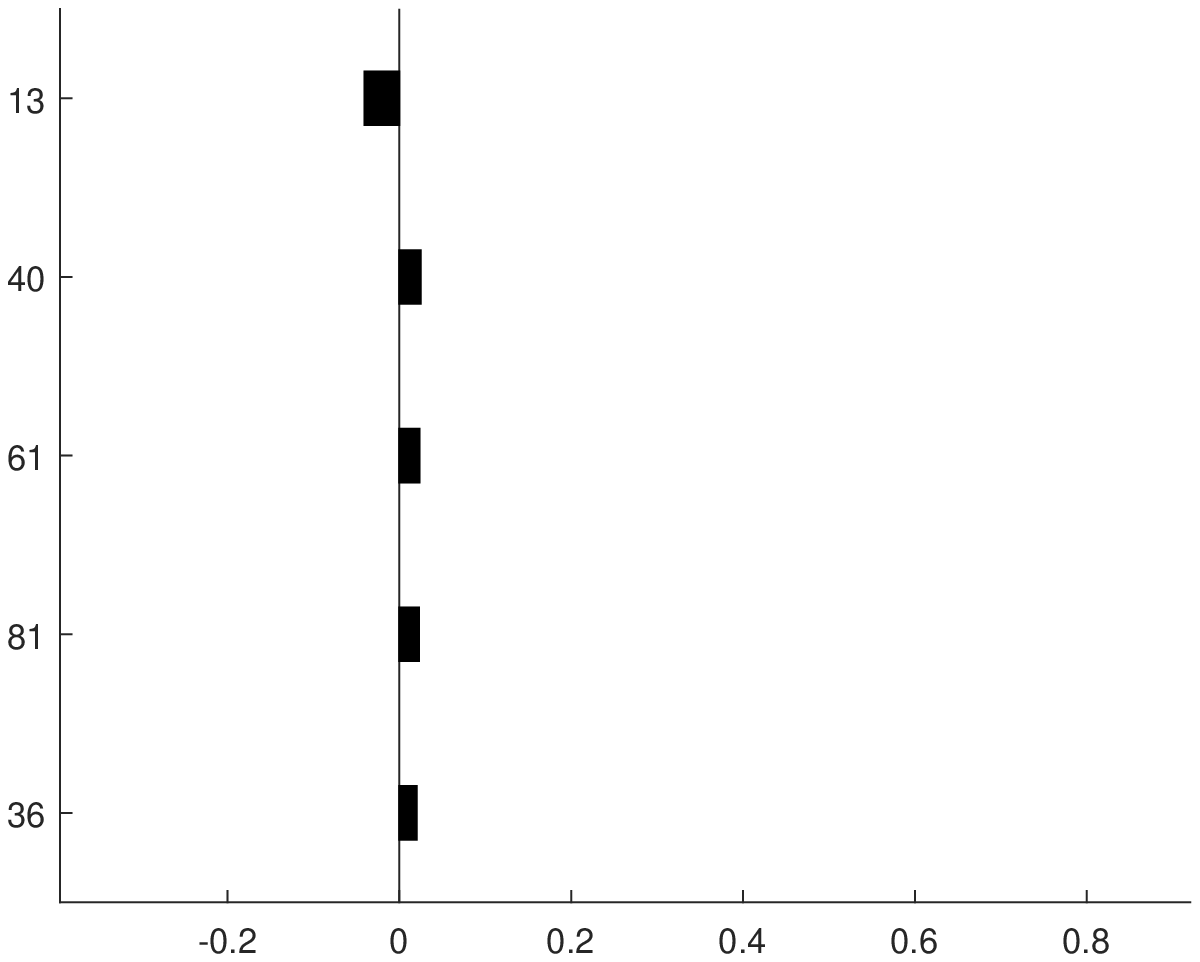}
\caption*{PCR}
\end{subfigure}\hspace*{\fill}
\begin{subfigure}{0.25\textwidth}
\includegraphics[width=\linewidth]{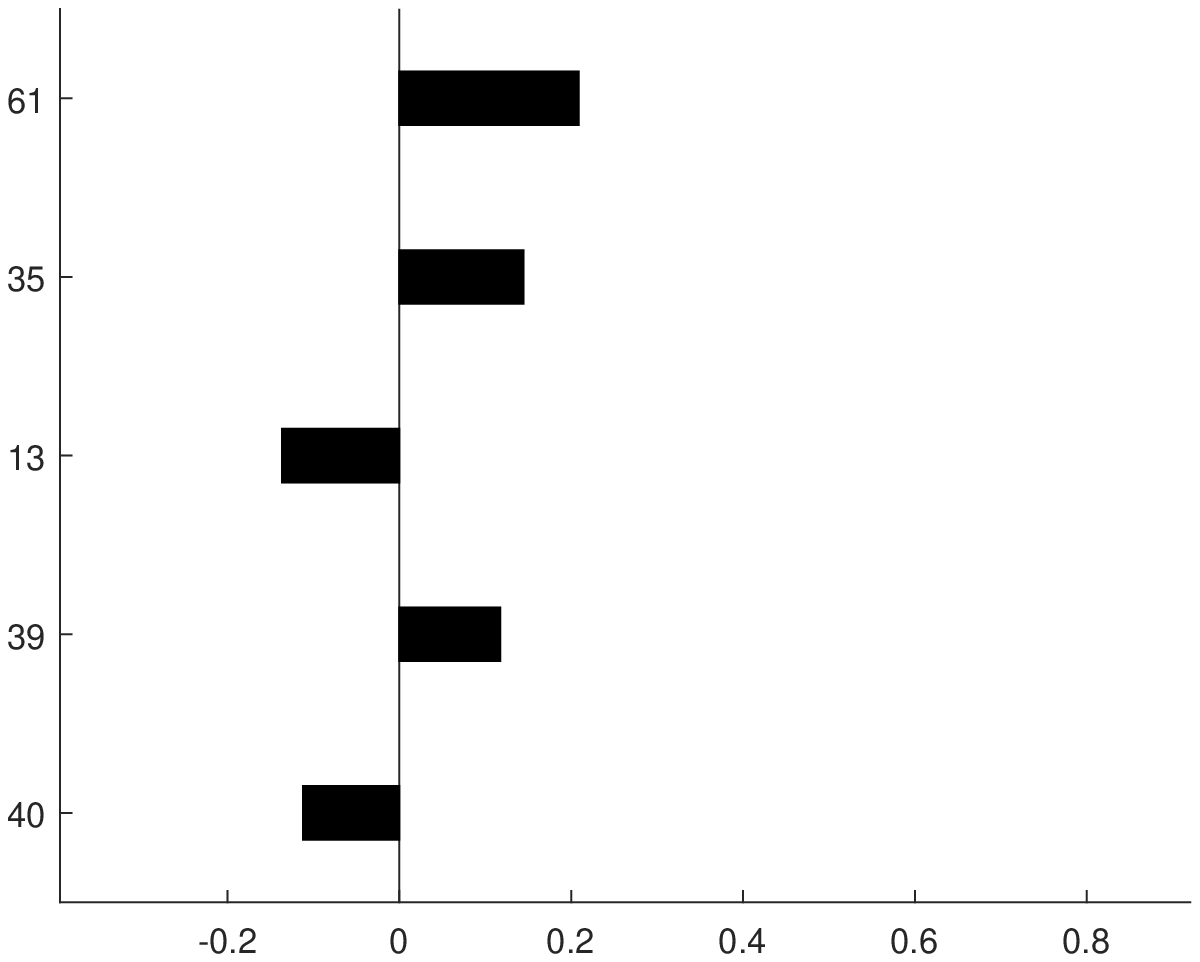}
\caption*{LASSO}
\end{subfigure}\hspace*{\fill}
\begin{subfigure}{0.25\textwidth}
\includegraphics[width=\linewidth]{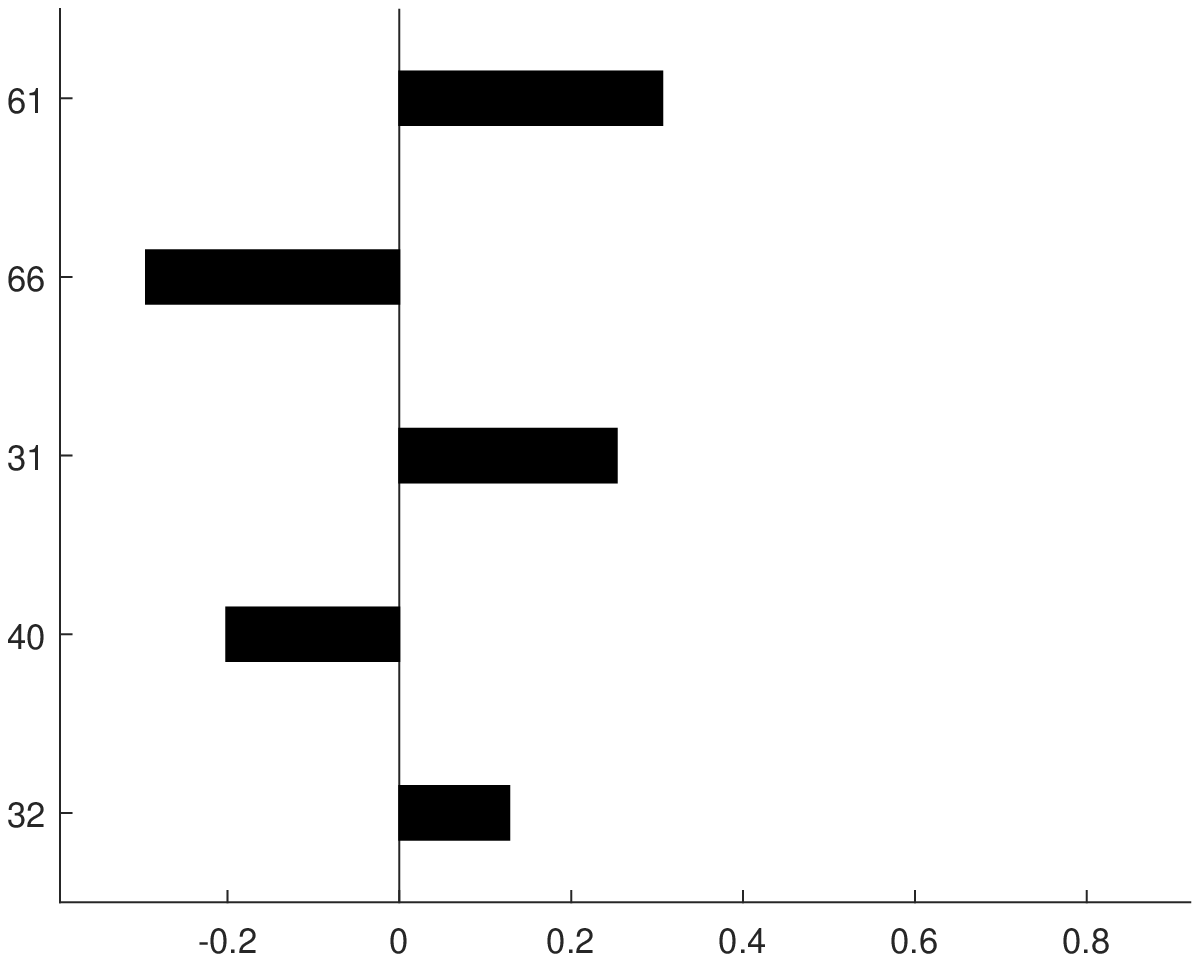}
\caption*{BSTS}
\end{subfigure}
\caption{\footnotesize The figure plots the development in capital ratio for the important bank and a synthetic version of the bank estimated using the original synthetic control (ADH), Principal Component Regression (PCR), regularization (LASSO) and Bayesian Structural Time Series (BSTS). The top 5 weights of each bank are reported along the counterfactual estimates.}
\label{fig_retail_app_banks_cf}
\end{minipage}
\end{figure}

\par Figure \ref{fig_retail_app_banks_cf} plots the capital ratio for the systematically important bank together with synthetic versions of this bank based on ADH, PCR, LASSO and BSTS. Clearly, ADH struggle to obtain a sensible fit in the pretreatment period, and consequently also a credible counterfactual. In technical terms, it is not possible to construct a convex combination of the control banks to produce the treated bank in the pretreatment period, so the ADH approach mainly rely on one unit close to the important bank. Indeed, ``bank 58'' in the sample receive a weight of 80\%, while all other banks are deemed less important, with weights below 10\%. Using negative weights, PCR, LASSO and BSTS all provide improved fits to the training data and indicate that the important bank has responded to the requirements as intended by the regulators.

\par I perform a placebo study and compare the estimated effects to the effect of the original study. PCR indicate highly robust results, with only 4 of 87 banks showing larger effects than the important bank. LASSO and BSTS also provide some robustness, with 10 and 19 larger effects, respectively. 

\subsection{The effect of changing retail prices} \label{sec_retail_applications_retail}
Scanner data is increasingly being used to answer macroeconomic problems. Such datasets may help assess price development, exchange-rate pass through or changing consumption patterns due to exogenous events. In this application I identify case studies in a large retail scanner dataset. Following an exogenous price change (a discount) on a specific product in a specific store, I assess the change in sold quantity resulting from the discount. The assessment is based on a counterfactual constructed from a large pool of other stores, offering the same product without any discount.

\par I use a novel transaction-level scanner dataset from all retail stores owned by a large Norwegian retailer. One transaction is defined as one particular product scanned by a cashier in any given store at any given time. The data spans from May 27th 2015 to September 13th 2016, giving rise to a large dataset with approximately 3 billion observations. The data contains a total of $J=576$ stores located all over Norway. The retailer had a market share of approximately 25\% during the sample period and sold a total of 27,207 different products from these stores during the $T=476$ days in the sample.

\begin{figure}
\centering
\begin{minipage}{1\textwidth}
\begin{subfigure}{0.32\textwidth}
\includegraphics[width=\linewidth]{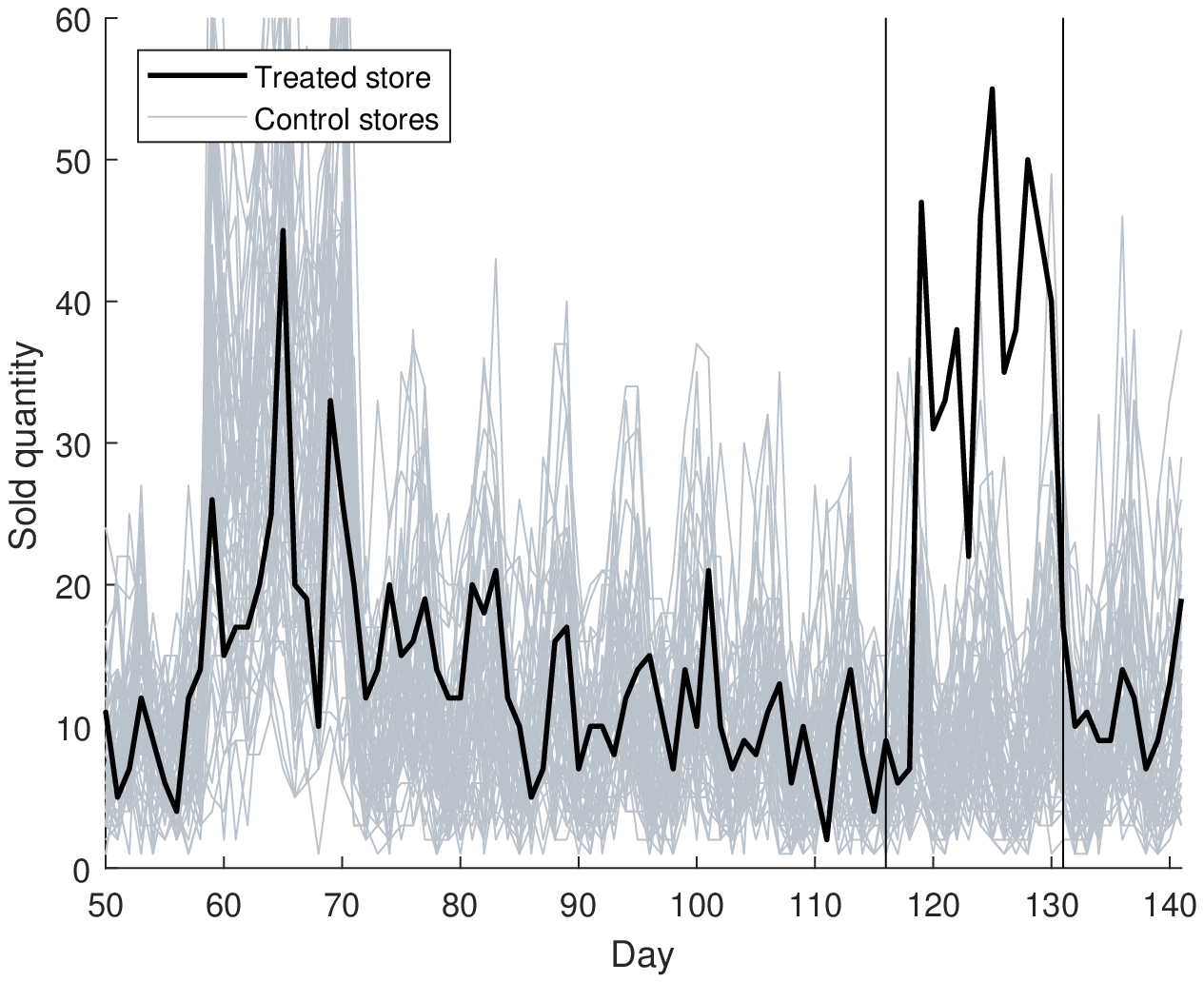}
\caption*{Apple Juice}
\end{subfigure}\hspace*{\fill}
\begin{subfigure}{0.32\textwidth}
\includegraphics[width=\linewidth]{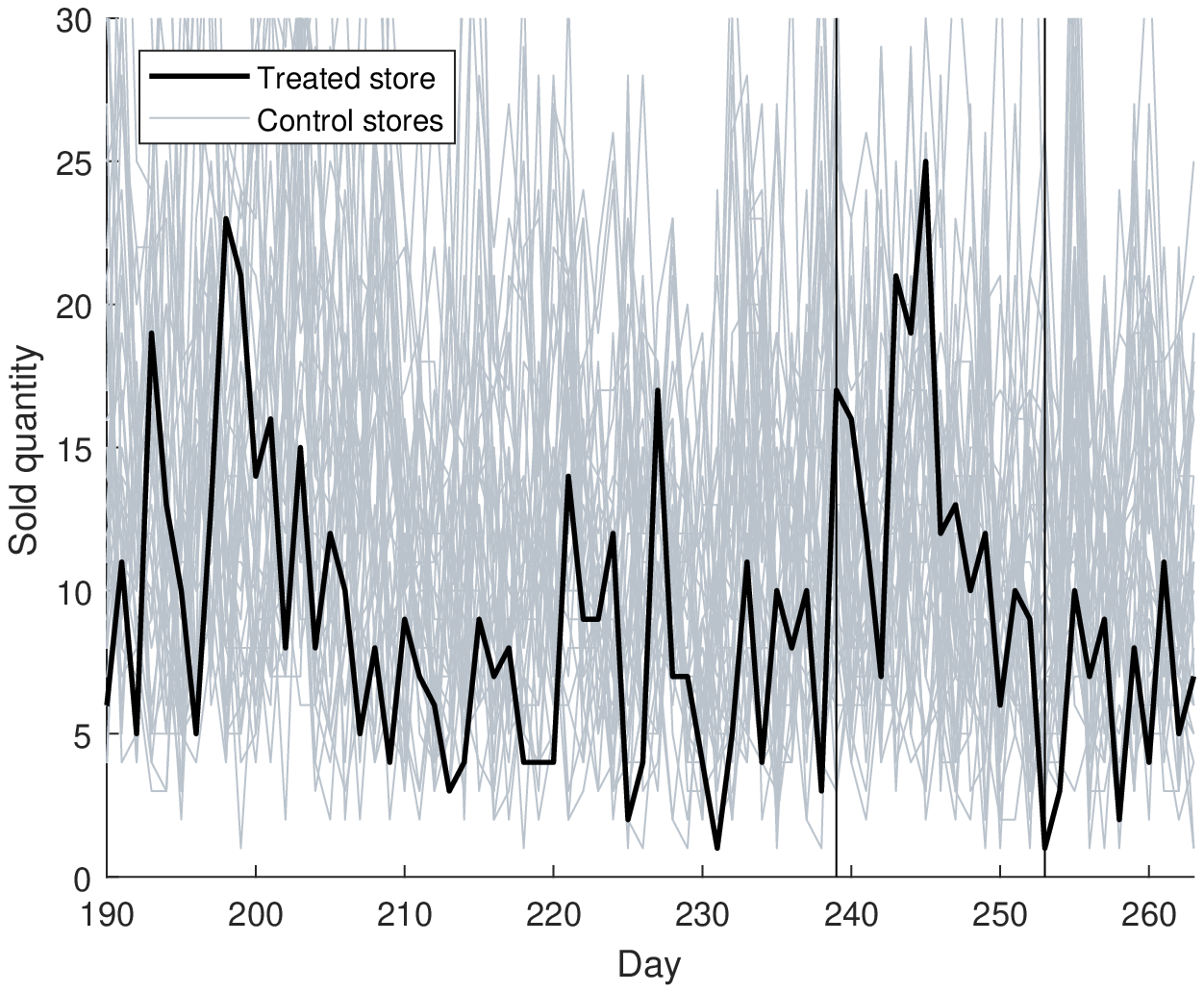}
\caption*{Yoghurt}
\end{subfigure}\hspace*{\fill}
\begin{subfigure}{0.32\textwidth}
\includegraphics[width=\linewidth]{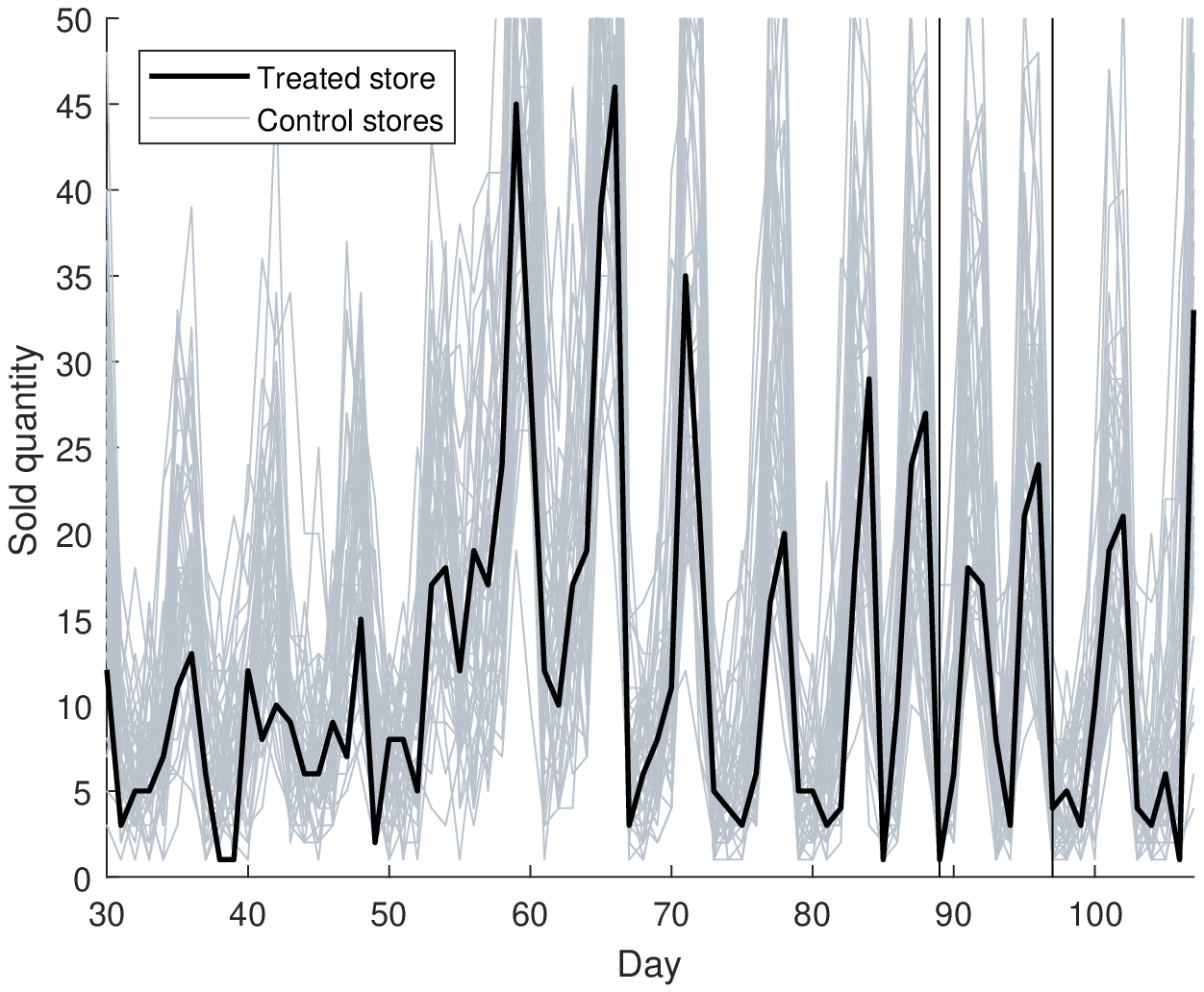}
\caption*{Chips}
\end{subfigure}
\caption{\footnotesize Daily observations of sold quantities for Apple Juice, Yoghurt and Chips for the store offering a discount and the pool of control stores not offering a discount. Discount periods are 15, 14 and 8 days, respectively.} 
\label{fig_retail_app_stores_data}
\end{minipage}
\end{figure}

\par The data is reduced to a set of daily observations by collapsing transactions based on days, stores and products. Summing quantities by these variables is straightforward, however prices are weighted by sold quantities to arrive at daily quantity-weighted prices for each product in each store. I identify three products from the pool of products with discounts. The products will be referred to as ``Apple Juice'', ``Yoghurt'' and ``Chips'' in the following analysis. For each of these products, let the outcome variable $y_{jt}$ be quantity sold in store $j$ at time $t$. The store changing the product price is the treated unit while all other stores keeping the price at the initial level are possible controls. The sold quantities of the three products are shown in Figure \ref{fig_retail_app_stores_data}. The treated store offers a discount in a short time window during the sample period, illustrated by the vertical bars. The data indicate an abrupt change in sold quantity for Apple Juice and Yoghurt following the discount, while there is no sign of changing sales for Chips.

\par
\begin{figure}
\centering
\begin{minipage}{1\textwidth}
\begin{subfigure}{0.32\textwidth}
\includegraphics[width=\linewidth]{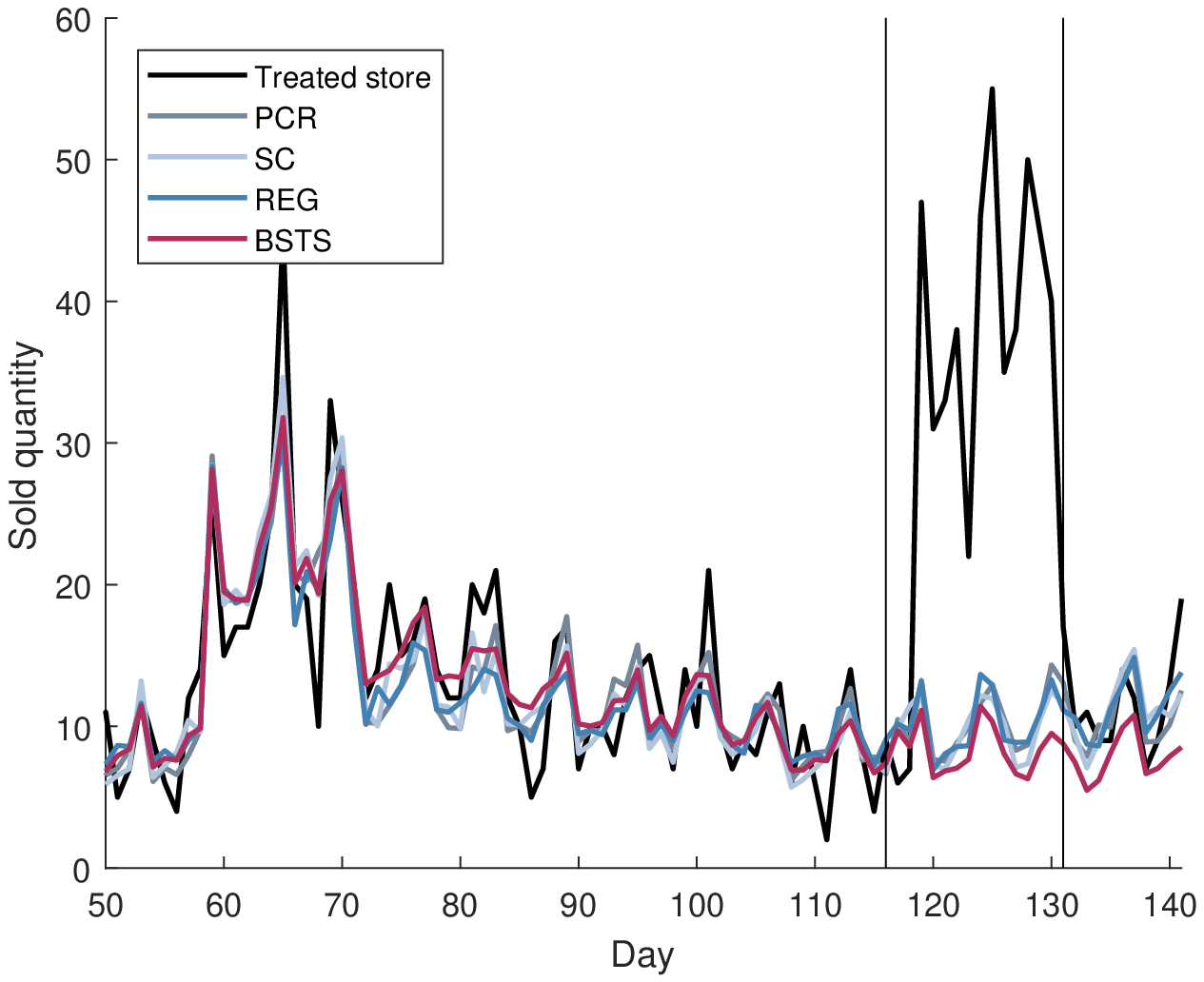}
\caption{Synthetic Controls}\label{fig_retail_app_stores_apple_sc}
\end{subfigure}\hspace*{\fill}
\begin{subfigure}{0.32\textwidth}
\includegraphics[width=\linewidth]{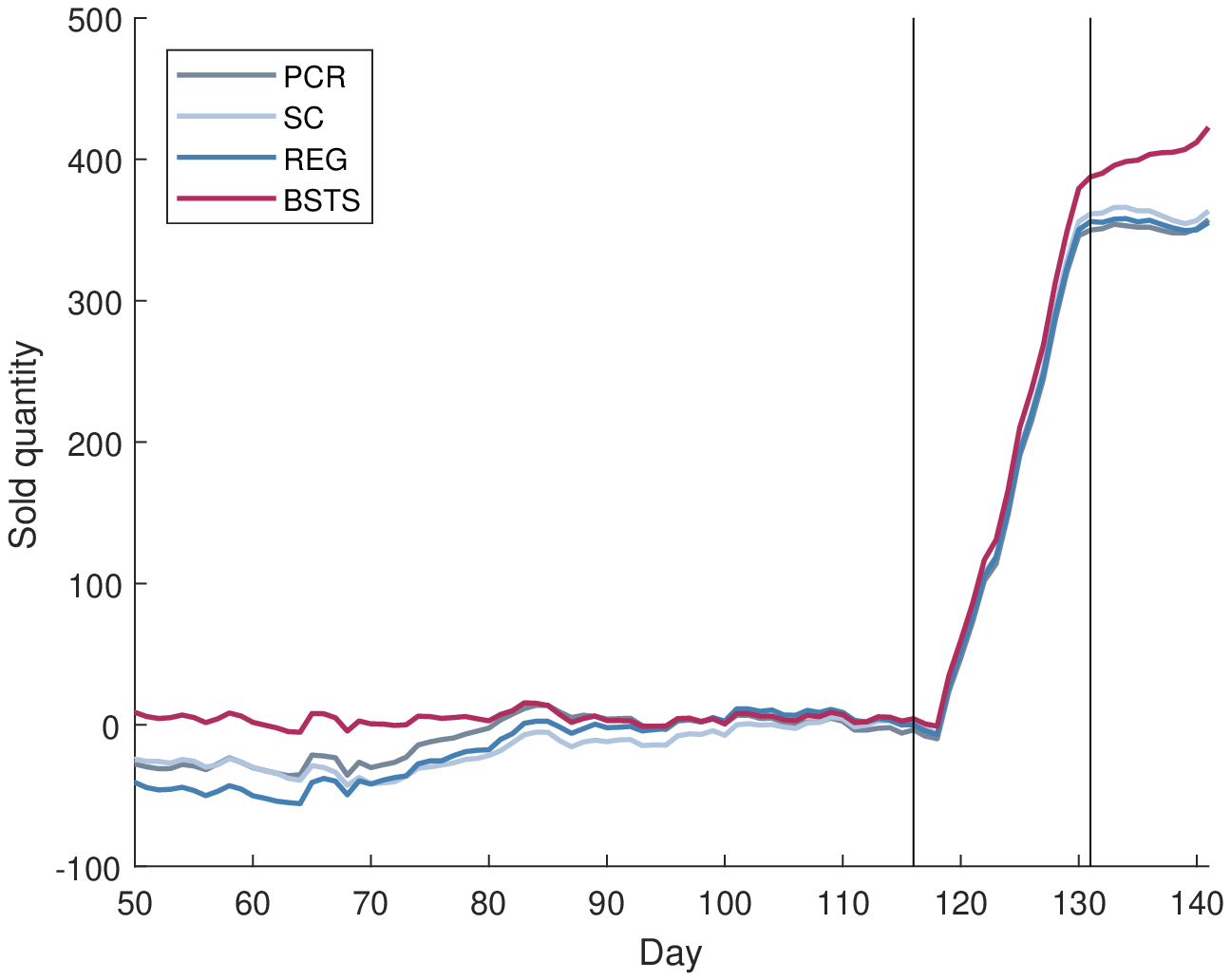}
\caption{Cumulative effects}\label{fig_retail_app_stores_apple_cum}
\end{subfigure}\hspace*{\fill}
\begin{subfigure}{0.32\textwidth}
\includegraphics[width=\linewidth]{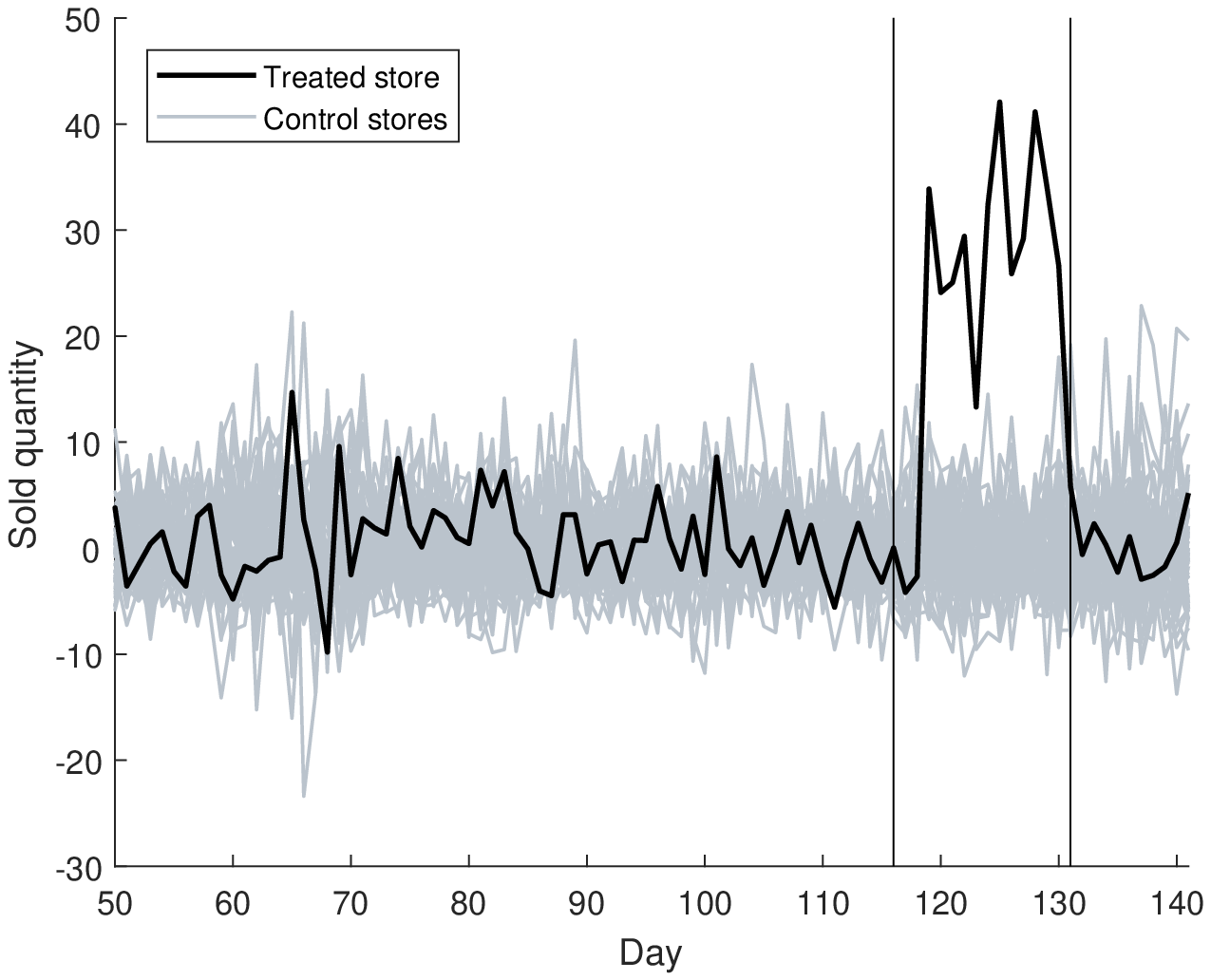}
\caption{Placebo studies}\label{fig_retail_app_stores_apple_placebo}
\end{subfigure}
\caption{\footnotesize Change in sold quantities of Apple Juice following a price change. Figure \ref{fig_retail_app_stores_apple_sc}: countercatual estimates using Principal Component Regression (PCR), the original synthetic control (ADH), Lasso regularization (LASSO) and Bayesian Structural Time Series (BSTS). Figure \ref{fig_retail_app_stores_apple_cum}: Cumulative effect of the price discount. Figure \ref{fig_retail_app_stores_apple_placebo}: Placebo studies based on LASSO.}
\label{fig_retail_app_stores_apple}
\end{minipage}
\end{figure}

 The counterfactual of the Apple Juice time series was obtained using ADH, PCR, LASSO and BSTS, and results are shown in Figure \ref{fig_retail_app_stores_apple} and Table \ref{tab_retail_app_stores_stats}. The results indicate a large, robust effect of increasing sales in the range 346 to 371 units, with none of the 61 identified controls giving higher treatment effects than the treated unit. Furthermore, the effect seem to die out post treatment, and the observed time series return to the counterfactual estimate. The story is similar for Yoghurt. The effect is ranging from approximately 40 to 103 units during the discount period, and is highly robust based on the placebo studies. As expected, the counterfactual estimates for Chips indicate no effect of the price discount.

\section{Conclusion}
The use of microecnomic data to assess macroeconomic policy is a promising, rapidly expanding area of reaseach. In this paper I have presented several tools for assessing macroeconomic policy from case studies and high dimensional data. Most of these tools are machine learning methods, recently proposed for synthetic control estimation by e.g. \cite{brodersen2015inferring} and \cite{doudchenko2016balancing}. I develop a unified framework, suitable for discussing how and why machine learning improve treatment effect estimation in high dimensional settings. In particular, I show that well known econometric methods such as difference-in-differences with matching and the original synthetic control by \cite{abadie2010synthetic} are restricted versions of the machine learning approaches, and argue that more flexibility can capture the data generating process to a larger extent.

\par I compare methods using simulation studies with several different data generating processes for the potential outcomes. Finally, these tools are used on policy questions such as the effect of liberalisation and bank behaviour following stricter capital requirements.

\clearpage
\bibliographystyle{aea}
\bibliography{reflist}

\clearpage
\begin{appendices}

\renewcommand{\thesection}{\Alph{section}}
\numberwithin{equation}{section}

\clearpage
\begin{figure}
\centering
\begin{minipage}{1\textwidth}
\begin{subfigure}{0.45\textwidth}
\includegraphics[width=\linewidth]{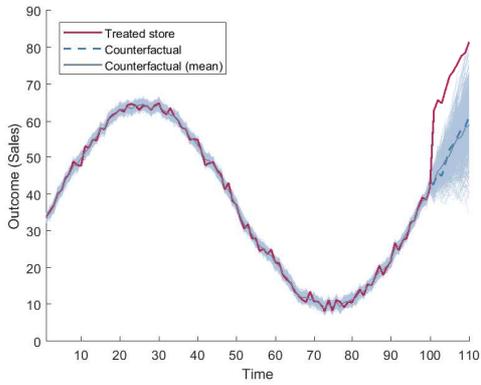}
\caption{Counterfactual estimates}
\end{subfigure}\hspace*{\fill}
\begin{subfigure}{0.45\textwidth}
\includegraphics[width=\linewidth]{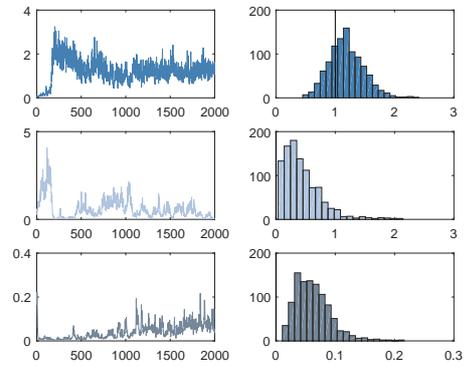}
\caption{Parameter distribution}
\end{subfigure}\\
\caption{\footnotesize Details on the BSTS Gibbs sampling procedure for one specific realization of the data for Scenario (B). Figure (a) plots all the Gibbs sampled trajectories of the treated unit, the mean value and the draw of the potential outcomes from the data generating process. Figure (b) shows the convergence of the parameter estimates and the distribution for $\sigma^2$, $\sigma_1^2$ and $\sigma_2^2$ from top to bottom, respectively.} 
\label{fig_retail_sim_bsts}
\end{minipage}
\end{figure}

\clearpage
\begin{table}
\caption{\textsc{Liberalisation and GDP Trends}}
\label{tab_retail_app_countries}
\scriptsize
\centering
\begin{threeparttable}
\begin{tabular}{llrrrr|rrrr}
\midrule
Country & Region & Treatment & $T_0$ & $J$ & BN & ADH & PCR & LASSO & BSTS\\
\midrule
\textbf{Group 1}\\
Singapore & Asia & 1965 & 3 & 69 & E* & E* & E* & E* & E* \\
South Korea & Asia & 1968 & 6 & 68 & E* & E* & E* & E* & E* \\
Indonesia & Asia & 1970 & 8 & 67 & E* & E & E & E & E\\
Colombia & Latin America & 1986 & 24 & 25 & E* & E* & E & E* & E\\
Costa Rica & Latin America & 1986 & 24 & 25 & E* & E* & E* & E* & E\\
Botswana & Africa & 1979 & 17 & 52 & E* & E* & E* & E* & E\\
Mauritius & Africa & 1968 & 6 & 68 & E & E* & E* & E* & E*\\
Ghana & Africa & 1985 & 23 & 29 & E & E & E & E & E\\
Guinea & Africa & 1986 & 24 & 25 & E & N & E & E & E\\
Benin & Africa & 1990 & 28 & 23 & E & E & E & E & E \\
Morocco & Middle East & 1984 & 22 & 33 & E & N & E & U & E\\ \\

\textbf{Group 2}\\
Chile & Latin America & 1976 & 14 & 60 & N & N & N & N & N\\
Gambia & Africa & 1985 & 23 & 29 & N & N & N & N & N\\
Cape Verde & Africa & 1991 & 29 & 23 & U & U & U & N & U \\
Guinea-Bissau & Africa & 1987 & 25 & 25 & U & N & U & U & N\\
Zambia & Africa & 1993 & 31 & 23 & N & N & N & N & N\\
Kenya & Africa & 1993 & 31 & 23 & N & N & N & N & N\\
Cameroon & Africa & 1993 & 31 & 23 & N & N & N & N & N\\
Niger & Africa & 1994 & 32 & 23 & N & N & U & U & U \\
Mauritania & Middle East & 1995 & 33 & 23 & N & N & N & N & U\\
Egypt & Middle East & 1995 & 33 & 23 & N & N & N & N & U\\ \\

\textbf{Group 3}\\
Barbados & Latin America & 1966 & 4 & 68 & E* & E* & N & N & N \\
Mexico & Latin America & 1986 & 24 & 25 & E* & E* & N & U & N \\
Nepal & Asia & 1991 & 29 & 23 & U & U & E & E & E\\
Mali & Africa & 1988 & 26 & 24 & U & U & E & E & E\\
Uganda & Africa & 1988 & 26 & 24 & U & U & E & E & E \\
Philippines & Asia & 1988 & 26 & 24 & N & U & U & E & E\\
South Africa & Africa & 1991 & 29 & 23 & N & N & E & E* & N\\
Ivory Coast & Africa & 1994 & 32 & 23 & N & E & E & E & E\\
Tunisia & Middle East & 1989 & 27 & 23 & N & E & E & E & E \\
\bottomrule
\end{tabular}
\begin{tablenotes}
\item The table summarieses the results from applying the original synthetic control (ADH), Principal Components Regression (PCR), Lasso (LASSO) and Bayesian Structural Time Series (BSTS) to each of the country liberalisation episodes discussed in \cite{billmeier2013assessing}. The results from the original study are summarised in column ``BN''. I use ``E$^*$'' to denote robust effects of the liberalisation 10 years post treatment. Effects are considered robust if less than 10\% of the placebo studies show larger treatment effects. Large, somewhat robust effects are indicated by ``E'', while ``U'' indicates that results are inconclusive. Results are deemed unclear if the effects are highly non-robust, if the pretreatment fit is poor or if the observed outcome only partly outperform the counterfactual. ``N'' is used to denote no effect.
\end{tablenotes}
\end{threeparttable}
\end{table}

\clearpage
\begin{figure}
\centering
\begin{minipage}{1\textwidth}
\begin{subfigure}{0.25\textwidth}
\includegraphics[width=\linewidth]{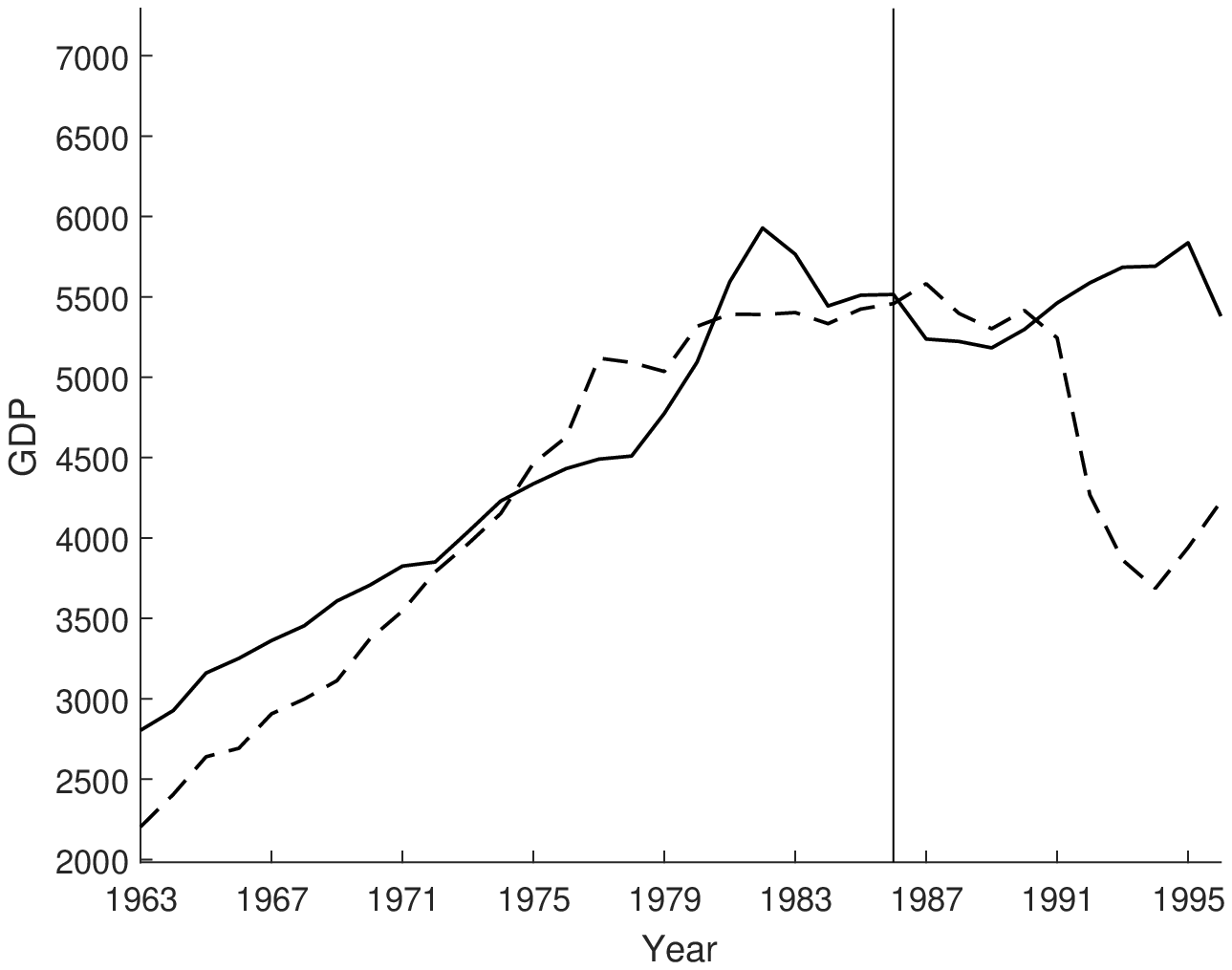}
\end{subfigure}\hspace*{\fill}
\begin{subfigure}{0.25\textwidth}
\includegraphics[width=\linewidth]{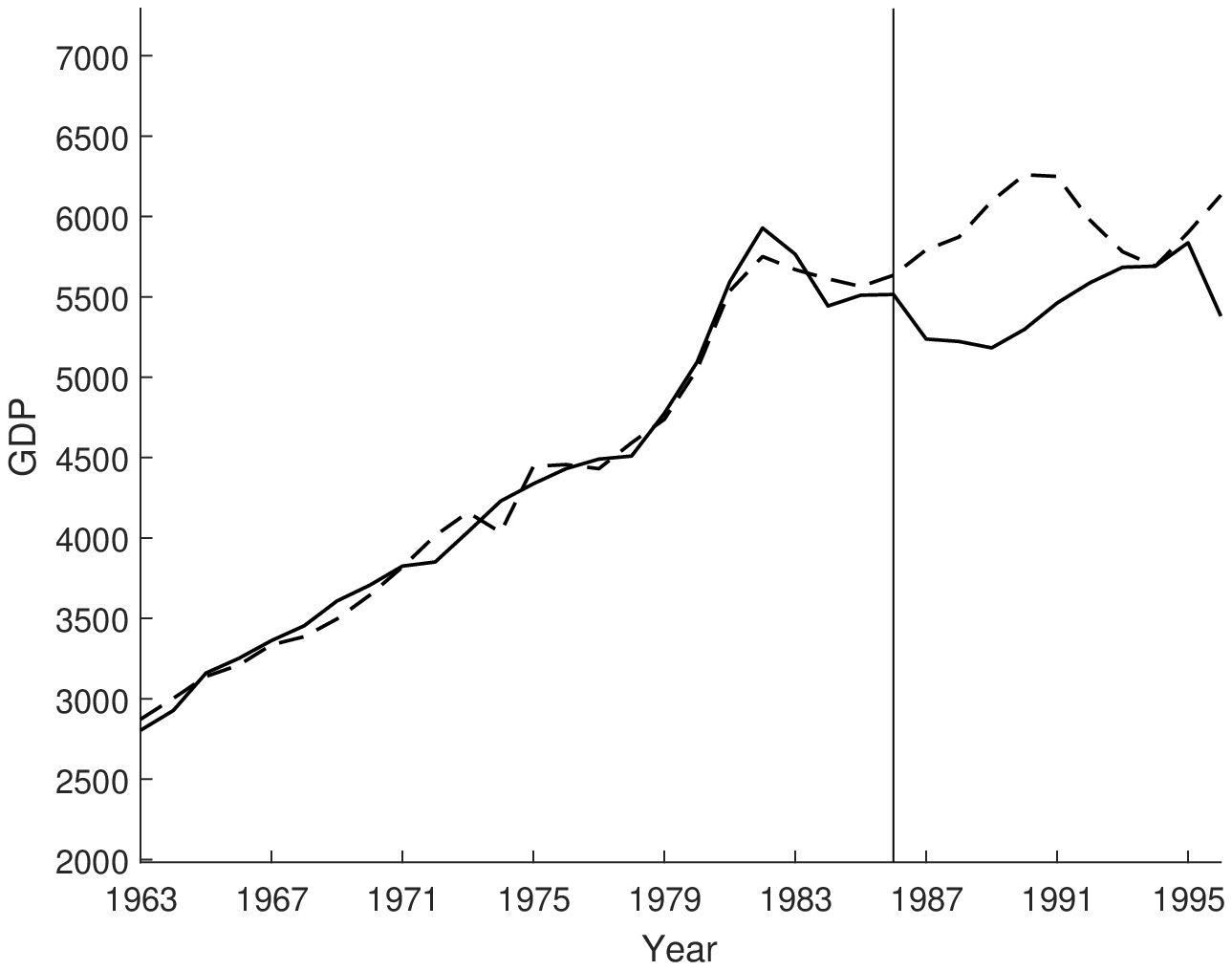}
\end{subfigure}\hspace*{\fill}
\begin{subfigure}{0.25\textwidth}
\includegraphics[width=\linewidth]{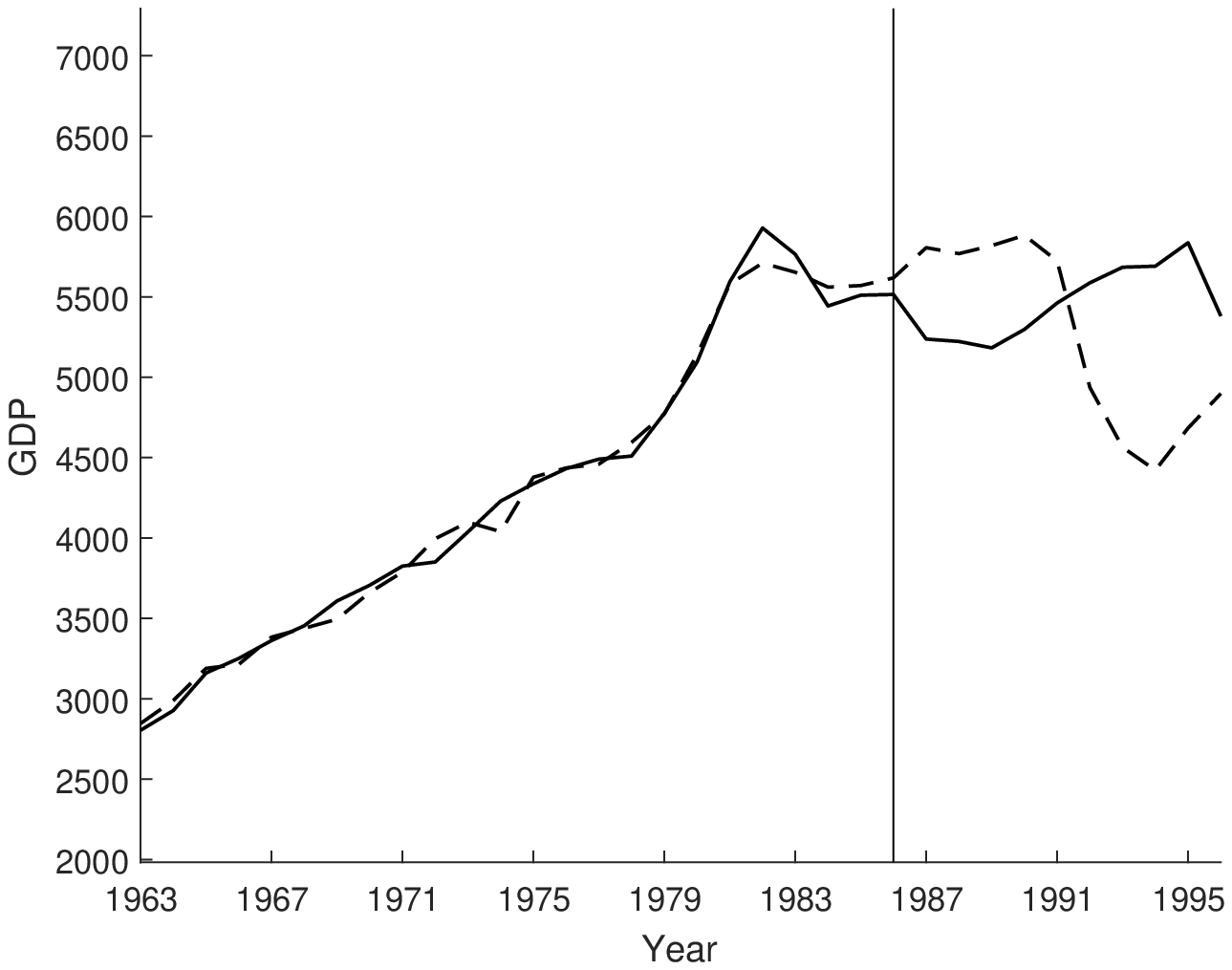}
\end{subfigure}\hspace*{\fill}
\begin{subfigure}{0.25\textwidth}
\includegraphics[width=\linewidth]{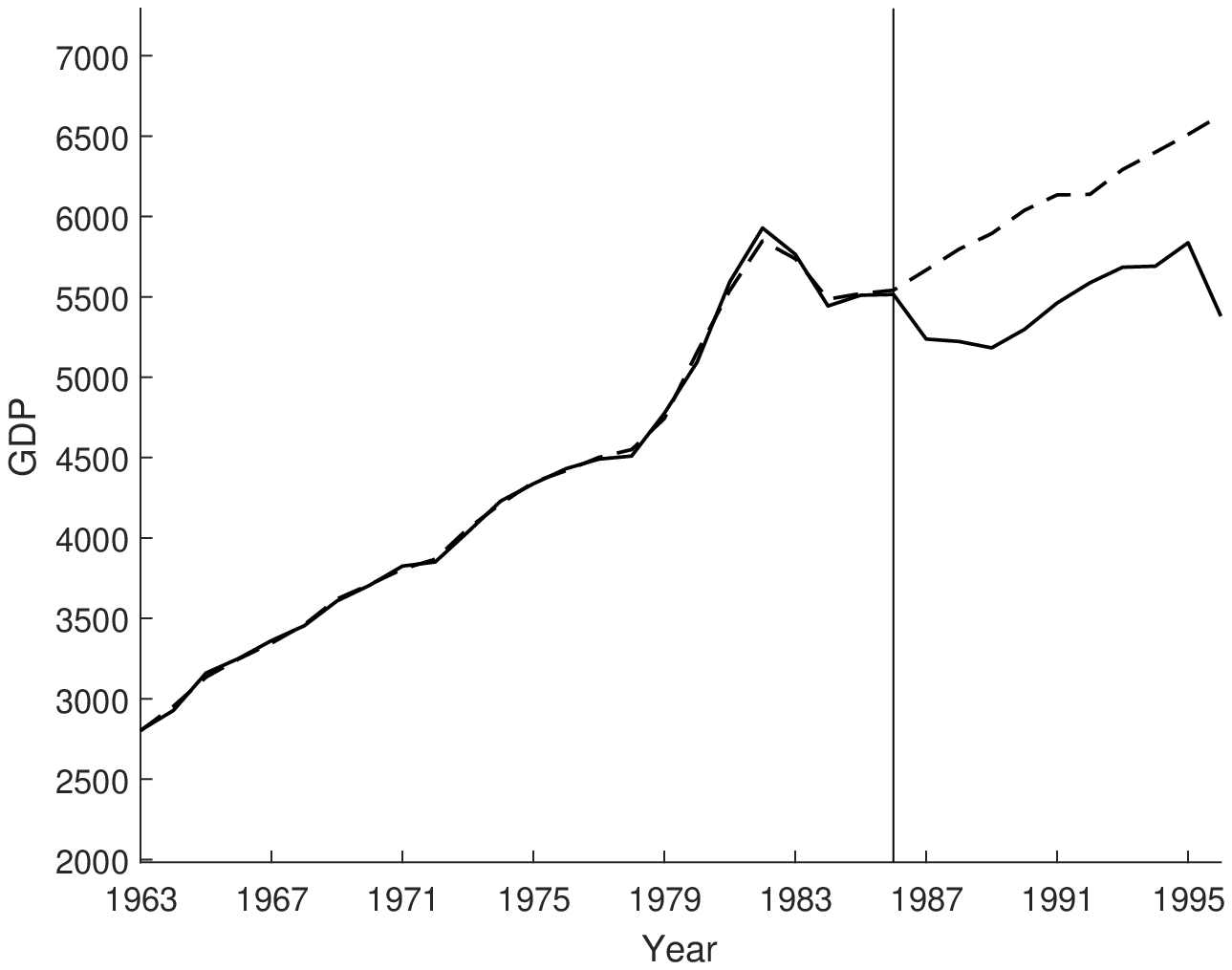}
\end{subfigure}
\medskip
\begin{subfigure}{0.25\textwidth}
\includegraphics[width=\linewidth]{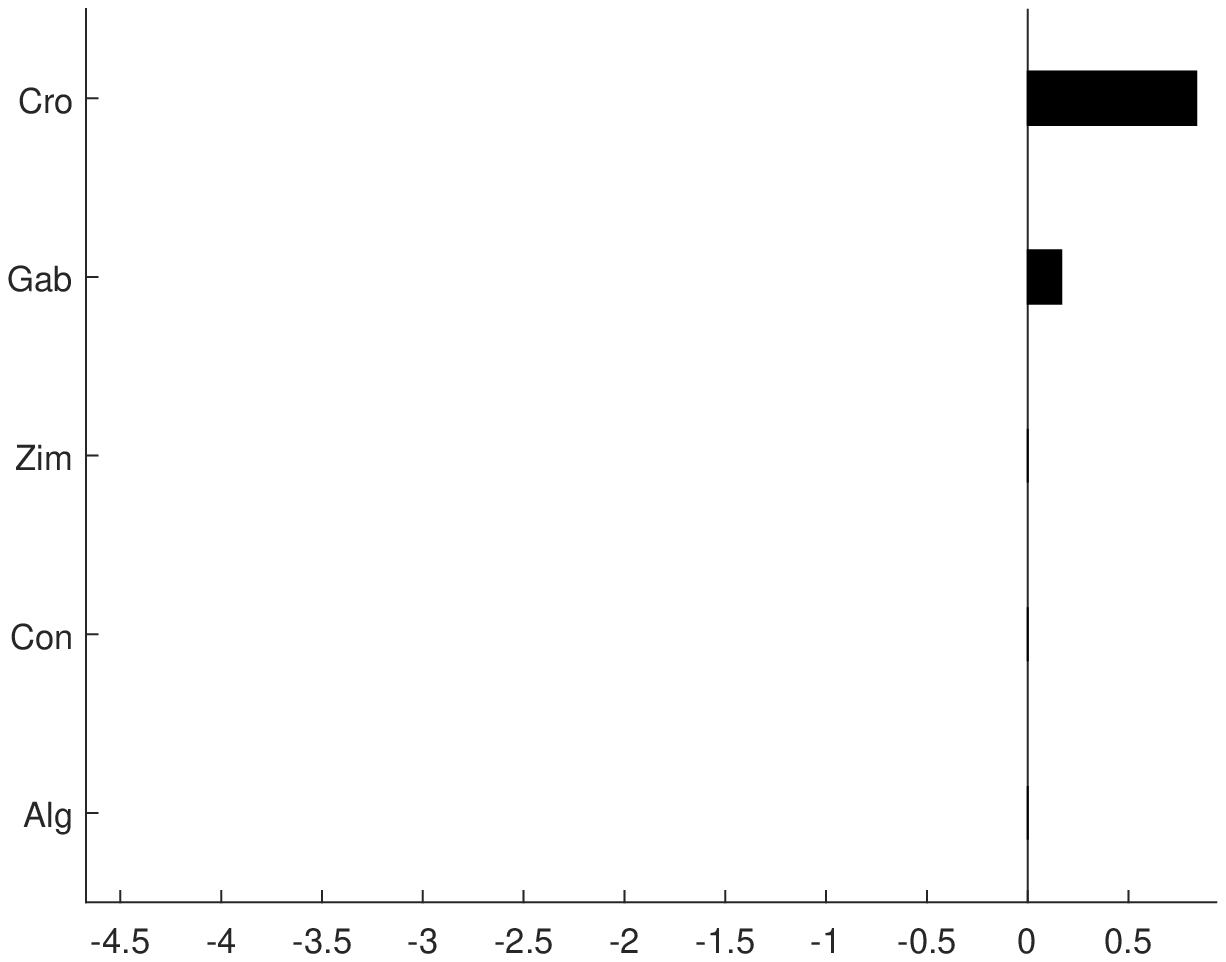}
\caption*{ADH}
\end{subfigure}\hspace*{\fill}
\begin{subfigure}{0.25\textwidth}
\includegraphics[width=\linewidth]{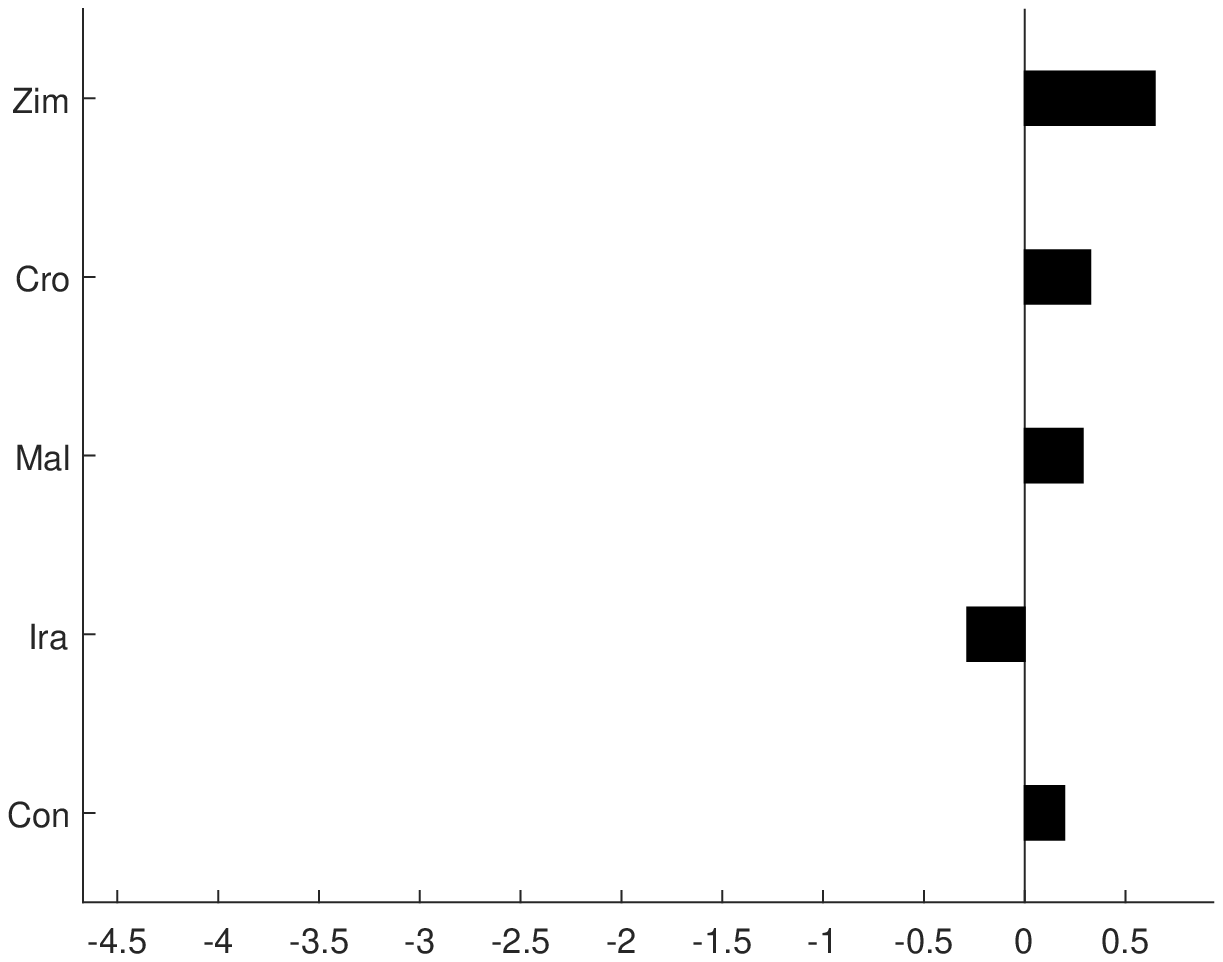}
\caption*{PCR}
\end{subfigure}\hspace*{\fill}
\begin{subfigure}{0.25\textwidth}
\includegraphics[width=\linewidth]{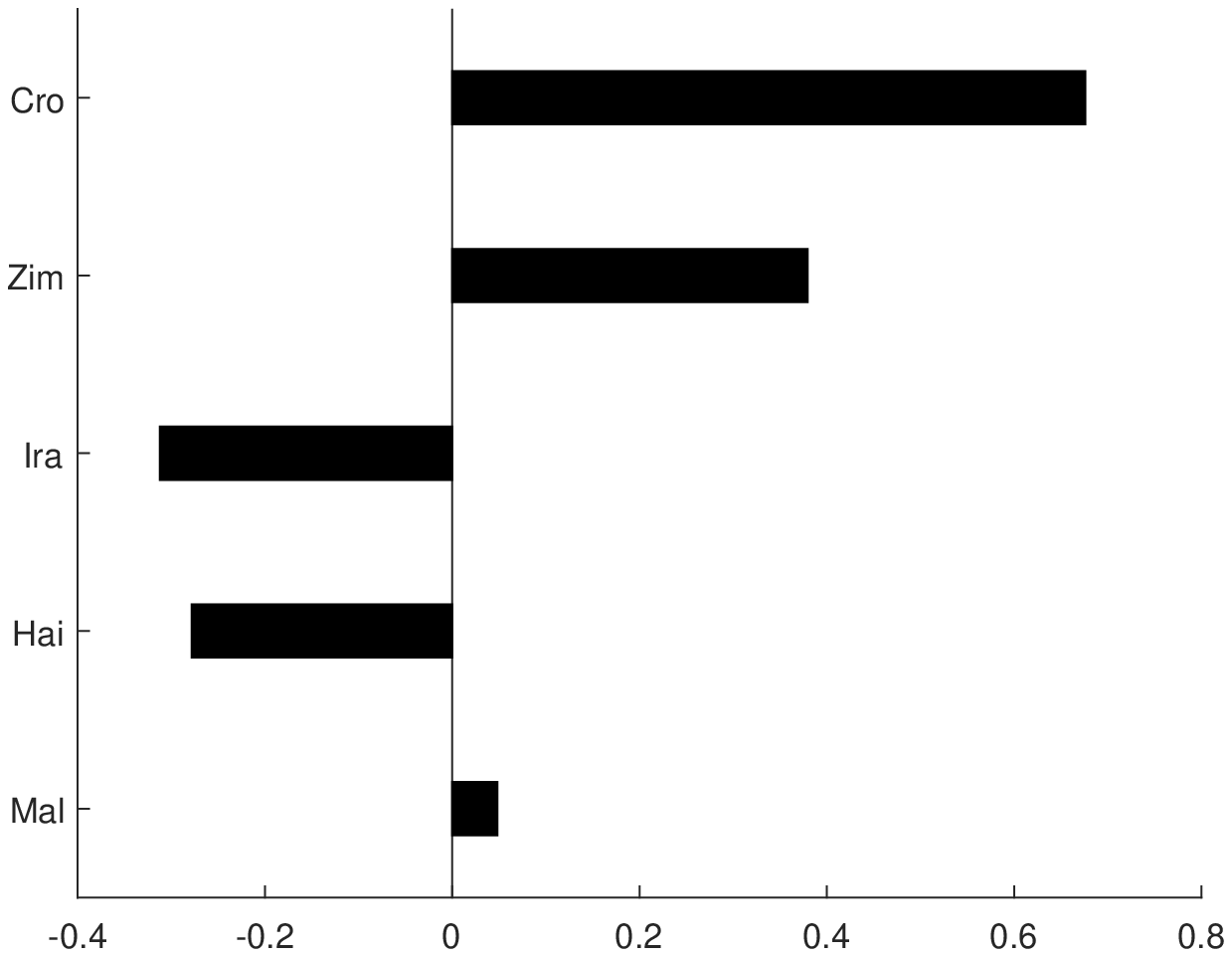}
\caption*{LASSO}
\end{subfigure}\hspace*{\fill}
\begin{subfigure}{0.25\textwidth}
\includegraphics[width=\linewidth]{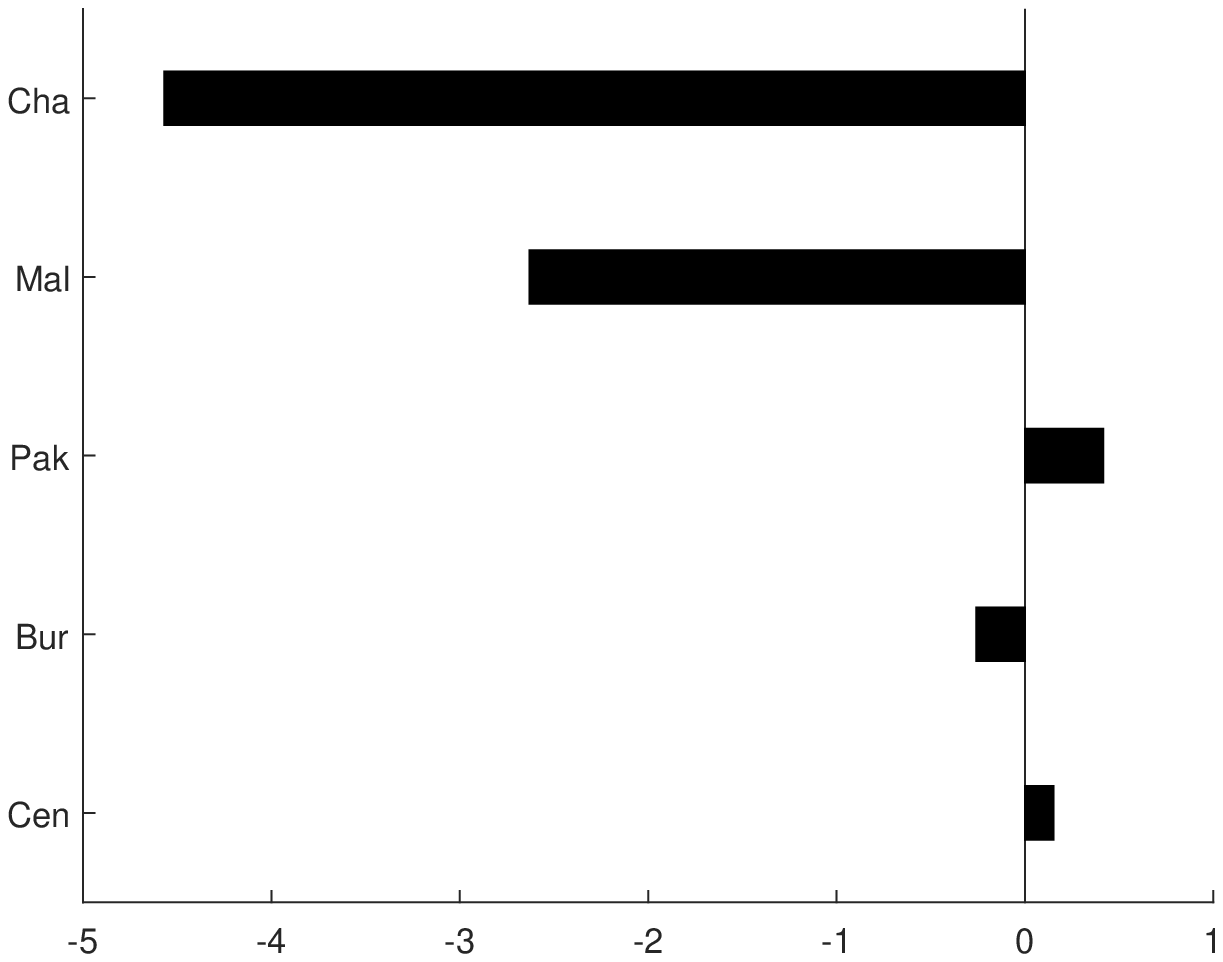}
\caption*{BSTS}
\end{subfigure}\\
\caption{\footnotesize Liberalisation of Mexico in 1986. Top row: Development in GDP per capita compared to synthetic controls estimates based on the original synthetic control (ADH), Principal Component Regression (PCR), Lasso-type regularization (LASSO) and Bayesian Structural Time Series (BSTS). The ADH results are robust as only 1 of 25 control units show a larger treatment effect 10 years after the liberalisation. For LASSO, 3/25 units show larger treatment effects. Bottom row: 5 largest weights.} 
\label{fig_retail_app_countries_mexico}
\end{minipage}
\end{figure}

\clearpage
\begin{figure}
\centering
\begin{minipage}{1\textwidth}
\begin{subfigure}{0.25\textwidth}
\includegraphics[width=\linewidth]{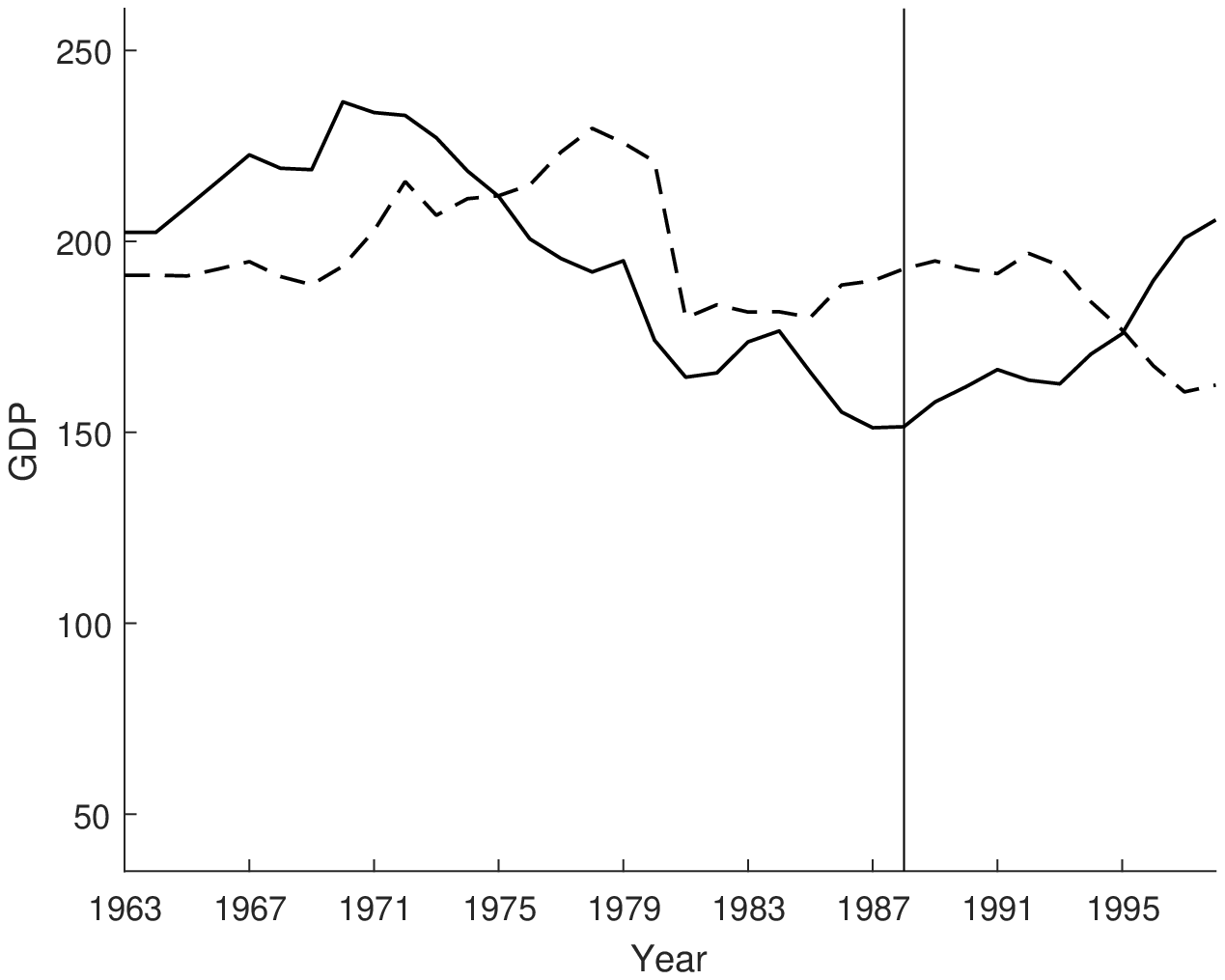}
\end{subfigure}\hspace*{\fill}
\begin{subfigure}{0.25\textwidth}
\includegraphics[width=\linewidth]{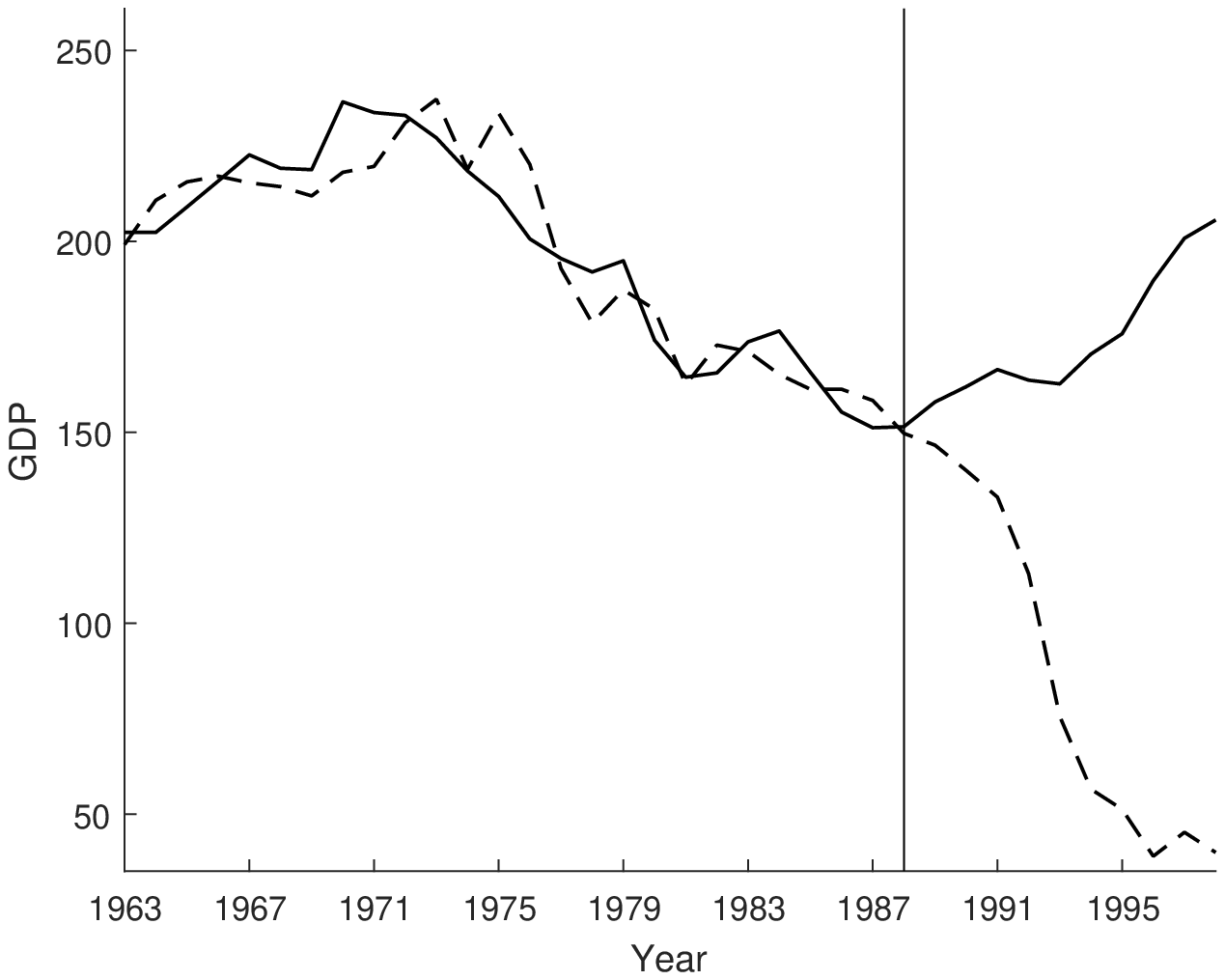}
\end{subfigure}\hspace*{\fill}
\begin{subfigure}{0.25\textwidth}
\includegraphics[width=\linewidth]{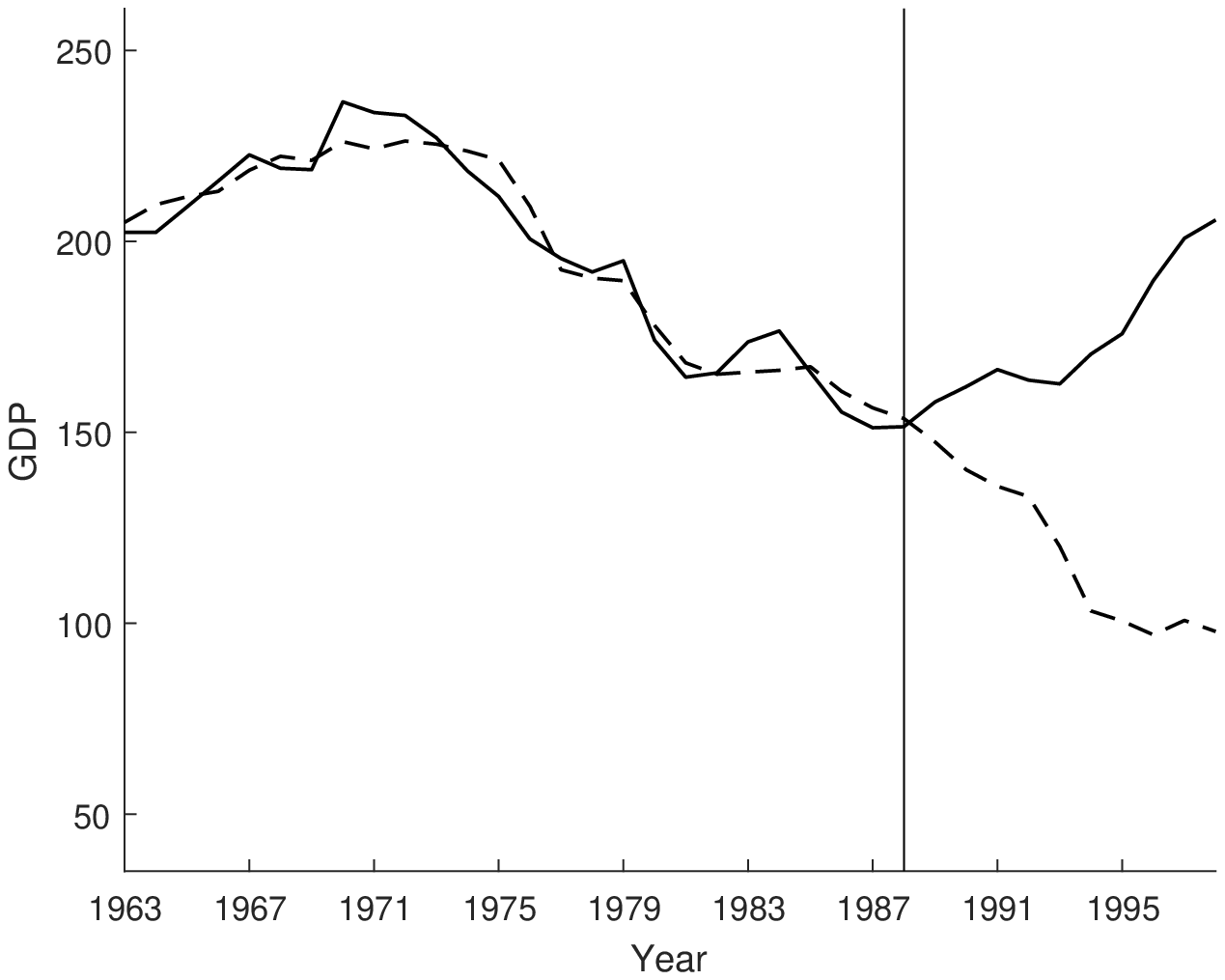}
\end{subfigure}\hspace*{\fill}
\begin{subfigure}{0.25\textwidth}
\includegraphics[width=\linewidth]{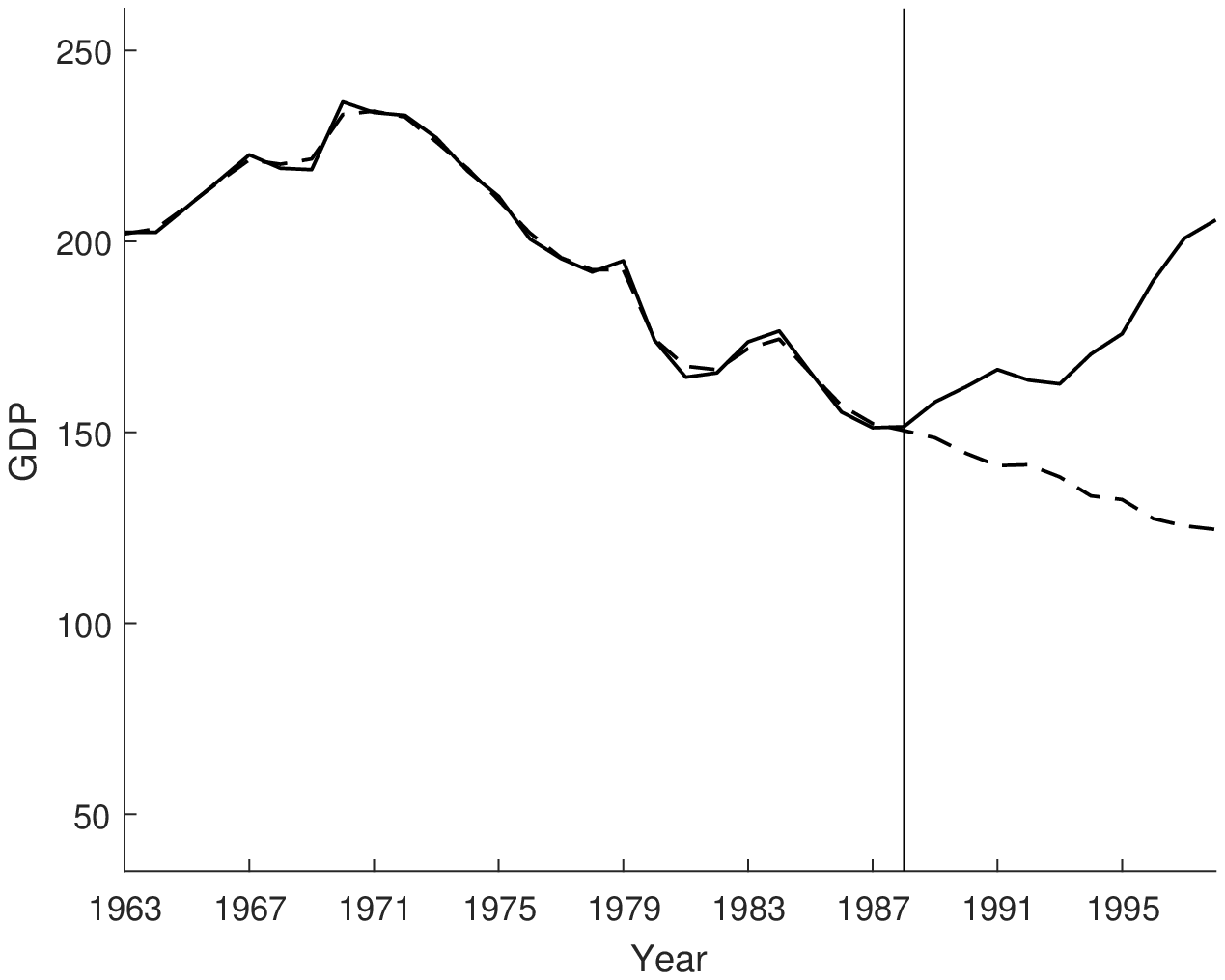}
\end{subfigure}
\medskip
\begin{subfigure}{0.25\textwidth}
\includegraphics[width=\linewidth]{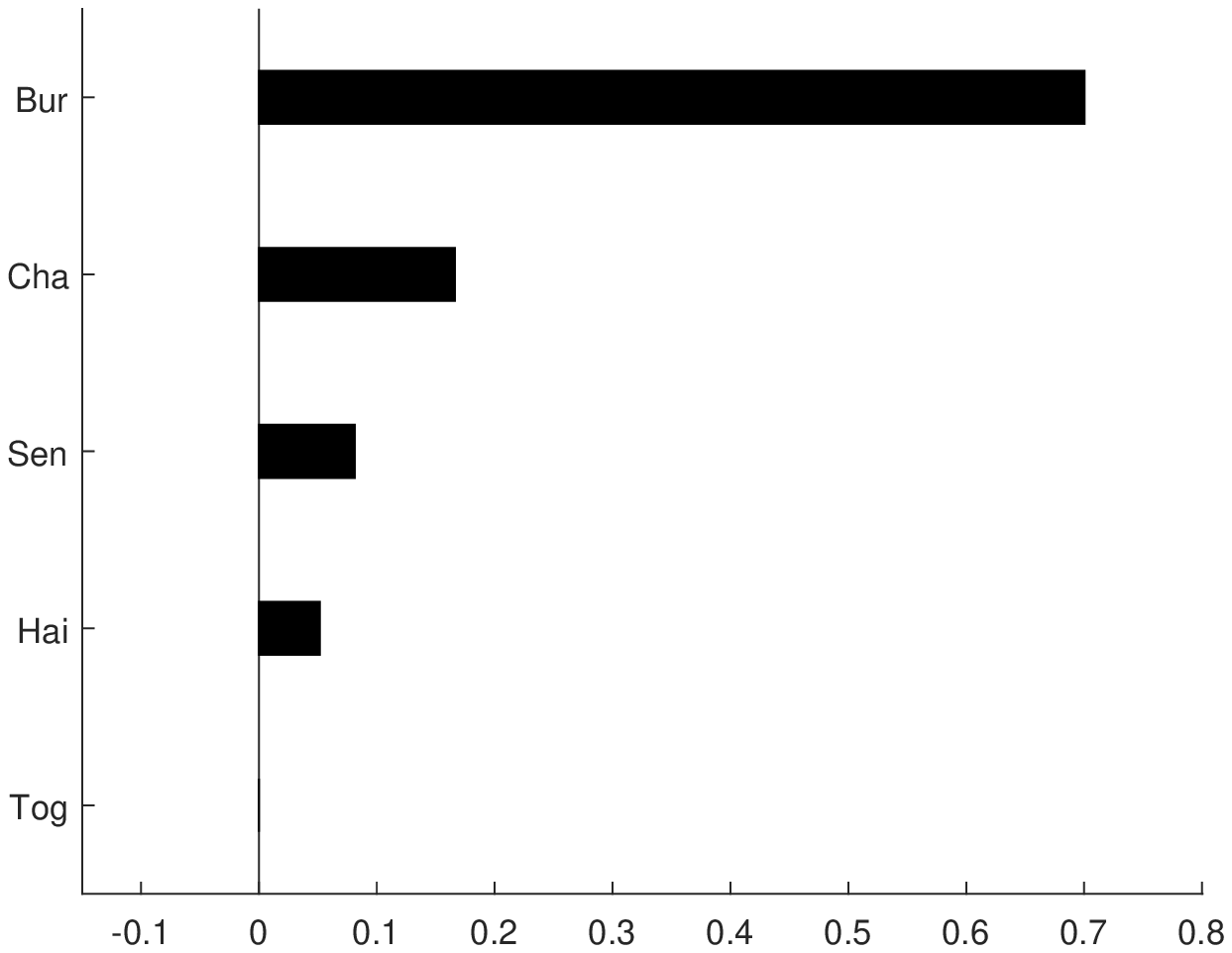}
\caption*{ADH}
\end{subfigure}\hspace*{\fill}
\begin{subfigure}{0.25\textwidth}
\includegraphics[width=\linewidth]{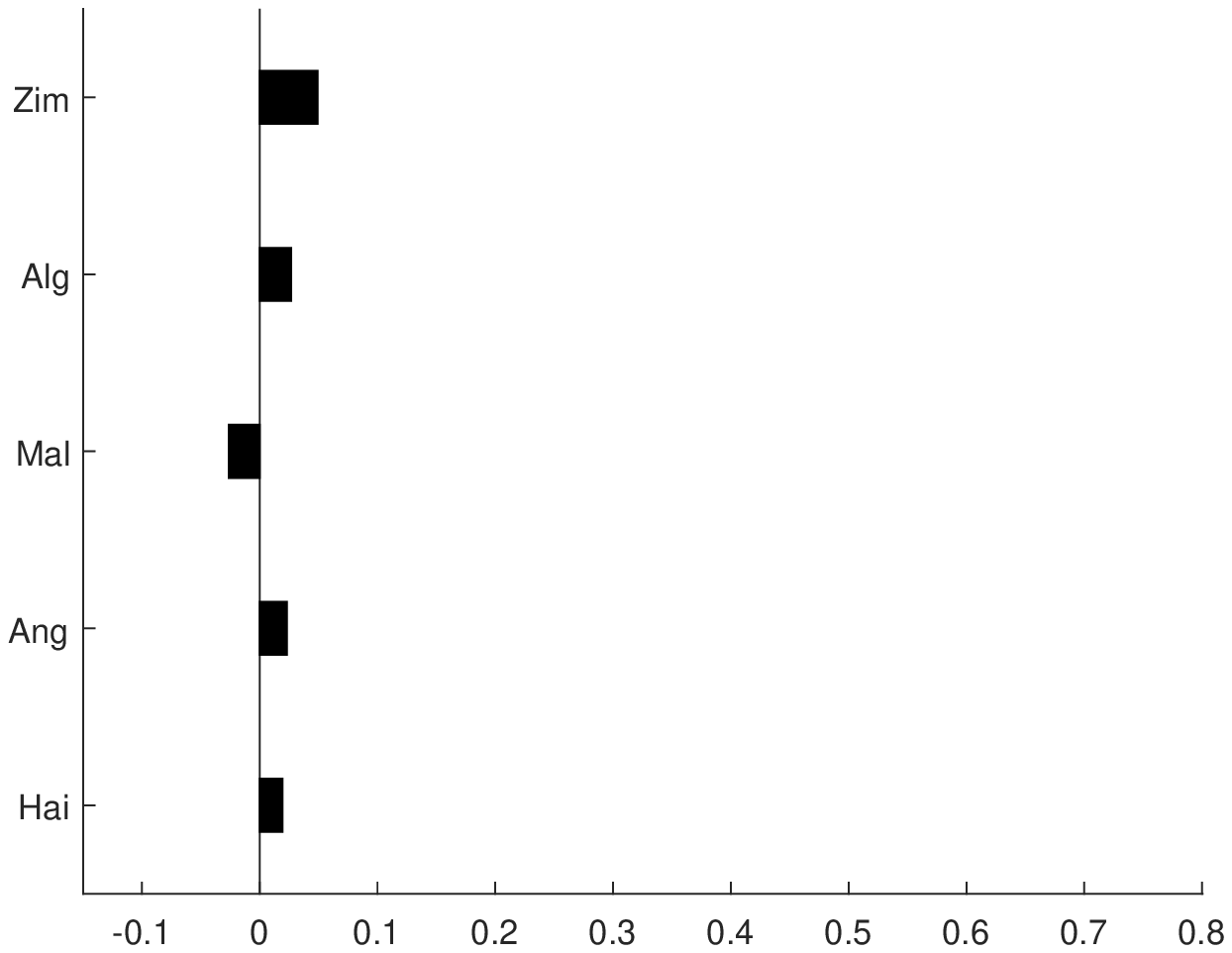}
\caption*{PCR}
\end{subfigure}\hspace*{\fill}
\begin{subfigure}{0.25\textwidth}
\includegraphics[width=\linewidth]{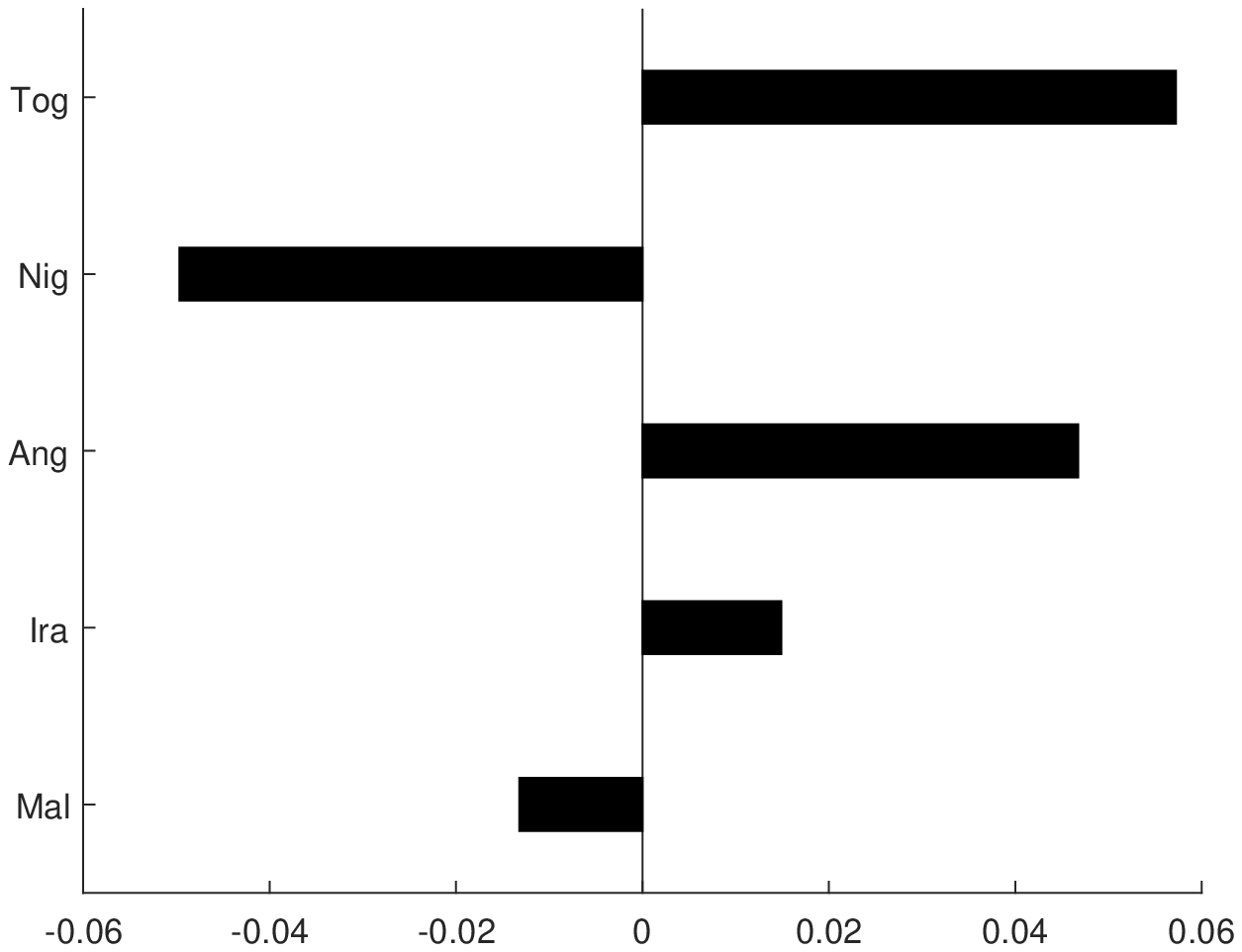}
\caption*{LASSO}
\end{subfigure}\hspace*{\fill}
\begin{subfigure}{0.25\textwidth}
\includegraphics[width=\linewidth]{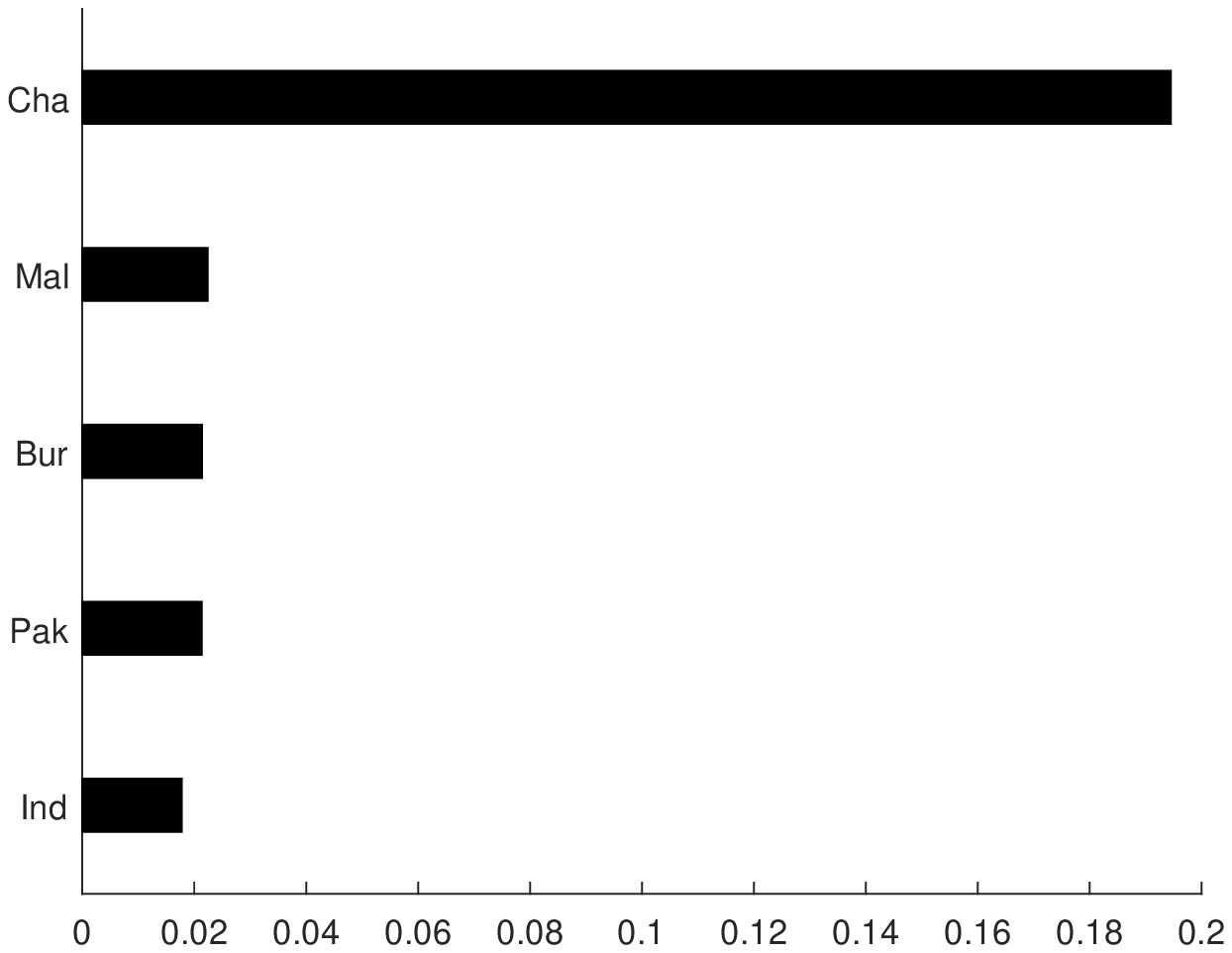}
\caption*{BSTS}
\end{subfigure}\\
\caption{\footnotesize Liberalisation of Uganda in 1988. Top row: Development in GDP per capita compared to synthetic controls estimates based on the original synthetic control (ADH), Principal Component Regression (PCR), Lasso-type regularization (LASSO) and Bayesian Structural Time Series (BSTS). The robustness test indicate large uncertainty as 7, 10, 8 and 8 of 24 control units show larger effects 10 years post liberalisation, respectively. Bottom row: 5 largest weights.} 
\label{fig_retail_app_countries_uganda}
\end{minipage}
\end{figure}

\clearpage
\begin{figure}
\centering
\begin{minipage}{1\textwidth}
\begin{subfigure}{0.25\textwidth}
\includegraphics[width=\linewidth]{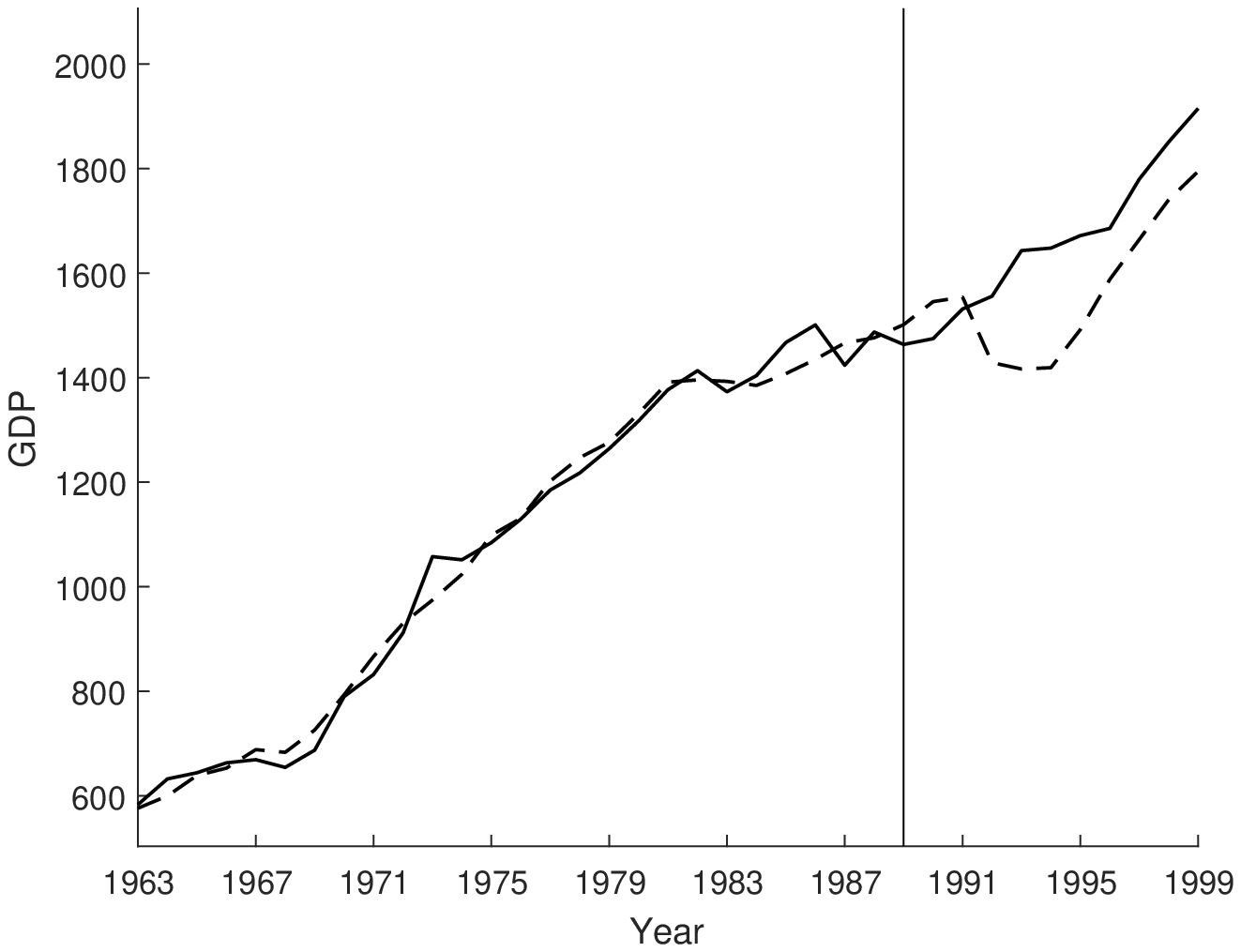}
\end{subfigure}\hspace*{\fill}
\begin{subfigure}{0.25\textwidth}
\includegraphics[width=\linewidth]{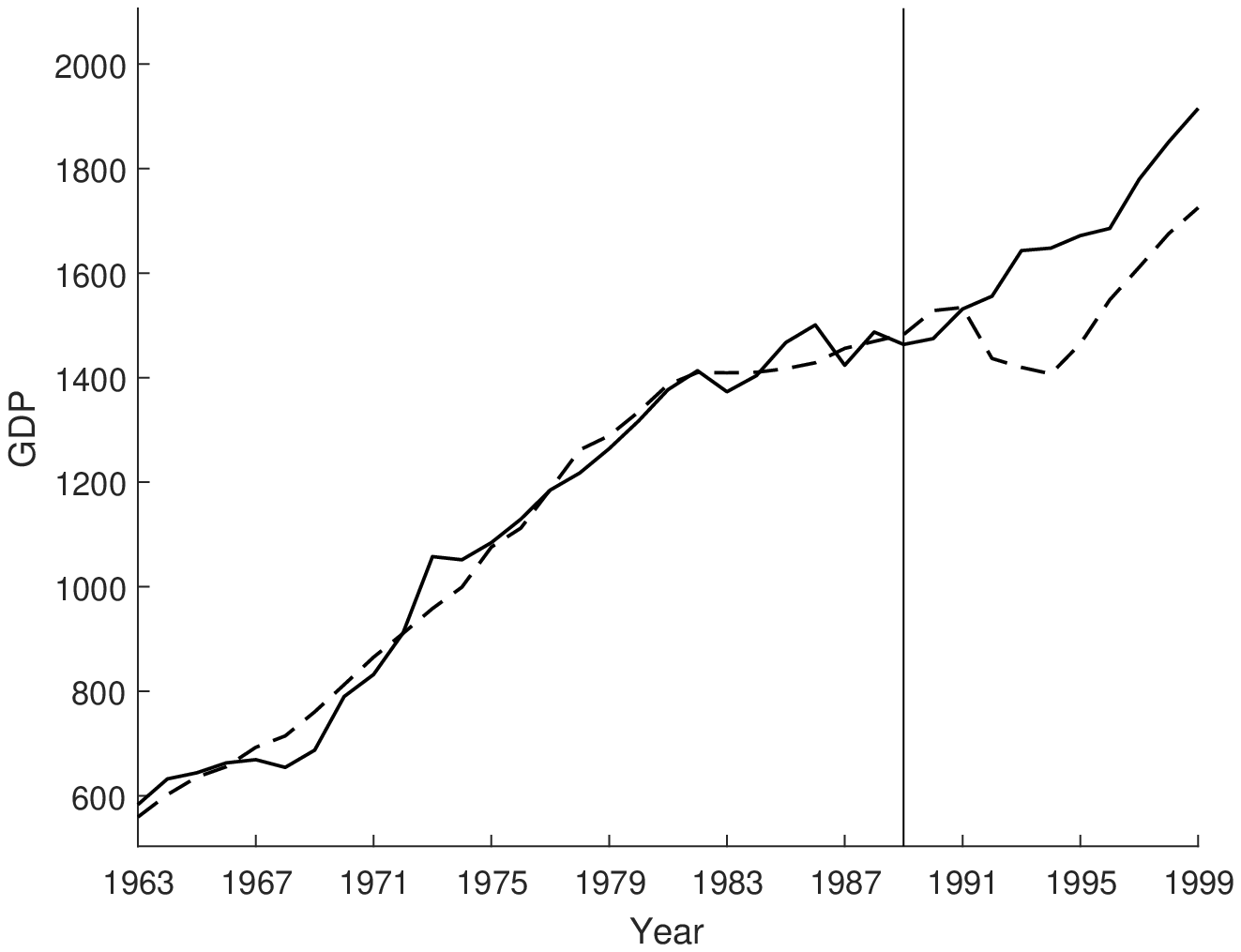}
\end{subfigure}\hspace*{\fill}
\begin{subfigure}{0.25\textwidth}
\includegraphics[width=\linewidth]{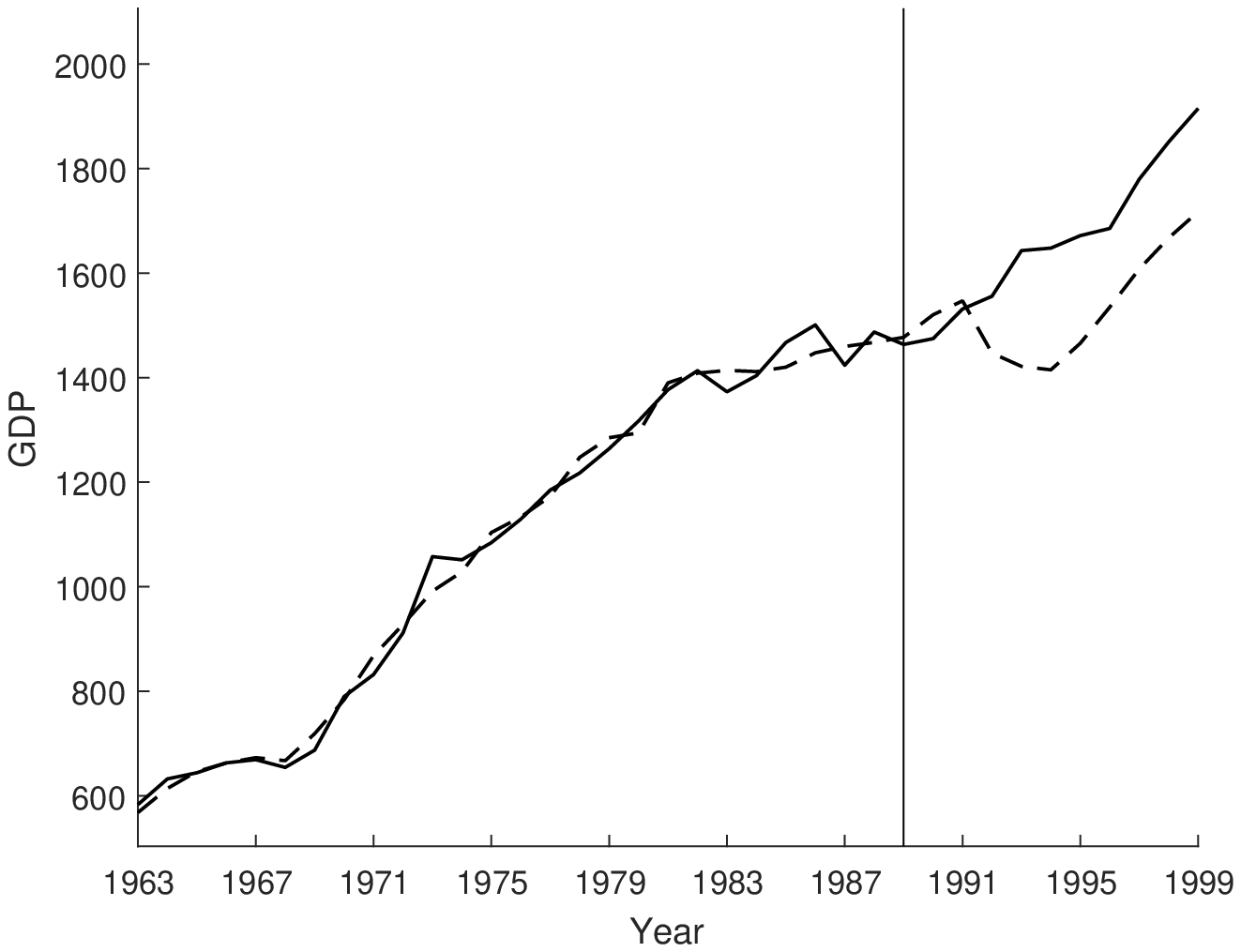}
\end{subfigure}\hspace*{\fill}
\begin{subfigure}{0.25\textwidth}
\includegraphics[width=\linewidth]{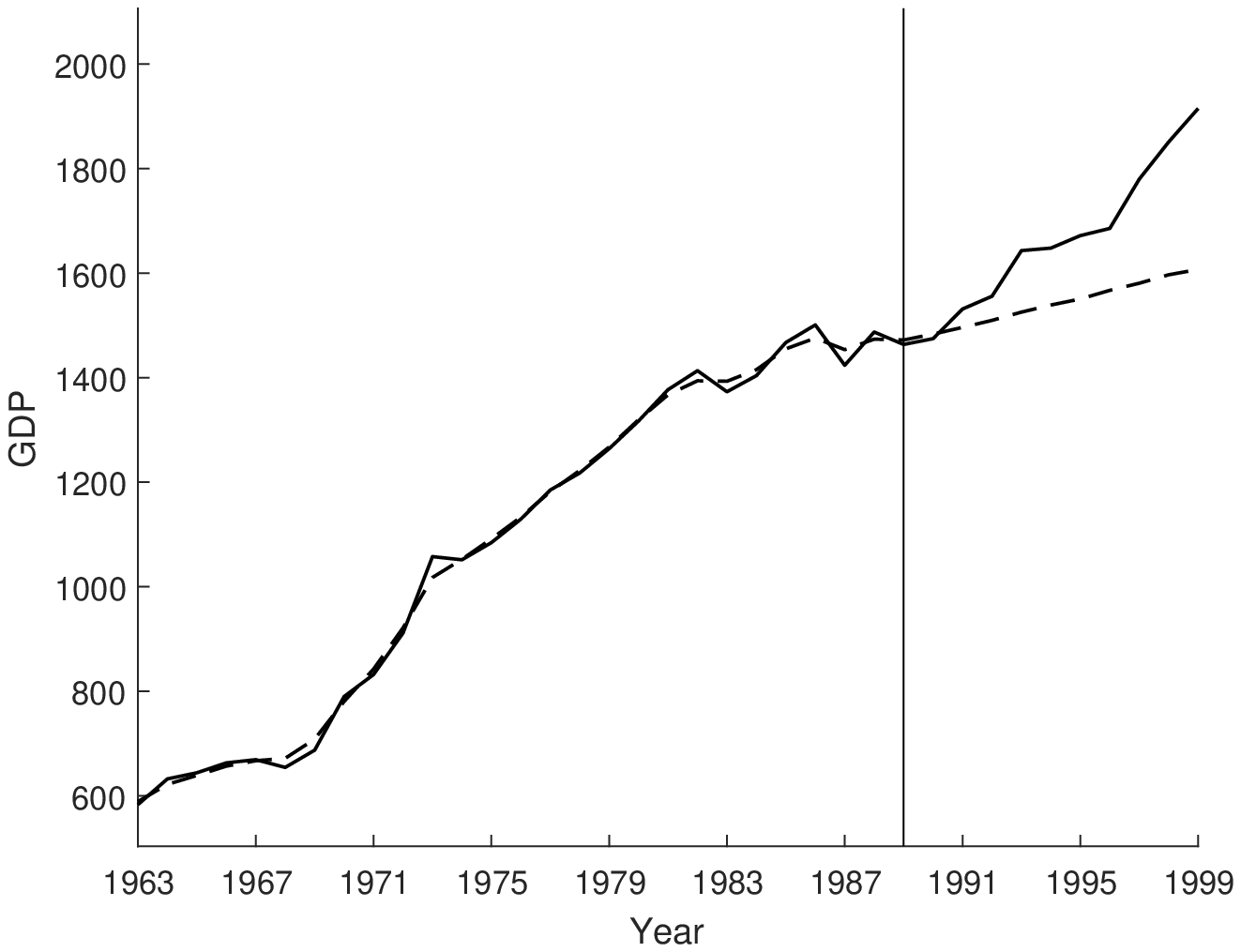}
\end{subfigure}
\medskip
\begin{subfigure}{0.25\textwidth}
\includegraphics[width=\linewidth]{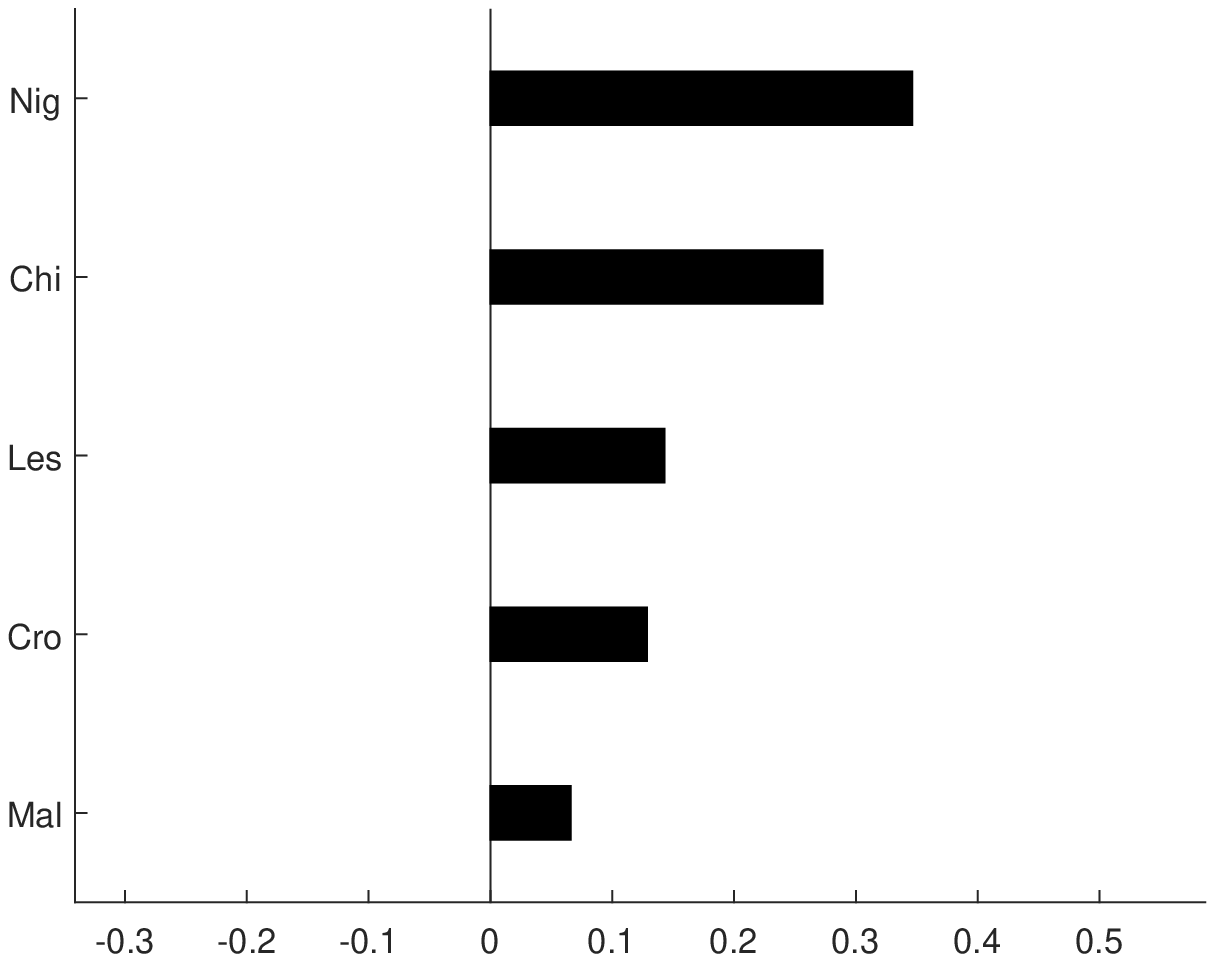}
\caption*{ADH}
\end{subfigure}\hspace*{\fill}
\begin{subfigure}{0.25\textwidth}
\includegraphics[width=\linewidth]{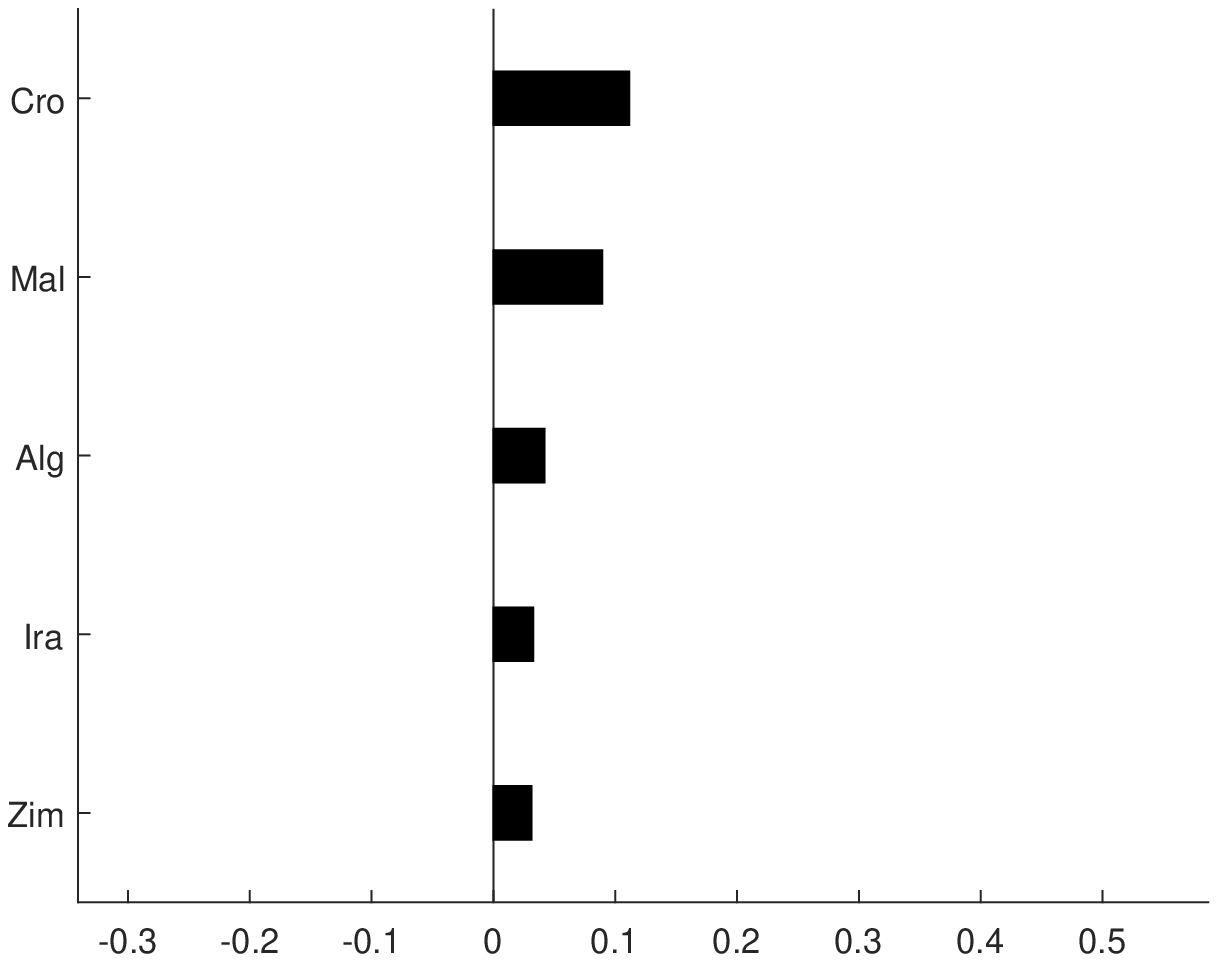}
\caption*{PCR}
\end{subfigure}\hspace*{\fill}
\begin{subfigure}{0.25\textwidth}
\includegraphics[width=\linewidth]{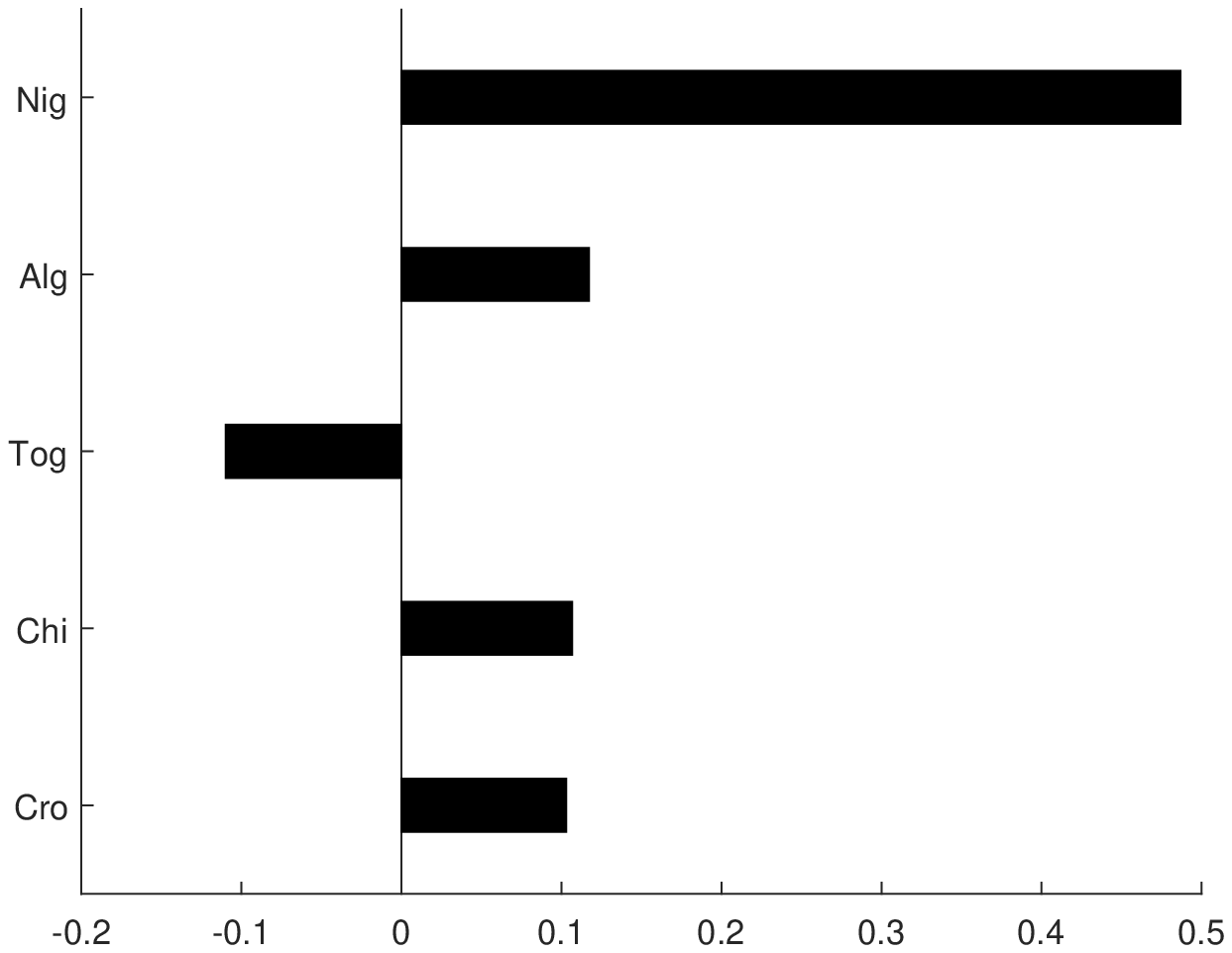}
\caption*{LASSO}
\end{subfigure}\hspace*{\fill}
\begin{subfigure}{0.25\textwidth}
\includegraphics[width=\linewidth]{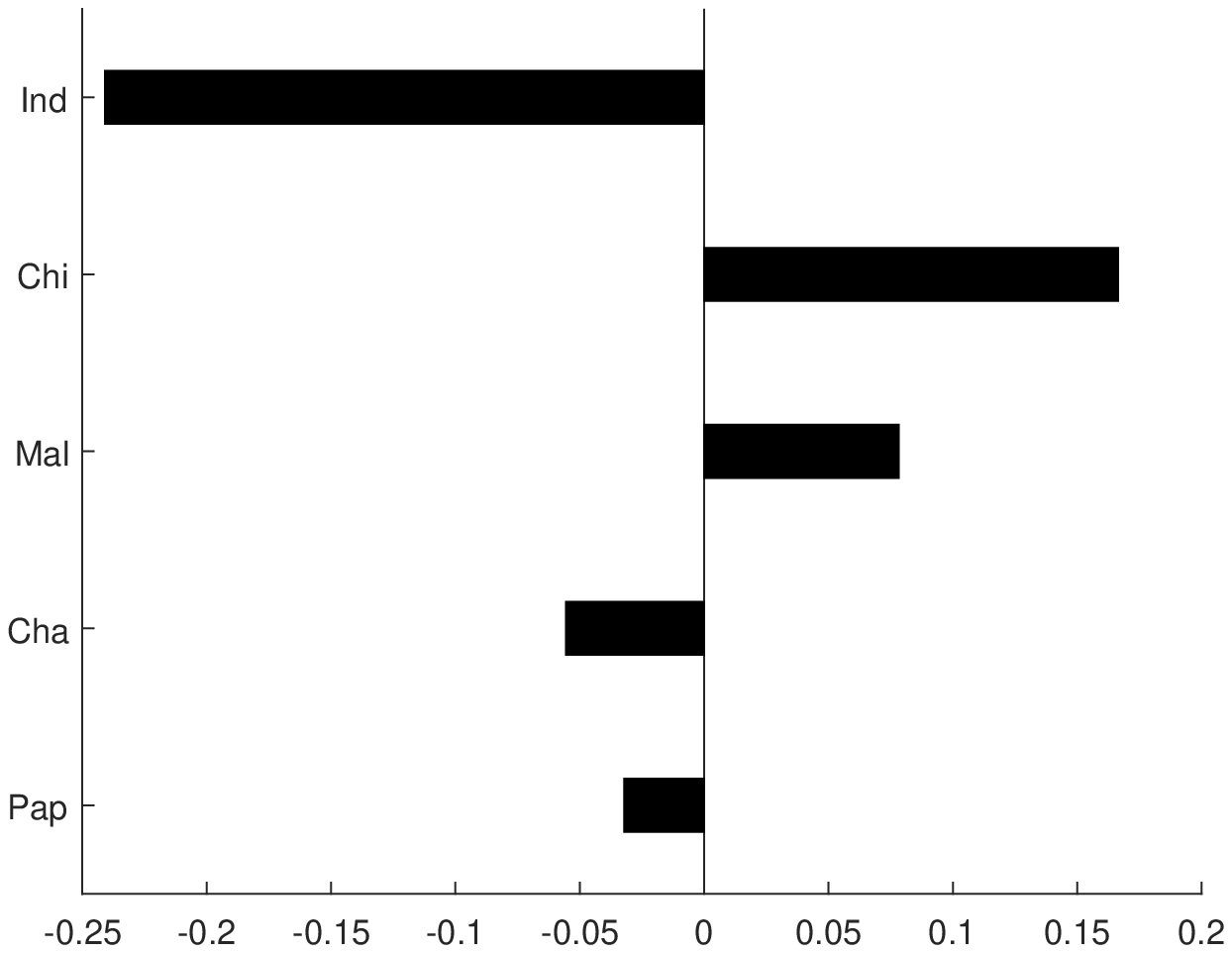}
\caption*{BSTS}
\end{subfigure}\\
\caption{\footnotesize Liberalisation of Tunisia in 1989. Top row: Development in GDP per capita compared to synthetic controls estimates based on the original synthetic control (ADH), Principal Component Regression (PCR), Lasso-type regularization (LASSO) and Bayesian Structural Time Series (BSTS). The robustness test indicate large uncertainty as 5, 9, 6 and 6 of 23 control units show larger effects 10 years post liberalisation, respectively. Bottom row: 5 largest weights.} 
\label{fig_retail_app_countries_tunisia}
\end{minipage}
\end{figure}

\clearpage
\begin{table}
\caption{\textsc{Treatment Effects and Robustness}}
\label{tab_retail_app_stores_stats}
\scriptsize
\centering
\begin{threeparttable}
\begin{tabular}{p{5cm}rrrrr}
\midrule
Description 		& MDD & PCR & ADH & LASSO & BSTS \\
\midrule
\textbf{Apple Juice}\\
Cumulative effect & 371.3 & 353.5 & 360.0 & 357.0 & 346.1\\
Standard error & 25.5 & 20.1 & 20.9 & 19.2 & 154.5\\
Placebo & 0/61 & 0/61 & 0/61 & 0/61 & 0/61\\ \\

\textbf{Yoghurt}\\
Cumulative effect & 103.2 & 39.5 & 36.1 & 67.4 & 62.8\\
Standard error & 65.1 & 42.0 & 33.2 & 32.7 & 301.5\\
Placebo & 0/38 & 0/38 & 1/38 & 0/38 & 0/38\\ \\

\textbf{Chips}\\
Cumulative effect & 3.8 & -0.7 & 3.6 & 9.8 & -12.5\\
Standard error & 30.1 & 29.8 & 27.4 & 27.2 & 63.5\\
Placebo & 0/46 & 5/46 & 6/46 & 4/46 & 8/46\\
\bottomrule
\end{tabular}
\begin{tablenotes}
\item The table presents the cumulative treatment effect following the price discount of Apple Juice, Yoghurt and Chips, respectively. The placebo study present the number of controls that yield a higher treatment effect (than the actual treated unit) of the total number of eligible control stores for each product.
\end{tablenotes}
\end{threeparttable}
\end{table}

\clearpage
\section{Bayesian Structural Time Series}\label{sec_retail_appendix_bsts}
\par The BSTS approach assumes the following specification for the likelihood of the observation equation $p(\tilde{\y}|\boldsymbol{\omega},\sigma^2)$ and the priors $p(\boldsymbol{\omega}|\sigma^2,\boldsymbol{\kappa})$, $p(\sigma^2|\boldsymbol{\kappa})$ and $p(\boldsymbol{\kappa})$, respectively
\begin{eqnarray*}
\tilde{\y}|\boldsymbol{\omega},\sigma^2 & \sim & \N(\Y\boldsymbol{\omega},\sigma^2\I)\\
\boldsymbol{\omega}|\sigma^2,\boldsymbol{\kappa} & \sim & \N(\boldsymbol{\omega}_0,\sigma^2 \mathbf{V}_0)\\
\sigma^2|\boldsymbol{\kappa} & \sim & \mathcal{G}^{-1}(s_0,r_0)\\
\boldsymbol{\kappa} & \sim & \mathcal{B}(\mathbf{q})
\end{eqnarray*}
where $\N$ denotes the normal distribution, $\mathcal{G}^{-1}$ is the inverse Gamma distribution and $\mathcal{B}$ is the Bernoulli distribution.
The objective of Bayesian variable selection is to compute the posterior
\begin{equation}
p(\boldsymbol{\kappa}|\tilde{\y}) \propto p(\tilde{\y}|\boldsymbol{\kappa})p(\boldsymbol{\kappa})
\end{equation}
The right hand side is given by the likelihood $p(\tilde{\y}|\boldsymbol{\omega},\sigma^2)$ times the joint prior of $\boldsymbol{\omega}$, $\sigma^2$ and $\boldsymbol{\kappa}$
\begin{eqnarray*}
p(\tilde{\y}|\boldsymbol{\kappa})p(\boldsymbol{\kappa}) &=& \int\int p(\tilde{\y}|\boldsymbol{\omega},\sigma^2)p(\boldsymbol{\omega},\sigma^2,\kappa)d\boldsymbol{\omega} d\sigma^2\\
&=& \int \int p(\tilde{\y}|\boldsymbol{\omega},\sigma^2) p(\boldsymbol{\omega}|\sigma^2,\boldsymbol{\kappa})d\boldsymbol{\omega}  p(\sigma^2|\boldsymbol{\kappa})d\sigma^2 p(\boldsymbol{\kappa})
\end{eqnarray*}
where the second equality follow by plugging in the joint prior defined in (\ref{eq_retail_methods_bsts_prior}) and rearranging. The objective is thus to integrate out $\boldsymbol{\omega}$ and $\sigma^2$ from the expression
\begin{equation}\label{eq_retail_appendix_methods_bsts_objective}
p(\boldsymbol{\kappa}|\tilde{\y})\propto \underbrace{\int \underbrace{\int p(\tilde{\y}|\boldsymbol{\omega},\sigma^2) p(\boldsymbol{\omega}|\sigma^2,\boldsymbol{\kappa})d\boldsymbol{\omega}}_{p(\tilde{\y}|\sigma^2,\boldsymbol{\kappa})}p(\sigma^2|\boldsymbol{\kappa})d\sigma^2}_{p(\tilde{\y}|\boldsymbol{\kappa})} p(\boldsymbol{\kappa})
\end{equation}
This can be done in three steps. First, integrate out $\boldsymbol{\omega}$ to obtain the posterior distribution of the regression coefficients $\boldsymbol{\omega}$, which is denoted by $p(\tilde{\y}|\sigma^2,\boldsymbol{\kappa})$. Second, integrate out $\sigma^2$ to obtain the posterior distribution of the observation noise parameter $\sigma^2$, which is denoted by $p(\tilde{\y}|\boldsymbol{\kappa})$. Finally, compute the posterior of the variable inclusion probabilities, $p(\boldsymbol{\kappa}|\tilde{\y}) \propto p(\tilde{\y}|\boldsymbol{\kappa})p(\boldsymbol{\kappa})$. Each step is described in detail below.

\subsection{Likelihood and prior distributions}
The probability density functions are given by
\begin{eqnarray*}
p(\tilde{\y}|\boldsymbol{\omega},\sigma^2) &=& (2\pi)^{-T_0/2}(\sigma^2)^{-T_0/2}\exp\left(-\frac{1}{2\sigma^2}(\tilde{\y}-\Y\boldsymbol{\omega})'(\tilde{\y}-\Y\boldsymbol{\omega})\right)\\
p(\boldsymbol{\omega}|\sigma^2,\boldsymbol{\kappa}) &=& (2\pi)^{-J/2}(\sigma^2)^{-J/2}|\mathbf{V}_0|^{-1/2}\exp\left(-\frac{1}{2}(\boldsymbol{\omega}-\boldsymbol{\omega}_0)'(\sigma^2\mathbf{V}_0)^{-1}(\boldsymbol{\omega}-\boldsymbol{\omega}_0)\right)\\
p(\sigma^2|\boldsymbol{\kappa}) &=& \frac{r_0^{s_0}}{\Gamma(s_0)} \left(\frac{1}{\sigma^2}\right)^{s_0+1}\exp\left(-\frac{r_0}{\sigma^2}\right)\\
p(\boldsymbol{\kappa}) &=& \prod_{j=1}^J q_j^{\kappa_j} (1-q_j)^{1-\kappa_j}
\end{eqnarray*}
Notice that the expression for the $\boldsymbol{\omega}$-prior above follow if we define $\Sigma=\text{diag}(\sigma^2)$. Then $|\Sigma\mathbf{V}_0|^{-1/2}=|\Sigma|^{-1/2}|\mathbf{V}_0|^{-1/2}=(\sigma^2)^{-J/2}|\mathbf{V}_0|^{-1/2}$

\subsection{Integrating out the regression coefficients}
Start with the integrand of the inner integral in (\ref{eq_retail_appendix_methods_bsts_objective})
\begin{eqnarray*}
p(\tilde{\y}|\boldsymbol{\omega},\sigma^2)p(\boldsymbol{\omega}|\sigma^2,\boldsymbol{\kappa}) &=& (2\pi)^{-(T_0+J)/2}|\mathbf{V}_0|^{-1/2}(\sigma^2)^{-(T_0+J)/2}\nonumber\\
&\times & \exp\left(-\frac{1}{2\sigma^2}\left\{(\tilde{\y}-\Y\boldsymbol{\omega})'(\tilde{\y}-\Y\boldsymbol{\omega}) + (\boldsymbol{\omega}-\boldsymbol{\omega}_0)'\mathbf{V}_0^{-1}(\boldsymbol{\omega}-\boldsymbol{\omega}_0)\right\}\right)\nonumber\\
&=& (2\pi)^{-T_0/2}|\mathbf{V}_0|^{-1/2}(\sigma^2)^{-T_0/2}\exp\left(-\frac{1}{2\sigma^2}\left\{c-\mathbf{m}'\mathbf{V}_1\mathbf{m}\right\}\right) \nonumber\\
&\times & (2\pi)^{-J/2}(\sigma^2)^{-J/2} \exp\left(-\frac{1}{2\sigma^2}(\boldsymbol{\omega}-\mathbf{V}_1 \mathbf{m})'\mathbf{V}_1^{-1}(\boldsymbol{\omega}-\mathbf{V}_1 \mathbf{m})\right)
\end{eqnarray*}
where we have defined $\mathbf{m} = \Y'\tilde{\y} + \mathbf{V}_0^{-1}\boldsymbol{\omega}_0$, $c=\tilde{\y}'\tilde{\y} + \boldsymbol{\omega}_0' \mathbf{V}_0^{-1}\boldsymbol{\omega}_0$ and $\mathbf{V}_1^{-1} = \Y'\Y + \mathbf{V}_0^{-1}$. The see that the second equality above must hold true, show that the term inside the bracket $\{.\}$ is equal to $c - \mathbf{m}'\mathbf{V}_1 \mathbf{m} + (\boldsymbol{\omega}-\mathbf{V}_1 \mathbf{m})'\mathbf{V}_1^{-1}(\boldsymbol{\omega}-\mathbf{V}_1 \mathbf{m})$. To see this, simplify as $c- \mathbf{m}'\mathbf{V}_1 \mathbf{m}+(\boldsymbol{\omega}'\mathbf{V}_1^{-1} - \mathbf{m}')(\boldsymbol{\omega}-\mathbf{V}_1 \mathbf{m})$ which further gives $c - \mathbf{m}'\mathbf{V}_1 \mathbf{m} + \boldsymbol{\omega}' \mathbf{V}_1^{-1}\boldsymbol{\omega} - 2 \boldsymbol{\omega}' \mathbf{m} + \mathbf{m}'\mathbf{V}_1 \mathbf{m}$ and finally $\boldsymbol{\omega}' \mathbf{V}_1^{-1}\boldsymbol{\omega} - 2 \boldsymbol{\omega}' \mathbf{m} + c$. Then plug in the definitions to get $\boldsymbol{\omega}'(\Y'\Y+\mathbf{V}_0^{-1})\boldsymbol{\omega} - 2\boldsymbol{\omega}'(\Y'\tilde{\y}+\mathbf{V}_0^{-1}\boldsymbol{\omega}_0) + \tilde{\y}'\tilde{\y} + \boldsymbol{\omega}_0'\mathbf{V}_0^{-1}\boldsymbol{\omega}_0$ which finally yields $(\tilde{\y}-\Y\boldsymbol{\omega})'(\tilde{\y}-\Y\boldsymbol{\omega}) + (\boldsymbol{\omega}-\boldsymbol{\omega}_0)\mathbf{V}_0^{-1}(\boldsymbol{\omega}-\boldsymbol{\omega}_0)$. We have thus proven the equality. To further simplify, define $\boldsymbol{\omega}_1=\mathbf{V}_1\mathbf{m}$ and $a=c-\mathbf{m}'\mathbf{V}_1 \mathbf{m}$.  Plugging in these definitions and multiplying and dividing by the determinant $|\mathbf{V}_1|^{1/2}$ gives the following
\begin{eqnarray*}
p(\tilde{\y}|\boldsymbol{\omega},\sigma^2)p(\boldsymbol{\omega}|\sigma^2,\boldsymbol{\kappa}) &=& (2\pi)^{-T_0/2}|\mathbf{V}_0|^{-1/2}|\mathbf{V}_1|^{1/2}(\sigma^2)^{-T_0/2}\exp\left(-\frac{a}{2\sigma^2}\right) \\
&\times & (2\pi)^{-J/2}(\sigma^2)^{-J/2} |\mathbf{V}_1|^{-1/2} \exp\left(-\frac{1}{2\sigma^2}(\boldsymbol{\omega}-\boldsymbol{\omega}_1)'\mathbf{V}_1^{-1}(\boldsymbol{\omega}-\boldsymbol{\omega}_1)\right)
\end{eqnarray*}
The trick of multiplying and dividing by the determinant yields the exact normal distribution of $\boldsymbol{\omega}$ with mean $\boldsymbol{\omega}_1$ and variance $\sigma^2\mathbf{V}_1$. We have thus found that for a given $\sigma^2$ and a given $\boldsymbol{\kappa}$, and after observing $\tilde{\y}$, the posterior distribution of $\boldsymbol{\omega}$ is
$$\boldsymbol{\omega}|\sigma^2,\boldsymbol{\kappa},\tilde{\y} \sim \text{N}(\boldsymbol{\omega}_1,\sigma^2\mathbf{V}_1)$$
where $\mathbf{V}_1 = (\Y'\Y + \mathbf{V}_0^{-1})^{-1}$ and $\boldsymbol{\omega}_1 = \mathbf{V}_1(\Y'\tilde{\y} + \mathbf{V}_0^{-1} \boldsymbol{\omega}_0)$. This is convenient, because integrating out $\boldsymbol{\omega}$ is now straightforward, since the integral of the above expression wrt $\boldsymbol{\omega}$ is a constant times the normal distribution of $\boldsymbol{\omega}$, which integrates to unity. We thus get the likelihood for a given $\sigma^2$ and $\boldsymbol{\kappa}$
\begin{eqnarray}\label{eq_retail_appendix_methods_bsts_posterior1}
p(\tilde{\y}|\sigma^2,\boldsymbol{\kappa}) &=& \int p(\tilde{\y}|\boldsymbol{\omega},\sigma^2)p(\boldsymbol{\omega}|\sigma^2,\boldsymbol{\kappa}) d\boldsymbol{\omega} \nonumber \\
&=& \frac{1}{(2\pi)^{n/2}} \frac{|\mathbf{V}_1|^{1/2}}{|\mathbf{V}_0|^{1/2}} \left(\frac{1}{\sigma^2}\right)^{T_0/2}\exp\left(-\frac{a}{2\sigma^2}\right)
\end{eqnarray}

\subsection{Integrating out the observation noise}
The integrand of the outer integral in (\ref{eq_retail_appendix_methods_bsts_objective}) is given by
\begin{eqnarray*}
p(\tilde{\y}|\sigma^2,\boldsymbol{\kappa})p(\sigma^2|\boldsymbol{\kappa})&=& \frac{1}{(2\pi)^{T_0/2}} \frac{|\mathbf{V}_1|^{1/2}}{|\mathbf{V}_0|^{1/2}} \left(\frac{1}{\sigma^2}\right)^{T_0/2}\exp\left(-\frac{a}{2\sigma^2}\right)
\times \frac{r_0^{s_0}}{\Gamma(s_0)} \left(\frac{1}{\sigma^2}\right)^{s_0+1}\exp\left(-\frac{r_0}{\sigma^2}\right)\\
&=& \frac{1}{(2\pi)^{T_0/2}} \frac{|\mathbf{V}_1|^{1/2}}{|\mathbf{V}_0|^{1/2}} \frac{r_0^{s_0}}{\Gamma(s_0)} \left(\frac{1}{\sigma^2}\right)^{s_0 + T_0/2 + 1}\exp\left(-\frac{r_0 + \frac{1}{2}a}{\sigma^2}\right)
\end{eqnarray*}
Define $s_1 = s_0 + \frac{1}{2}T_0$ and $r_1 = r_0 + \frac{1}{2}a$. Plugging in these definitions and multiplying and dividing by $r_1^{s_1}$ and $\Gamma(s_1)$ gives
\begin{equation}
p(\tilde{\y}|\sigma^2,\boldsymbol{\kappa})p(\sigma^2|\boldsymbol{\kappa})=\left[\frac{1}{(2\pi)^{n/2}} \frac{|\mathbf{V}_1|^{1/2}}{|\mathbf{V}_0|^{1/2}} \frac{\Gamma(s_1)}{\Gamma(s_0)} \frac{r_0^{s_0}}{r_1^{s_1}}\right]\frac{r_1^{s_1}}{\Gamma(s_1)}\left(\frac{1}{\sigma^2}\right)^{s_1 + 1}\exp\left(-\frac{r_1}{\sigma^2}\right)
\end{equation}
Again, the multiply/divide trick gives the exact inverse gamma distribution with parameters $s_1$ and $r_1$. This means that the Gamma distribution above is the posterior of $\sigma^2$ given $\boldsymbol{\kappa}$ and observations $\tilde{\y}$. In other words,
\begin{equation}
\sigma^2|\boldsymbol{\kappa},\tilde{\y} \sim \mathcal{G}^{-1}(s_1,r_1)
\end{equation}
where $s_1 = s_0 + \frac{1}{2}T_0$ and $r_1 = r_0 + \frac{1}{2}a= r_0 + \frac{1}{2}(c-\mathbf{m}'\mathbf{V}_1\mathbf{m})= r_0 + \frac{1}{2}(\tilde{\y}'\tilde{\y} + \boldsymbol{\omega}_0' \mathbf{V}_0^{-1}\boldsymbol{\omega}_0 - \boldsymbol{\omega}_1'\mathbf{V}^{-1}\boldsymbol{\omega}_1)$, where the final equality follow since $\mathbf{V}_1\mathbf{m}=\boldsymbol{\omega}_1$ and $\mathbf{m}=\mathbf{V}^{-1}\boldsymbol{\omega}_1$. Again, it is convenient to integrate out $\sigma^2$ since the integral of the above expression wrt $\sigma^2$ is a constant times the inverse gamma distribution of $\sigma^2$, which integrates to unity. We thus get the likelihood for a given $\boldsymbol{\kappa}$ from
\begin{equation}
p(\tilde{\y}|\boldsymbol{\kappa})= \int p(\tilde{\y}|\sigma^2,\boldsymbol{\kappa})p(\sigma^2|\boldsymbol{\kappa})d\sigma^2=\frac{1}{(2\pi)^{n/2}} \frac{|\mathbf{V}_1|^{1/2}}{|\mathbf{V}_0|^{1/2}} \frac{\Gamma(s_1)}{\Gamma(s_0)} \frac{r_0^{s_0}}{r_1^{s_1}}
\end{equation}

\subsection{Deriving the variable inclusion posterior}
The posterior of $\boldsymbol{\kappa}$ given the observations $\tilde{\y}$ is given from Bayes rule as
$$p(\boldsymbol{\kappa}|\tilde{\y}) = \frac{p(\boldsymbol{\kappa},\tilde{\y})}{p(\tilde{\y})} = \frac{p(\tilde{\y}|\boldsymbol{\kappa})p(\boldsymbol{\kappa})}{p(\tilde{\y})}$$
We do not have the closed form expression for $p(\tilde{\y})$, i.e. we cannot integrate out $\boldsymbol{\kappa}$ from $p(\tilde{\y}|\boldsymbol{\kappa})$ analytically. We thus use $p(\boldsymbol{\kappa}|\tilde{\y}) \propto p(\tilde{\y}|\boldsymbol{\kappa})p(\boldsymbol{\kappa})$. Plugging in the distributions give
$$p(\boldsymbol{\kappa}|\tilde{\y}) \propto \frac{1}{(2\pi)^{T_0/2}} \frac{|\mathbf{V}_1|^{1/2}}{|\mathbf{V}_0|^{1/2}} \frac{\Gamma(s_1)}{\Gamma(s_0)} \frac{r_0^{s_0}}{r_1^{s_1}} \prod_{j=1}^J q_j^{\kappa_j} (1-q_j)^{1-\kappa_j}$$
This is the final posterior shown in (\ref{eq_retail_methods_bsts_posterior_idx}). As an illustration of how to implement the above framework, let $\boldsymbol{\kappa}_{-j}$ denote all values of $\boldsymbol{\kappa}$ except $\kappa_j$, i.e. $\boldsymbol{\kappa}_{-j}=(\kappa_1,\hdots,\kappa_{j-1},\kappa_{j+1},\hdots,\kappa_J)$. Taking the ratio of the values for $\kappa_j=1$ and $\kappa_j=0$ removes the normalizing constant. Thus we can calculate the posterior probability of some variable being included in the model as
$$\text{Prob}(\kappa_j=1,\boldsymbol{\kappa}_{-j}|\tilde{\y}) = \frac{p(\kappa_j=1,\boldsymbol{\kappa}_{-j}|\tilde{\y})}{p(\kappa_j=1,\boldsymbol{\kappa}_{-j}|\tilde{\y})+p(\kappa_j=0,\boldsymbol{\kappa}_{-j}|\tilde{\y})}$$
In practice the log-likelihood is computed and then converted back before computing the inclusion probability. The log-likelihood is given as follows. Assume that every covariate has the same prior probability of being included or excluded from the model, i.e. $q_j=0.5$ for $j=1,\hdots,J$. The Bernoulli distribution simplifies to $\prod_{j=1}^J 0.5^{\kappa_j}0.5^{1-\kappa_j}=0.5^J$. The log-likelihood is
\begin{eqnarray*}
\text{loglik} &=& -\frac{T_0}{2} \log 2\pi + \frac{1}{2}\log |\mathbf{V}_1| - \frac{1}{2}\log |\mathbf{V}_0| + \log \Gamma(s_1) - \log \Gamma(s_0)\\
&+& s_0 \log r_0 - s_1 \log r_1 + J \log 0.5
\end{eqnarray*} 

\end{appendices}

\end{document}